\documentclass[11pt]{article}
\usepackage[a4paper,margin=2.5cm,bottom=3cm]{geometry}

\usepackage[utf8]{inputenc}
\usepackage{amsmath}
\usepackage{siunitx}
\usepackage[export]{adjustbox}
\usepackage{lipsum}
\usepackage{xcolor}
\definecolor{darkred}{rgb}{0.5,0,0}
\definecolor{darkblue}{rgb}{0,0,0.5}
\definecolor{firebrick}{rgb}{0.75,0.125,0.125}
\definecolor{darkgreen}{rgb}{0,0.5,0}
\usepackage[colorlinks=true,linkcolor=firebrick,citecolor=darkgreen,urlcolor=darkblue]{hyperref}

\title{Chapter 6: \\ Radio Detection of High Energy Neutrinos in Ice}

\vspace{10cm}

\author{
  S. Barwick\\
  Department of Physics and Astronomy,\\
University of California,\\
Irvine, CA 92617, USA \\
sbarwick@uci.edu
  \and
  C. Glaser\\
  Department of Physics and Astronomy,\\
Uppsala University,\\
Uppsala, Sweden, \\
christian.glaser@physics.uu.se
}
\begin{document}

\begin{titlepage}
% \maketitle
% \blfootnote{To be published in Neutrino Physics and Astrophysics, edited by F. W. Stecker, in the Encyclopedia of Cosmology II, edited by G. G. Fazio, World Scientific Publishing Company, Singapore, 2022.}

\centering 
\scshape 
\vspace*{\baselineskip} 

\rule{\textwidth}{1.6pt}\vspace*{-\baselineskip}\vspace*{2pt} 
\rule{\textwidth}{0.4pt} 

\vspace{0.75\baselineskip} 

{\LARGE Chapter 6: \\ Radio Detection of High Energy Neutrinos in Ice\\} % Title

\vspace{0.75\baselineskip} 

\rule{\textwidth}{0.4pt}\vspace*{-\baselineskip}\vspace{3.2pt} 
\rule{\textwidth}{1.6pt}

\vspace{2\baselineskip} 

\upshape
To be published in \textsc{Neutrino Physics and Astrophysics}, edited by F. W. Stecker,\\ in the \textsc{Encyclopedia of Cosmology II}, edited by G. G. Fazio, \\World Scientific Publishing Company, Singapore, 2022. 
	
\vspace*{3\baselineskip}

	\vspace{0.5\baselineskip} 
	
	\begin{minipage}{0.45\textwidth}
		\begin{centering}
% 			\large
{\Large \textsc{Steven~W. Barwick}}\\
Department of Physics and Astronomy\\
University of California\\
Irvine, CA 92617, USA \\
sbarwick@uci.edu\\
		\end{centering}
	\end{minipage}
	~
	\begin{minipage}{0.45\textwidth}
		\begin{centering}
% 			\large
{\Large \textsc{Christian Glaser}}\\
Department of Physics and Astronomy\\
Uppsala University\\
Uppsala, Sweden \\
christian.glaser@physics.uu.se\\
		\end{centering}
	\end{minipage}
	
% 	{\scshape\Large Steven Barwick \\ Christian Glaser \\} % Editor list
	
	\vspace{0.5\baselineskip}

% \markboth{Barwick and Glaser}{Radio Detection of High Energy Neutrinos}

% \author[S. Barwick and C. Glaser]{Steven W. Barwick }
%\index[aindx]{Author, F.} % or \aindx{Author, F.}
%\index[aindx]{Author, S.} % or \aindx{Author, S.}

% \address{Department of Physics and Astronomy,\\
% University of California,\\
% Irvine, CA 92617, USA \\
% sbarwick@uci.edu}

% \author{Christian Glaser}

% \address{
% Department of Physics and Astronomy,\\
% Uppsala University,\\
% Uppsala, Sweden, \\
% christian.glaser@physics.uu.se }

\vspace{1cm}

\begin{abstract}
\large
Radio-based detection of high-energy particles is growing in maturity. In this chapter, we focus on the detection of neutrinos with energies in excess of 10 PeV that interact in the thick, radio-transparent ice found in the polar regions. High-energy neutrinos interacting in the ice generate short duration, radio-frequency flashes through the Askaryan effect that can be measured with antennas installed at shallow depths. The abundant target material and the long attenuation lengths of around \SI{1}{km} allow cost-effective instrumentation of huge volumes with a sparse array of radio detector stations. This detector architecture provides sufficient sensitivity to the low flux of ultra-high-energy neutrinos to probe the production of ultra-high-energy cosmic rays whose origin is one of the longest-standing riddles in astroparticle physics. 
We describe the signal characteristics, propagation effects, detector setup, suitable detection sites, and background processes. We give an overview of the current experimental landscape and an outlook into the future where almost the entire sky can be viewed by a judicious choice of detector locations. 
\end{abstract}

\end{titlepage}

%\markboth{Even Page Header}{Odd Page Header} % Customized running heads
\markboth{Barwick and Glaser}{Radio Detection of High Energy Neutrinos}

% \body

\newpage
\setcounter{tocdepth}{2}
\tableofcontents

\newpage

\section{Introduction and historical review of the development of radio detection of high energy particles}

A radically new kind of astronomy emerged in the decade between 2010 and 2020.  For the first time, telescopes were able to observe gravitational waves and high-energy neutrinos \cite{PhysRevLett.116.061102, IceCube2013-PeVNu}, both of which provide unique information (and with luck, insight) on the complex mechanisms that create the high-energy universe. The emergence of these new technologies creates opportunities to view astronomical sources with an integrative suite of tools that are sensitive to far more of the broadcast information than before. The term "multi-messenger astronomy" has been coined to describe the nascent process of complementing electromagnetic and cosmic-ray observations with the detection of neutrinos and gravitational waves.  

The emergence of multi-messenger astronomy relied on the painstaking development of the underlying technologies in prior decades.  In particular, the success of the high energy neutrino facility IceCube \cite{IceCube2016} began with AMANDA at the same South Pole location, which established the feasibility of deep ice drilling, the operational reliability of electronic and sensor systems, calibration procedures, and perhaps most critically, a solid understanding of background signatures, which vastly outnumbered the expected number of signal events. As a consequence,  simulation tools were vetted with data, and the efficacy of a particular analysis approach could be extrapolated to larger detector volumes with confidence.

It was soon realized that the great potential of neutrino astronomy could be expanded by increasing the sensitivity of neutrino detectors at the most extreme energies, called the extreme high energy regime (EHE). However, this was not going to be easy. At these energies, the flux of neutrinos decreases dramatically, which necessitates the development of new technologies to complement the optical techniques utilized by IceCube and others. The challenge:  increase the detection volume by a factor of 10 over the current state of the art,  while reducing the cost by a factor of 10 at the targeted energies.  The astrophysical community rapidly converged on developing techniques that depended on a relatively obscure effect that generates coherent radio pulses as neutrinos interact in dense materials.  

In this article, we describe the physical mechanism responsible for the radio emission, the propagation through natural ice, and the development of radio-based detectors for extreme energy neutrinos, highlighting the similarities and differences relative to the successful road map created by optical neutrino telescopes.

\subsection{A brief introduction of in-ice neutrino detection}
We start by giving a brief introduction of ultra-high energy neutrino detection with an in-ice radio detector. The intent is to provide a high-level overview of the important physics and concepts at the start of this chapter before we discuss each aspect in more detail in later sections. The detection concept can be divided into four modular steps.
\subsubsection{The neutrino interaction}
Let's set the stage for detector design by noting a few germane neutrino properties. Because neutrinos only interact via the weak force, we can't measure neutrinos directly. When Pauli first postulated the neutrino he even thought that neutrinos will never be detectable. Luckily he was proven wrong and we can detect neutrinos by looking for the particles that are produced in a neutrino interaction. The probability that such a neutrino interacts with the medium it passes through (i.e. its cross-section) is small but is proportional to the amount of matter it passes through. Thus, if only a large enough amount of matter is instrumented a few neutrinos will interact and we can observe the products of this interaction. A dense medium, such as ice, is obviously preferred compared for example to air which is a thousand times thinner and would require a thousand times larger volume to yield the same number of neutrino interactions. 

When a high-energy neutrino interacts in the ice it creates a cascade of millions of secondary particles which is often referred to as particle shower. These showers consist almost exclusively of photons, electrons and positrons. Comparing the neutrino energy of interest of around \SI{e18}{eV} with the rest mass energy of an electron or positron of \SI{511}{keV} reveals that about a trillion of electrons-positron pairs could be created if all kinetic energy of the neutrino is transferred into the creation of new particles, i.e., the particle shower. 

\subsubsection{Radio emission from in-ice showers}
When a particle shower develops in a dielectric medium, such as ice, it develops a charge asymmetry. As the cascade develops, the number of electrons increasingly exceeds the number of positrons. This time-varying negative charge excess generates coherent radiation in the radio frequency range. The radiation is typically referred to as \emph{charge-excess radiation} or \emph{Askaryan radiation} as it was postulated by Gurgen Askaryan in 1962 \cite{Askaryan1962}.

\begin{figure}[t]
    \centering
    \includegraphics[width=\textwidth]{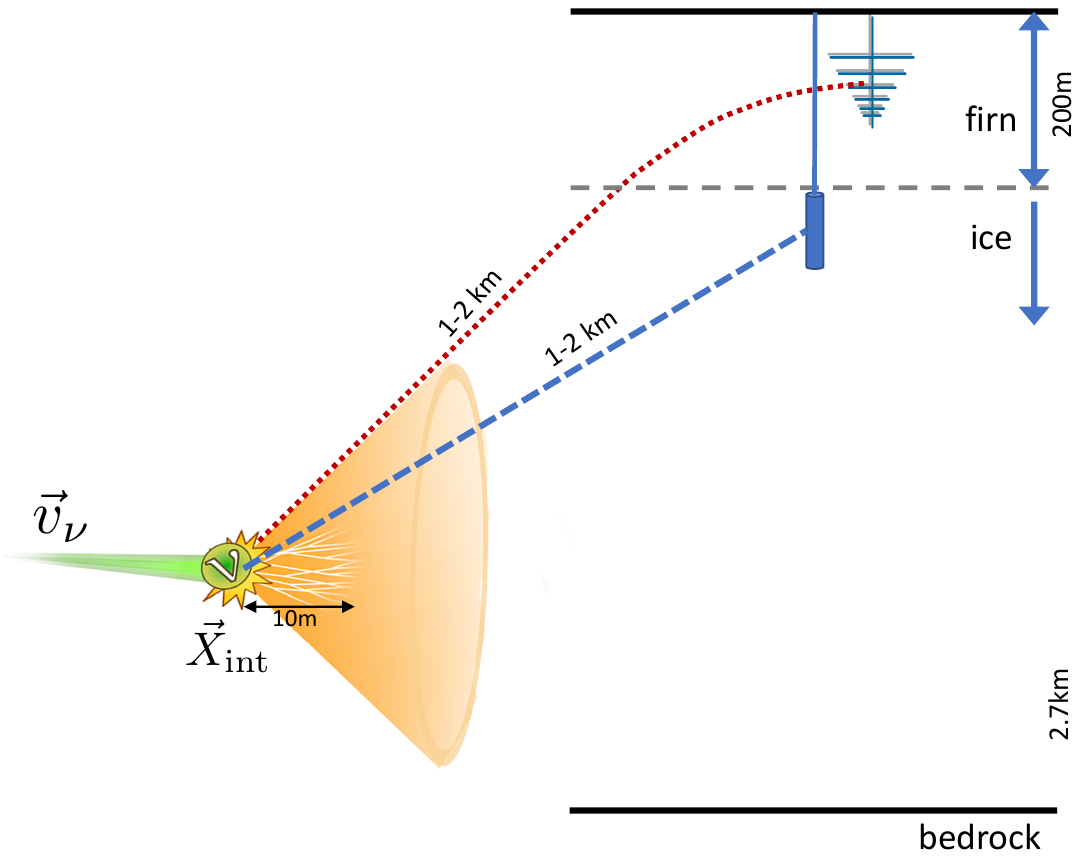}
    \caption{Sketch of Askaryan radiation in ice. The dimensions are shown for a typical case at the South Pole.}
    \label{fig:sketch_askaryan_radiation}
\end{figure}

The Askaryan radiation exhibits a characteristic feature of being emitted on a cone around the shower direction as illustrated in Fig.~\ref{fig:sketch_askaryan_radiation}. The shower propagates with the vacuum speed-of-light $c_0$ whereas the radio signal propagates at a slower speed-of-light of $c_0/n$ where the index-of-refraction of deep ice is $n = 1.78$. This leads to constructive interference of all emission points along the $\mathcal{O}(\SI{10}{m})$ long shower track if the shower is observed at the \emph{Cherenkov} angle of \SI{56}{\degree}. The further one goes away from the Cherenkov cone, the weaker the emission gets. The typical width of the Cherenkov cone is a few degrees. 

\subsubsection{Propagation of radio waves in ice}

As described earlier, the current limits on the flux of neutrinos is so low, and the interaction cross-section is so small, that the active volume of neutrino telescopes must encompass hundreds to thousands of cubic kilometers of dielectric material. To keep costs under control, the material must be free (though access to the material, as we will see, is far from free). 

While the neutrino interaction probability benefits from instrumenting a medium with a density as large as possible, this is not the entire story. The radio signal must propagate from the interaction site to the radio receiver. Radio power is absorbed quickly by a small admixture of liquid water, so the natural medium should be dry, such as salt domes, or better yet, frozen to a solid state.  

The main reason for using cold ice as the detector medium is its transparency to radio waves, as measured by the frequency-dependent attenuation length, which is the distance over which the signal amplitude of the electric field is reduced by $1/e$. It depends mostly on the temperature and purity of the medium. The colder the medium, the larger the attenuation length.  The maximum value on earth has been observed for ice near the surface of the South Pole in Antarctica, reaching \SI{2}{km} \cite{Barwick2005-PoleAtten, ARA-LongBaseline2020}. This is an order-of-magnitude improvement over the scattering and absorption length of optical \emph{Cherenkov} light \cite{IceCube-IceAttenuation} of $\mathcal{O}(\SI{100}{m})$  and the reason why large volumes can be instrumented relatively sparsely, increasing the cost-efficiency. Furthermore, the scattering of radio waves is negligible which allows a single compact detector station more than a kilometer away to still determine the direction and energy of the neutrino. 

The sea-level site at Moore's Bay, located near the coast of Antarctica, has shorter attenuation lengths due to warmer ice.  However, at \SI{500}{m}, it is still substantially larger than optical, and this site has a unique reflective surface at the bottom of the ice due to the underlying ocean water. A third site that has been considered by the community is on the ice plateau in Greenland, where attenuation lengths are between the South Pole and Moore's Bay. 

The attenuation length tends to decrease with increasing depth due to the fact that the ice temperatures tend to rise with increasing depth. The reason for this feature - heat from radioactive elements in the earth's core is trapped by the ice overburden, which acts as an insulating thermal blanket. The poor thermal conductivity of cold ice generates a temperature gradient between the land or ocean beneath the ice and the air at the surface.  This feature suggests that instruments can remain near the surface, which simplifies the installation procedures. 

To summarize, the excellent transparency of cold, polar ice to radio frequency radiation allows Askaryan signals to propagate kilometers without significant reduction. Consequently, a much sparser array of radio detector stations is required compared to optical detectors.

What is the optimal depth of a radio sensor?  This turns out to have a rather complicated answer.  We start by noting that, unlike optical techniques, it is possible to install the sensors within a few meters of the surface, which has obvious advantages in terms of installation and maintenance. The sensors of in-ice optical Cherenkov detectors are inserted at depths below \SI{1500}{m} in the Antarctic ice to (1) avoid air bubbles that scatter the optical photons, and (2) reduce the generation of background light by muons produced by cosmic-ray interactions in the atmosphere. Neither of these two issues is relevant for radio-based neutrino detectors, so there is no requirement to deploy radio receivers at great depth. The diameter of air bubbles is much smaller than the wavelengths of the radio pulses, so scattering by air bubbles is not large and is generally ignored.   Also, at the relevant energies (E \textgreater $10^{16}$ eV), the small flux of cosmic rays generates less muon background  (though not negligible, as discussed in Section \ref{sec:Background}). While there are obvious cost, deployment and maintenance advantages to near-surface neutrino stations, there is one important reason to consider installing radio sensors several hundred meters below the surface, which we discuss next.

All natural occurring ice sheets exhibit the property that their density increases as a function of increasing depth in their upper $\mathcal{O}(\SI{100}{m})$, known as the firn layer. The firn layer is made of granular snow that has not yet been compressed into ice. It is created by the compaction of low-density snow that falls onto the surface or that is blown onto it from another location by the strong winds. As the snow overburden increases with time, its weight will slowly compress the snow, with its density asymptotically reaching that of pure water ice. 

To a good approximation, the firn density is linearly related to the index-of-refraction of the medium, $n_i$. The density decreases near the surface of the glacial sheets due to trapped air within the ice matrix. The air is mixed homogeneously with pure ice on length scales small compared to the wavelengths of the radio emission so the propagation velocity of the electromagnetic pulse is linearly related to the average optical pathlength (and therefore, the density). Within a few meters of the firn-air surface,  $n_\mathrm{i} = 1.35$. As a consequence of the nearly continuous change in the index of refraction in the firn layer, for a range of upward propagation directions, radio signals will roll over and propagate downwards. This limits the observable volume for receivers in the firn as signals emitted from certain positions cannot reach the receiver as the signal trajectory is bent downwards. Receivers can be placed below the firn to alleviate this effect,  which increases the neutrino sensitivity per detector station but also increases both the installation challenges and hardware costs of the station.

The recent discovery that the radio signal may propagate into "forbidden" regions of firn \cite{Barwick2018} provides a reason to consider a near-surface location (i.e., within about 20 meters of the surface) for a radio neutrino station.  It is related to the physics of annual snow accumulation, which creates layers that deviate from a perfectly continuous transition in the firn density.  The subtle variation has been shown experimentally and in simulation \cite{Deaconu2018} to alter the propagation of radio pulses so that signal may arrive from the classically forbidden region (if one naively assumes a monotonically changing density in the firn ice), expanding the observable volume of ice. These additional signals are distorted in time and attenuated to about 1\% of the field amplitude. It is possible that future detectors can exploit this feature. For example, relatively nearby, high-energy neutrinos may produce electric fields that are too large to be well measured by the data acquisition system.  However, the attenuated signals originating in the shadow zone may fall within the operating parameters, thereby increasing the effective volume of the detector.

\subsubsection{Measurement of radio pulses}
Finally, the radio pulses need to be measured which is achieved by inserting antennas into the ice. The best antenna type for this purpose is Log Periodic Dipole Antennas (LPDA antennas) which consists of several dipoles of varying length which are separated such that the individual signals interfere constructively in the waveguide. This gives the antenna a directional gain and a broadband frequency response. Near surface, nearly 2D arrays use this feature to help reject downward traveling background signals since neutrino signals propagate upward from the vertex in the bulk ice to the surface.  The disadvantage is that these antennas are fairly big and can only be installed close to the surface. 

The other type of antennas typically used are bicone and slot antennas which fit into a narrow borehole of $\mathcal{O}(\SI{20}{cm})$ diameter and thus allow for an installation deeper in the ice. The bicone antennas are sensitive to vertically polarized signals and the slot antennas are sensitive to vertically polarized signals. These antennas are symmetrically sensitive in zenith angle, so background rejection is accomplished by measuring propagation time in a 3D array. Another strategy to reject downgoing background signals, as discussed in section \ref{sec:Background}, requires precise relative timing by a vertically oriented linear array of closely spaced dipoles to measure the propagation direction of the RF signal.  The slot antenna type has the smallest gain,  which follows our intuition that it is difficult to design an antenna that mostly extends along the vertical dimension to observe signals that are horizontally polarized. 

\subsection{Overview of detection sites}
The main requirement for a detector site is a lot of cold ice. Three suitable sites have been established around the world. The ice sheet underneath the South Pole, which is already home to the optical IceCube neutrino detector \cite{IceCube2016}, is \SI{2.7}{km} deep and has attenuation lengths of more than a kilometer, the longest of all suitable sites. The South Pole is place of a large research station that provides excellent logistical support. The RICE \cite{RICE2003-Performance} experiment was the first to deploy a radio neutrino detector at this site, though co-located with the AMANDA optical detector. AMANDA was the predecessor of the IceCube high energy neutrino detector.  Due to noise generated by the photomultiplier tubes and distortions created by the long metal cables, the RICE configuration was not optimal.  More recently, the ARA and ARIANNA collaborations have operated detectors at locations several kilometers or more from the research station at the South Pole to reduce the contributions of radio noise by ongoing human activity and other operating experiments at or near the Amundsen-Scott research station.

An alternative site in central Greenland provides similar properties as the South Pole location with an ice sheet of \SI{3}{km} thickness and attenuation lengths of around \SI{1}{km}. The Summit research station \cite{SummitGreenland}, although much smaller than the research station at the South Pole, can provide logistical support for constructing and running a neutrino detector. 

The third established site is Moore's Bay on the Ross Ice Shelf close to the coast of Antarctica. The thickness of the ice shelf is \SI{576}{m} and the attenuation lengths are around \SI{450}{m} due to the warmer ice temperatures \cite{ARIANNA-Atten}. Though the vertical depth and radio transparency of ice is less than at other sites, the Ross Ice Shelf has several unique advantages: (1) The ice/water interface at the bottom of the ice shelf reflects radio waves which increases the observable ice volume as downward propagating radio waves get reflected and can be picked up by antennas close to the surface. Most importantly, this property increases the sky coverage substantially, (2) the sea-level location reduces the energy spectrum of cosmic ray air showers that reach the snow surface.  The highest energy cores may lead to an insidious background created by reflection layers internal to the ice fabric (for sites are higher elevations) or by the water-ice interface at Moore's Bay.

Moore's Bay is not home to a research station but is fairly close ($\sim$\SI{1}{h} helicopter flight) to McMurdo, the largest research base in Antarctica, and the warmer temperatures in the summer simplify summer camp operations and detector deployment. In contrast to Greenland and South Pole, interior sites that require long-distance logistical operations, the relative short distance (\SI{110}{km}) to the Moore's Bay site may allow access by short-haul tracked vehicles or helicopters to ferry equipment and establish a near field summer camp. 

All three sites nominally satisfy the basic requirements - (1) large volumes of dielectric medium that need not be purchased, (2) radio signal must be able to propagate with little distortion or attenuation over length scales of a kilometer or more, and (3) background emission from anthropogenic sources is small.  Nevertheless, the relative differences in these properties at the various sites lead to different design optimizations that are under active study.  None of the sites are easy to get to, but each one has distinct transportation and logistical advantages.

\section{Science Goals}
 The science case for extremely high energy (EHE) neutrino astronomy has been made by other authors in this book, and we will not repeat those arguments here, but only make one point. In 2013, IceCube reported the first evidence of an isotropic flux of astrophysical neutrinos in the TeV-PeV energy range. While the flux is by now observed with high significance, its astrophysical origin is unknown. Only recently, IceCube was able to report the first compelling evidence of neutrino emission from the gamma-ray blazar TXS 0506+056. The present paucity of neutrino point source detections at energies below 10 PeV suggests that the observed isotropic flux is dominated by relatively weak extragalactic sources. Most likely, the neutrino sky is complex and several source classes may contribute. In fact, the EeV sky probed by radio neutrino detectors may be as different from the IceCube sky as optical observations are different from x-ray observations, which too is 3 order of magnitude difference in energy.  In particular, transient and impulsive sources may rise to prominence as the energies become more extreme.  We advocate an experimental approach that balances the search for point and diffuse emission over a broad range of energies.

\section{Neutrino interactions in dense media/ice}
\label{sec:interaction}
\begin{figure}[t]
    \centering
    \includegraphics[width=0.49\textwidth]{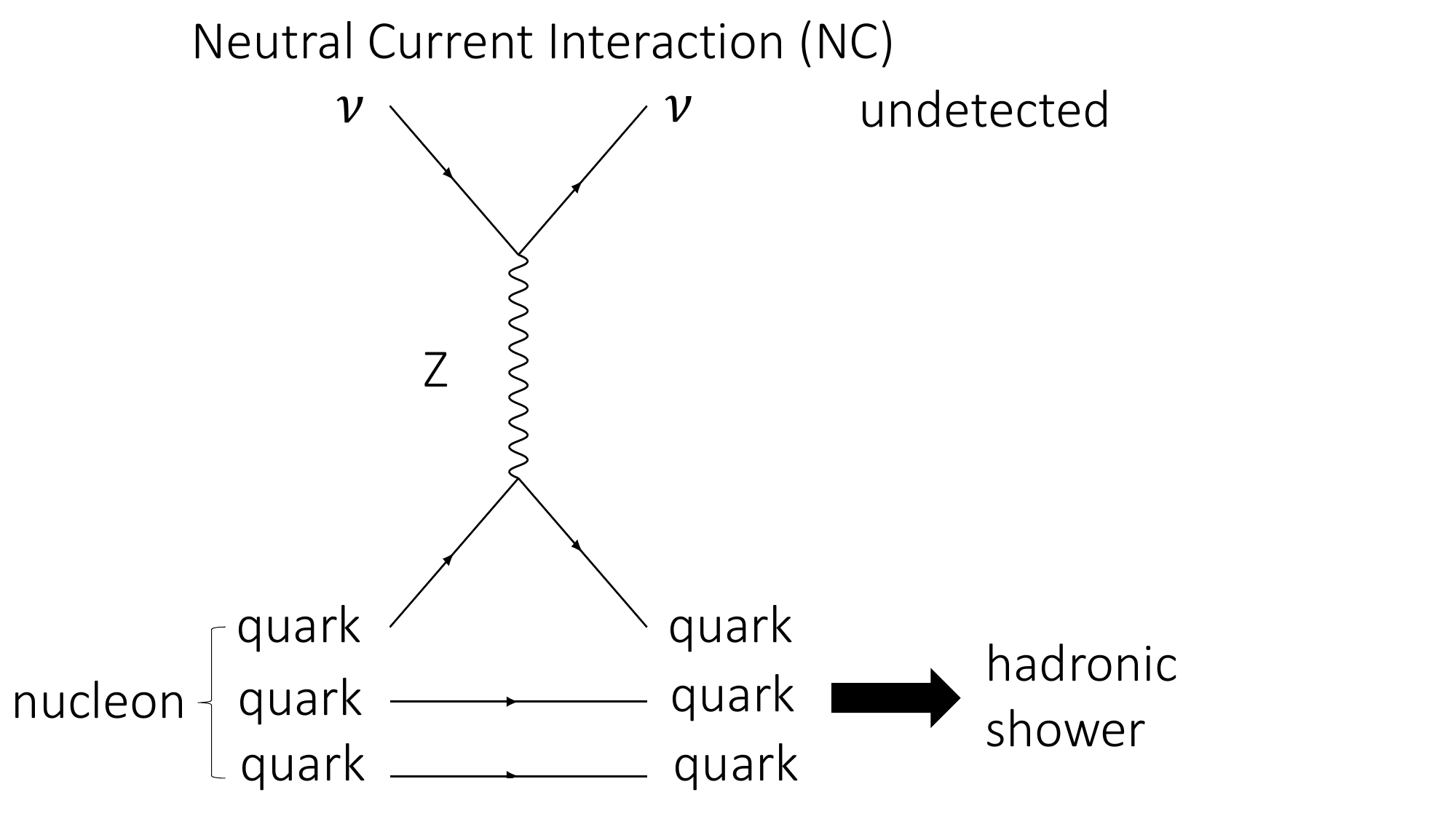}
    \includegraphics[width=0.49\textwidth]{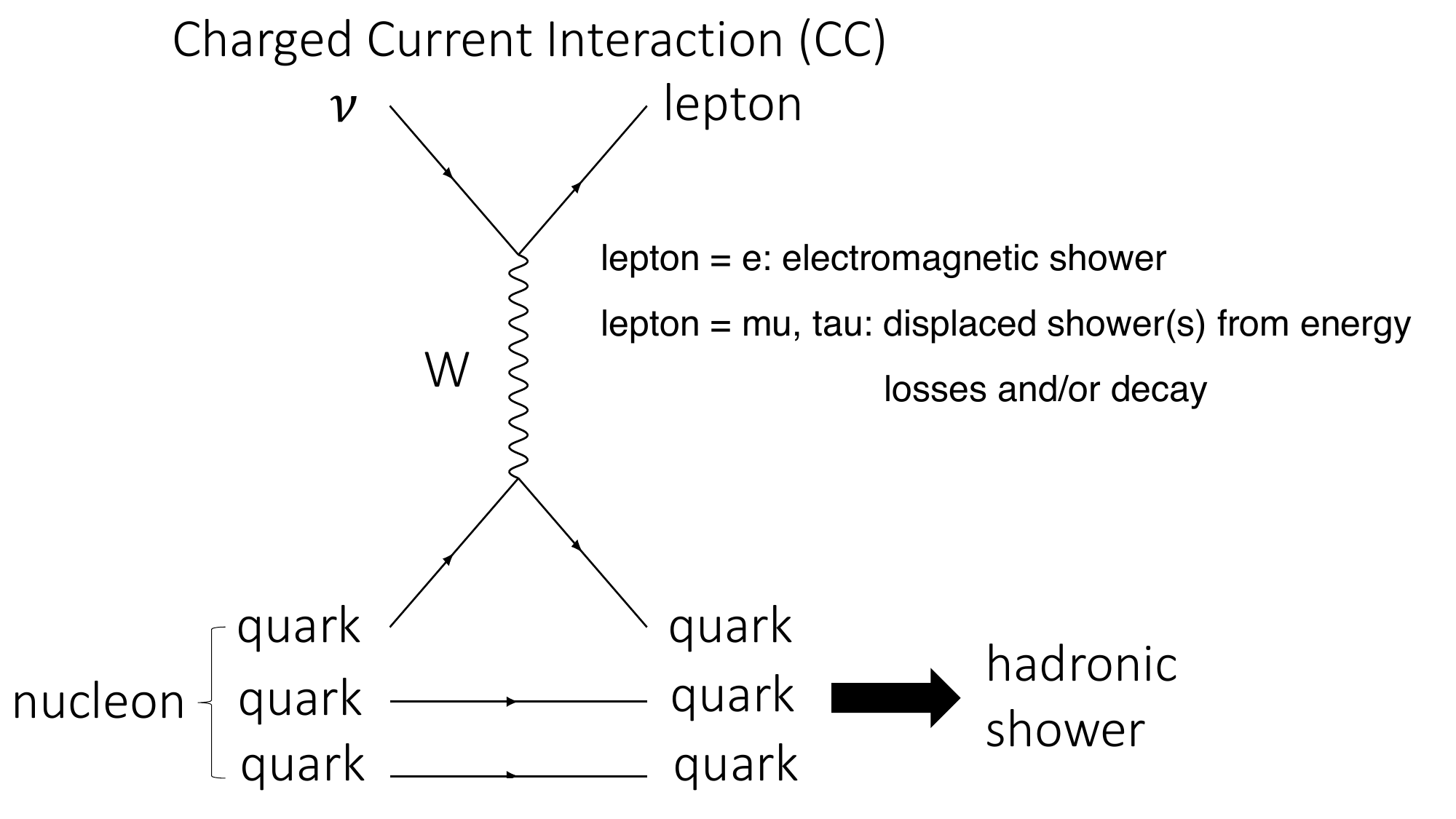}
    \caption{Feynman diagrams of a neutral current and charged current neutrino interaction. Figure adapted from \cite{NuRadioMC2019}.}
    \label{fig:feynman_diagram}
\end{figure}

\begin{figure}[t]
    \centering
    \includegraphics[width=0.49\textwidth]{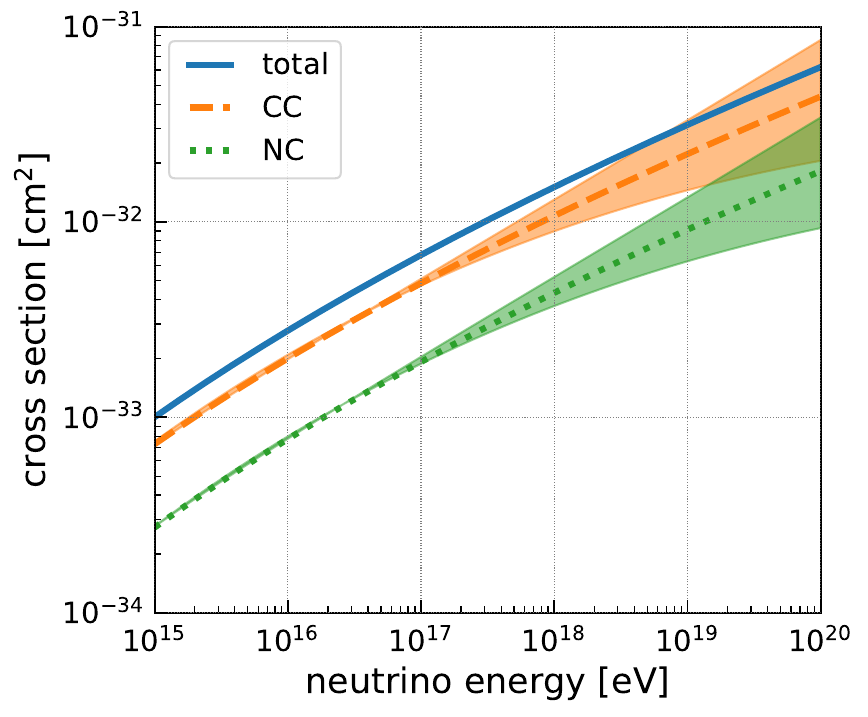}
    \includegraphics[width=0.49\textwidth]{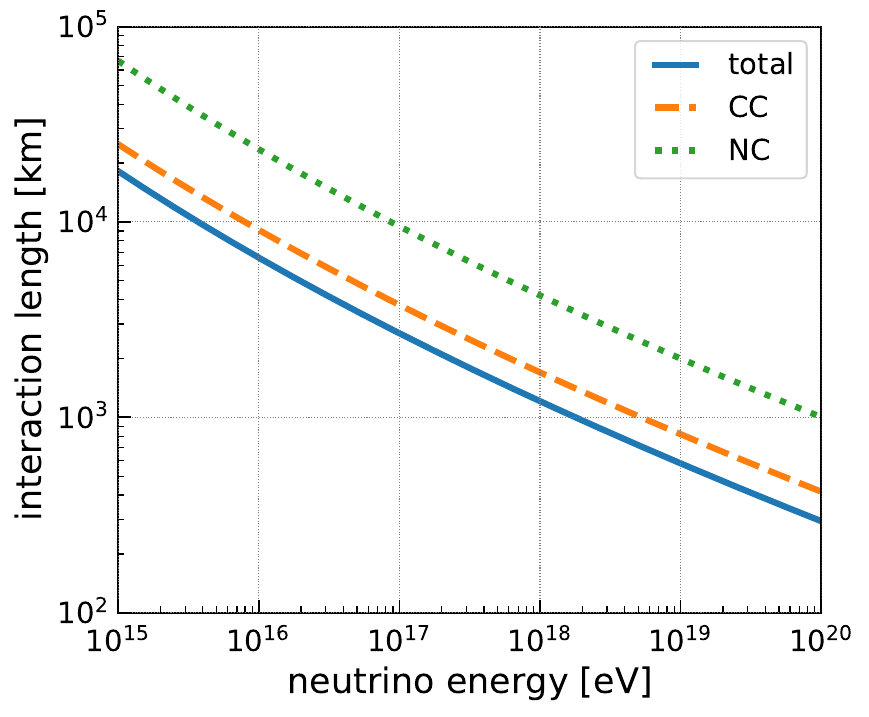}
    \caption{(left) Neutrino-nucleon cross section using the CTW \cite{Connolly2011} calculation. The shaded band represents the uncertainties. (right) Corresponding interaction length in ice.}
    \label{fig:cross_section}
\end{figure}

Neutrinos can only interact via the weak force\footnote{Due to the small mass of at least two of the three neutrino flavors, neutrinos -- of course -- also interact via gravity. However, because of the small magnitude of the gravitational force it can be neglected here.}. The dominant two interaction channels are neutral current (NC, exchange of a $Z$ boson) and charge current (CC, exchange of a $W$ boson) interactions with a nucleon, i.e., with a proton or neutron of the ice molecules. The corresponding Feynman diagrams are depicted in Fig.~\ref{fig:feynman_diagram}. The cross-section -- a measure of the interaction probability -- of the neutrino-nucleon interaction can be calculated from the Parton Distribution Functions (PDFs) of the nucleon \cite{Gandhi1998, CooperSarkar2011, Connolly2011, Bertone:2018dse} and is shown in Fig.~\ref{fig:cross_section} left. The increasing uncertainty at high energies results from missing experimental constraints on the Parton Distribution Functions (PDFs) at low ($< \num{e-5}$) Bjorken x.

A more useful parameter is to convert the cross-section $\sigma$ into the interaction length via 
\begin{equation}
    L_\mathrm{int} = \frac{m_n}{\sigma \rho} \, ,
\end{equation}
with $m_n$ being the nucleon mass, which we approximate with the proton mass, and $\rho$ = \SI{0.917}{g/cm^3} being the density of ice. The interaction length is the mean distance traveled by the neutrino before undergoing an interaction and is shown in Fig.~\ref{fig:cross_section} right. We find interaction lengths of hundreds of kilometers to several thousand kilometers which is much larger than the size of the typical detection volume of a single radio detector station. For an average trajectory of the neutrino through the volume surrounding one detector station of \SI{6}{km}, the fraction of neutrinos interacting in the detector can be calculated via $P_\mathrm{int} = 1 - \exp(-\SI{6}{km}/L_\mathrm{int})$ and is shown in Tab.~\ref{tab:cross_section}. The interaction probabilities are 1\% or smaller, thus, most neutrinos pass through the ice without interacting.
\begin{table}[t]
\centering
\caption{Neutrino-nucleon cross-section using the CTW \cite{Connolly2011} calculation, corresponding interaction length in ice, and interaction probability for a \SI{6}{km} long path through the ice.}
    {
    \begin{tabular}{c c c c}
    \hline 
        neutrino energy & cross section &$L_\mathrm{int}$ & $P_\mathrm{int}$  \\ \hline
         $10^{17}~$eV & \SI{6.8e-33}{cm^2} & \SI{2693}{km} & 0.22\% \\
        $10^{18}~$eV & \SI{1.5e-32}{cm^2} & \SI{1209}{km} & 0.49\% \\
        $10^{19}~$eV & \SI{3.1e-32}{cm^2} & \SI{581}{km} & 1.03\% \\ \hline 
    \end{tabular}}
    
    \label{tab:cross_section}
\end{table}

\begin{figure}[t]
    \centering
    \includegraphics[width=0.5\textwidth]{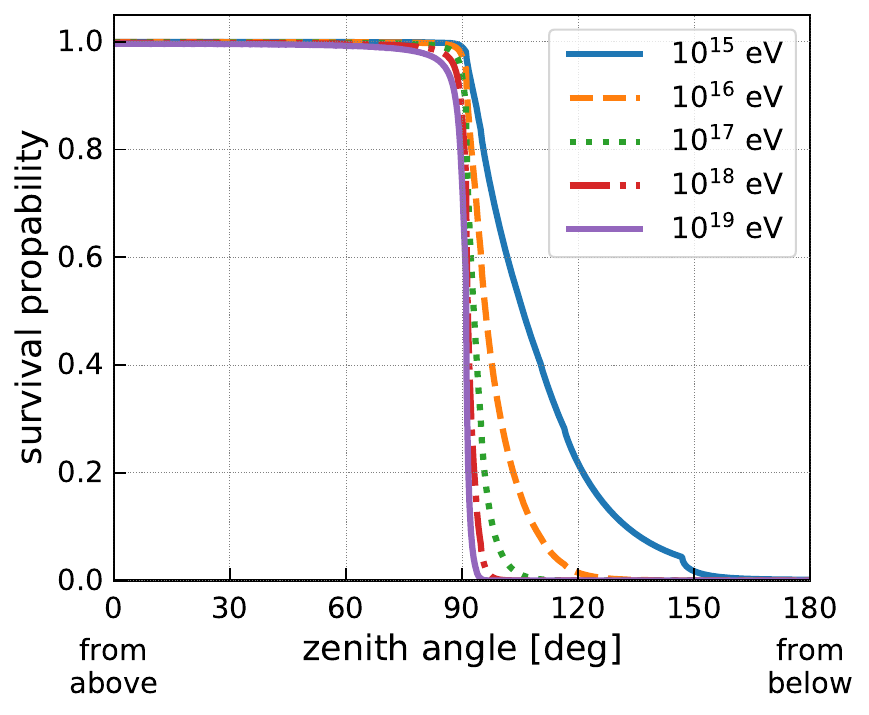}
    \caption{Probability of a neutrino to reach a point \SI{2}{km} below the ice surface, i.e., the probability to not get absorbed by the Earth. The calculation is based on the PREM Earth model \cite{Verhoeven2005} using the implementation in \emph{NuRadioMC} \cite{NuRadioMC2019}. It does not include tau regeneration.}
    \label{fig:earth_absorption}
\end{figure}

On the other hand, the interaction length is much smaller than the radius of the Earth of \SI{6371}{km}, so the Earth is opaque to high energy neutrinos. In Fig.~\ref{fig:earth_absorption}, the probability of neutrinos reaching the detector is shown. At the relevant energies, attenuation by the Earth is rather strong even for trajectories only a few degrees below the local horizon. The figure also illustrates that the atmosphere has very little impact on the neutrino flux except for horizontal trajectories. Though the effect is small, with enough statistics and precise angular reconstruction, it is possible to constrain the neutrino cross-section from the angular dependence near the horizon \cite{Valera:2022ylt, Esteban:2022uuw}. 

However, we also should mention a secondary effect not shown in this figure. It has been shown that tau neutrinos will propagate through the earth with less attenuation than the other flavors due to the rapid decay of tau leptons back to tau neutrinos \cite{HalzenSaltzberg-TauRegen}.  The overall effect reduces the energy to the point where the earth becomes transparent, but the average energy after propagating through the earth is small (below 10 PeV) and difficult to observe by radio neutrino detectors unless the source flux is much flatter than $E^{-2}$.  

\subsection{Interaction products and particle showers}
A neutrino interaction at such high neutrino energies generates a huge cascade of secondary particles of mostly electrons and positrons. A neutrino energy of \SI{e18}{eV} equals the rest-mass energy of \num{e12} electron-positron pairs ($m_e$ = \SI{511}{keV}). Not all of the neutrino energy is converted into the creation of new particles but already a small fraction is sufficient to create a huge particle shower. 

We start with describing the interaction of an electron that results from a charged-current (CC) interaction of an electron neutrino. A large fraction of the neutrino energy is transferred to the electron and at these energies radiation losses dominate. The electron continuously radiates photons where each time about half of the electron energy is transferred to the photon. The photon will decay into an electron-positron pair which will again radiate photons. In the simplified Heitler model \cite{Heitler, CarlsonOppenheimer1937}, each interaction happens after a fixed propagation length and the energy is equally split in each interaction. Hence, after $n$ splittings, $2^n$ particles are present where the initial energy is distributed equally over all particles. The interactions continue until the electrons (positrons) reach a critical energy $\zeta_c^e$ where the collisional energy losses exceed the radiative energy losses. That is, particles are more likely to lose energy kinetically than by radiating into electron-positron pairs. This transition occurs at approximately \SI{80}{MeV} in ice. At this point, the particle cascade has reached its maximum number of constituents and begins to decay as the shower of particles continues its propagation. These cascades are referred to as electromagnetic (EM) showers. 

The other type of particle shower are hadronic (HAD) showers that are started in the breakup of the nucleon which happens in NC as well as CC interaction for all neutrino flavors. The name \emph{hadronic shower} is confusing because most of the shower energy ends up in electromagnetic cascades as well and only the first few interactions are hadronic interactions. Most of the energy is quickly transferred into pions \cite{Matthews2005387}. One third of them are neutral pions that directly decay into photon pairs and thus transfer 1/3 of the energy to the electromagnetic sub-showers in each step. The other two thirds of the new particles are charged pions that interact further and produce a new generation of pions, where a third of them are again neutral pions, until their energy is below their critical energy. Hence, after $n$ interactions the energy in the electromagnetic shower component is $\left[1 - \left(\frac{2}{3}\right)^n\right]$ and between 90\% (at a shower energy of \SI{e15}{eV}) and 95\% (at shower energy of \SI{e19}{eV}) of the shower energy ends up in electromagnetic sub showers \cite{Alvarez-Muniz1998}.

From the description above we would conclude that hadronic and electromagnetic showers behave very similarly, which is indeed the case at shower energies below \SI{e15}{eV}. At higher energies, the Landau–Pomeranchuk–Migdal (LPM) effect becomes important \cite{Landau:1953um,Migdal:1956tc} which reduces the bremsstrahlung and pair production cross-sections of electrons/positrons because the characteristic length of the interaction becomes larger than atomic spacing and collective effects of the atomic fields have to be considered \cite{Konishi1991,Stanev1982,Alvarez-Muniz1997b,Alvarez-Muniz1998,SpencerLPMReview1999}. Therefore, the first few particles of an electromagnetic shower, where the particle energies are above a PeV, have a considerably larger mean free path length. This results in multiple spatially-displaced electromagnetic showers that depend on how the energy is distributed among the first few particles and also results in significant fluctuations between showers (we will see an example later). The LPM effect is significantly reduced in hadronic showers as the energy of the first EM particles is typically below a \si{PeV} \cite{Alvarez-Muniz1998}. 

\begin{figure}[t]
    \centering
    \includegraphics[width=0.32\textwidth]{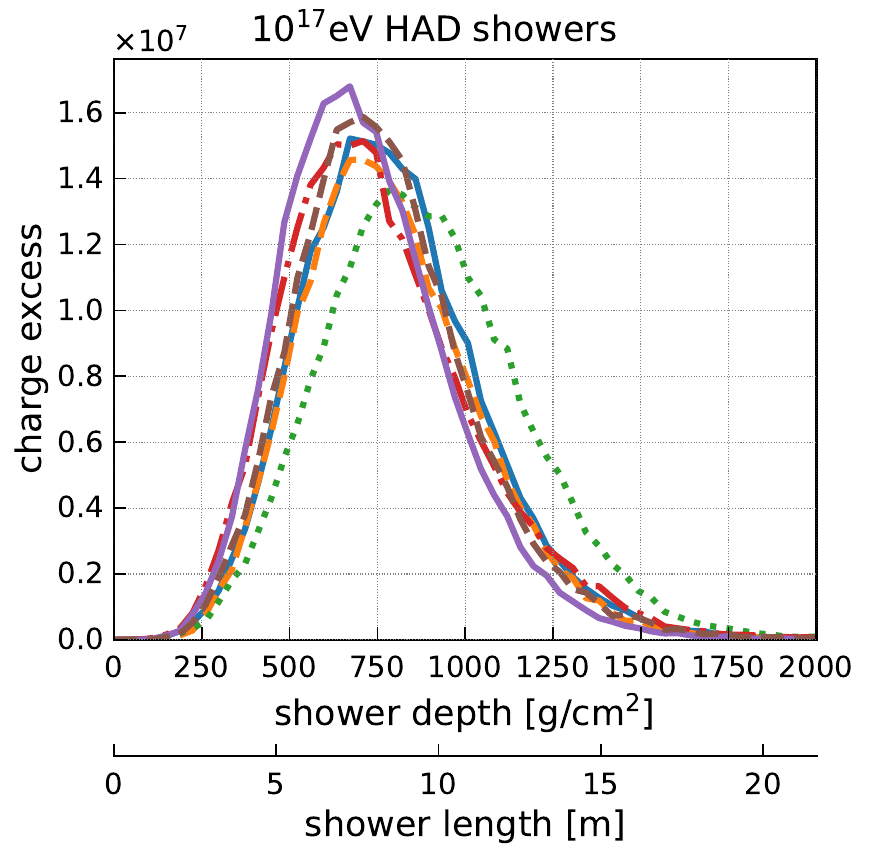}
    \includegraphics[width=0.32\textwidth]{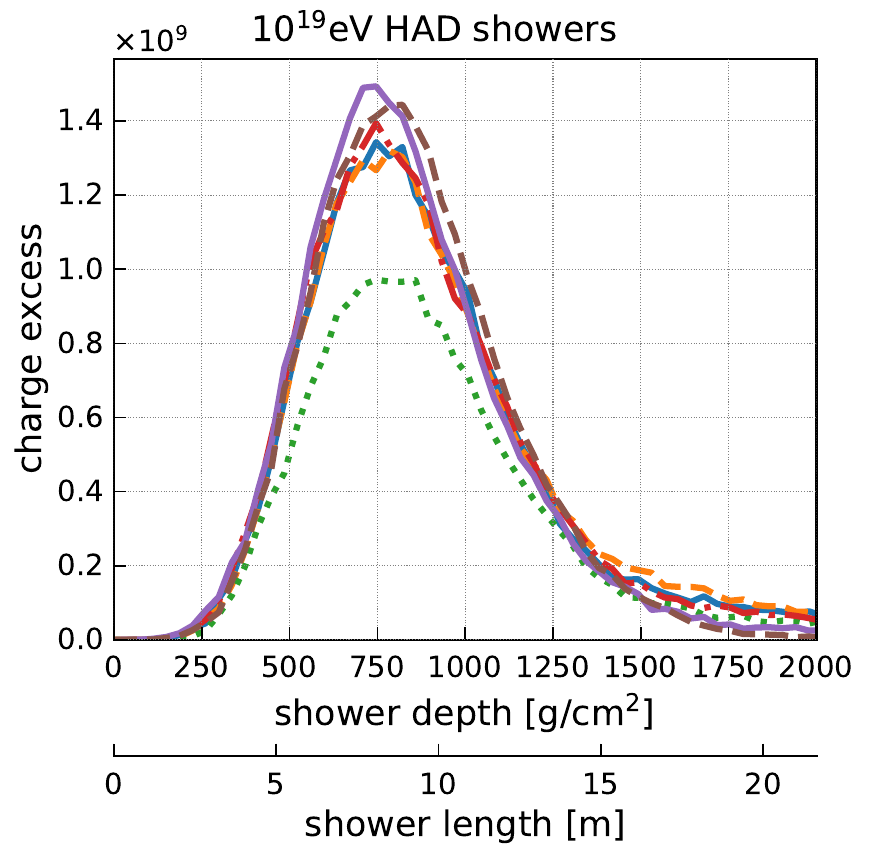}
    \includegraphics[width=0.32\textwidth]{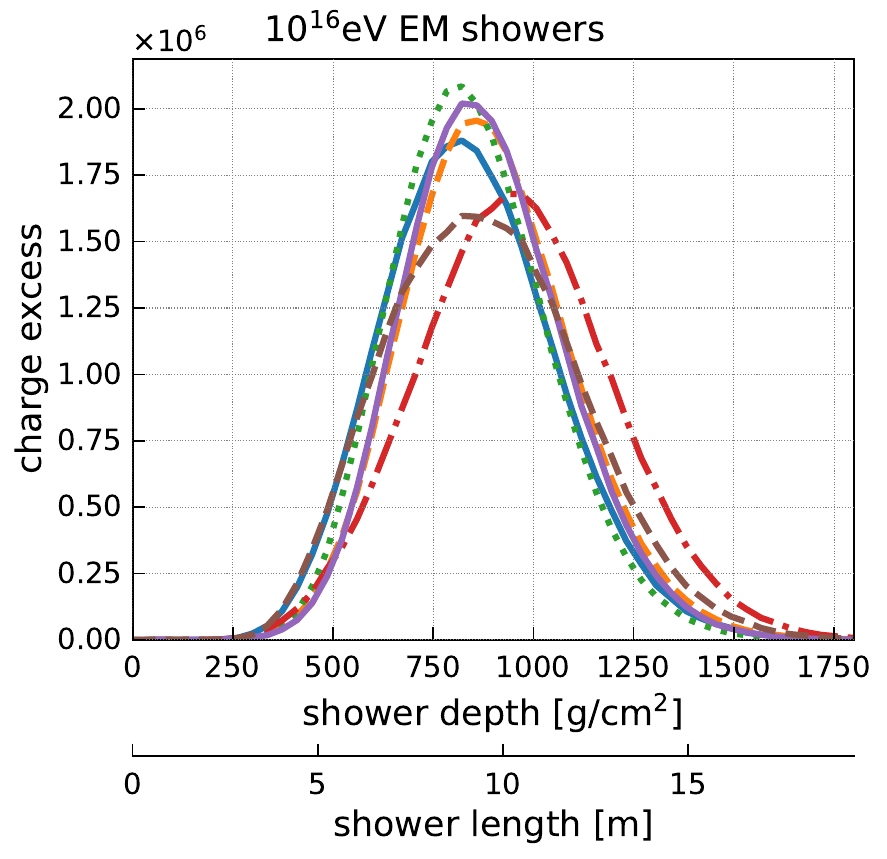}
    \includegraphics[width=0.32\textwidth]{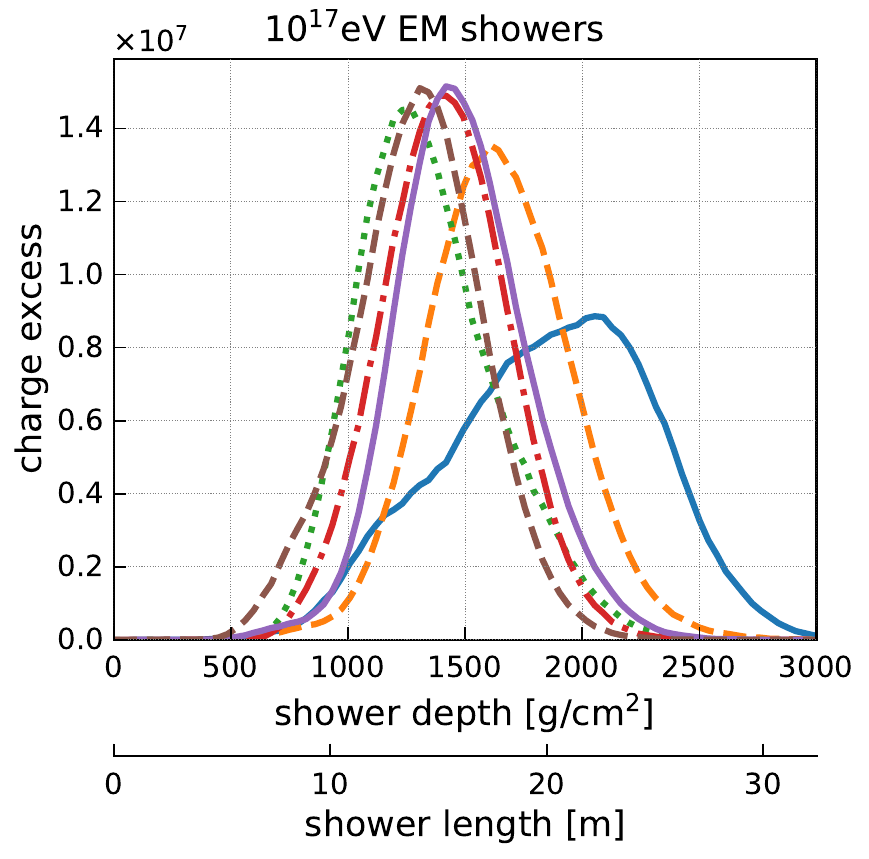}
    \includegraphics[width=0.32\textwidth]{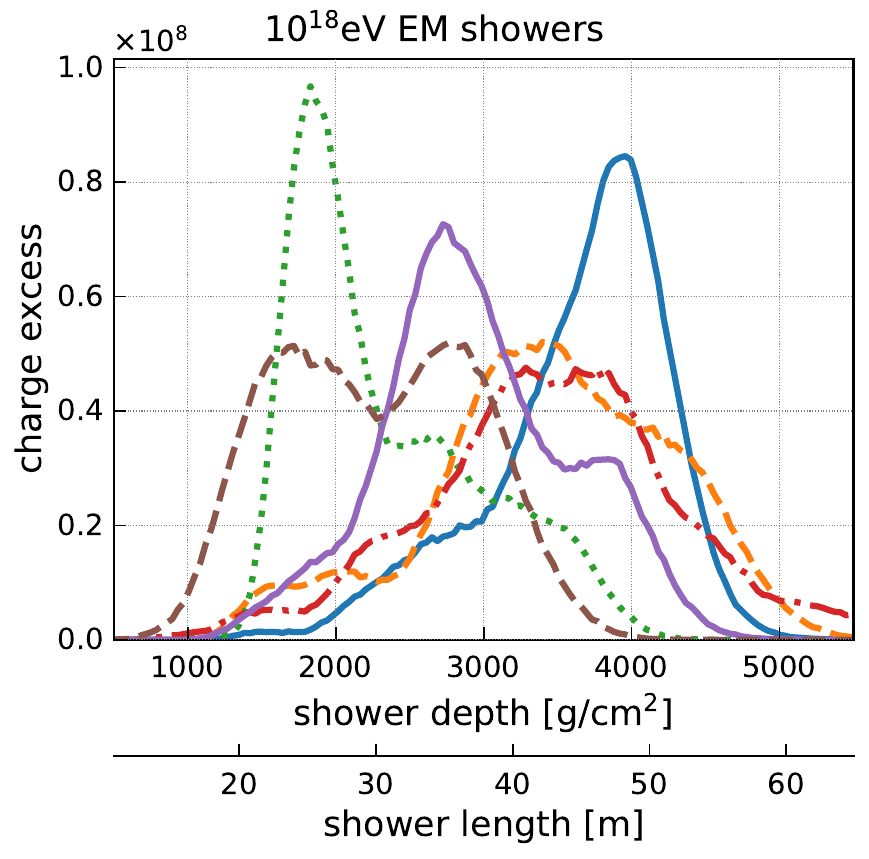}
    \includegraphics[width=0.32\textwidth]{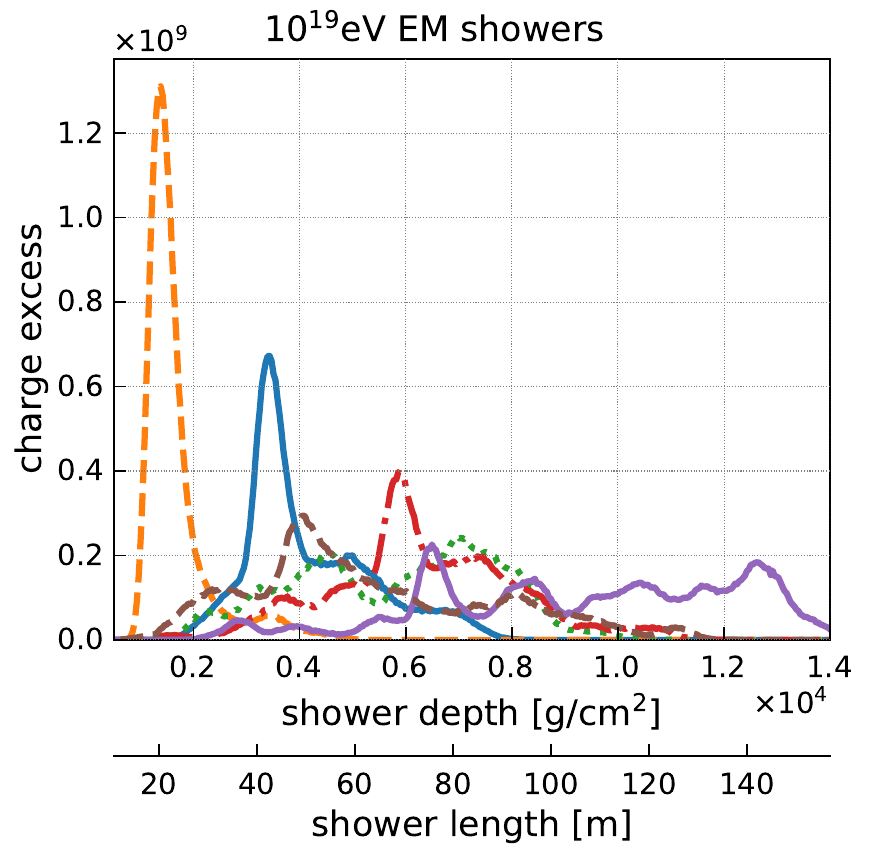}
    \caption{Examples of charge-excess profiles for different shower types and energies as indicated by the figure title. Each panel shows six random realizations of a shower for the same shower energy and type. The y-axis shows the charge-excess, i.e., the difference between the number of electrons and positrons. Two x-axes are shown, once in shower depth measured in grammage, and once in the corresponding depth in deep ice. The showers were taken from the shower library of NuRadioMC \cite{NuRadioMC2019} and were simulated using HERWIG \cite{herwig} for the simulation of the ﬁrst neutrino nucleon interaction, and ZHAireS \cite{AlvarezMuniz:2011bs} for the subsequent simulation of the particle shower in ice.}
    \label{fig:ceprofiles}
\end{figure}

The shower parameter relevant for radio emission, as we discuss in the next section in detail, is the charge-excess, i.e., the number of electrons minus the number of positrons. In Fig.~\ref{fig:ceprofiles} we show charge-excess profiles as a function of shower length, for hadronic (HAD) and electromagnetic (EM) showers at different energies for a variety of different shower realizations. The showers were simulated using HERWIG  \cite{herwig} for the simulation of the first neutrino nucleon interaction, and ZHAireS \cite{AlvarezMuniz:2011bs} for the subsequent simulation of the particle shower in ice and taken from the shower library of the NuRadioMC simulation code \cite{NuRadioMC2019}. The absolute value of the charge excess increases roughly linearly with the shower energy. The charge-excess profiles of different random realizations of hadronic showers are similar and the width of the shower increases only slightly with energy. 
The situation is drastically different for electromagnetic showers due to the LPM \cite{Landau:1953um,Migdal:1956tc} effect. At shower energies of \SI{e16}{eV} and below, different shower realizations are still quite similar as the electron/position energies are still close to or below the LPM threshold energy. At larger energies, the LPM effect becomes more and more pronounced. At a shower energy of \SI{e17}{eV}, the LPM effect mostly delays the start of the shower as the electron/positron energies are only above the LPM threshold at the very beginning of the shower development. At higher energies, the electron/position energies are above the LPM threshold in several of the first shower stages (remember that the energy is approximately halved after each shower stage). This leads to large shower-to-shower fluctuations and the development of multiple spatially displaced sub-showers. 

The remaining possible products of a high-energy neutrino interaction that we haven't discussed so far are muons and taus that are created in a charged-current interaction of the respective neutrinos (cf. Fig.~\ref{fig:feynman_diagram}) \cite{GarciaFernandez2020}. The decay length, i.e., the average distance until the tau decays (which will produce a measurable particle shower), ranges between a few hundred meters at an energy of \SI{e16}{eV} and a few tens of kilometers at \SI{1e19}{eV} \cite{NuRadioMC2019}. Thus, depending on the tau energy, there is a good chance that the tau decays within the ice and can be observed. Especially for neutrino directions close to the horizon where the tau propagates a long distance through the ice before reaching the bedrock. For example, at an elevation angle of \SI{5}{\degree} above the horizon and a thickness of the ice sheet of \SI{2.7}{km}, the trajectory through the ice is $\SI{2.7}{km} / \sin(\SI{5}{\degree}) \approx \SI{30}{km}$. 

\begin{figure}[t]
    \centering
    \includegraphics[width=0.6\textwidth]{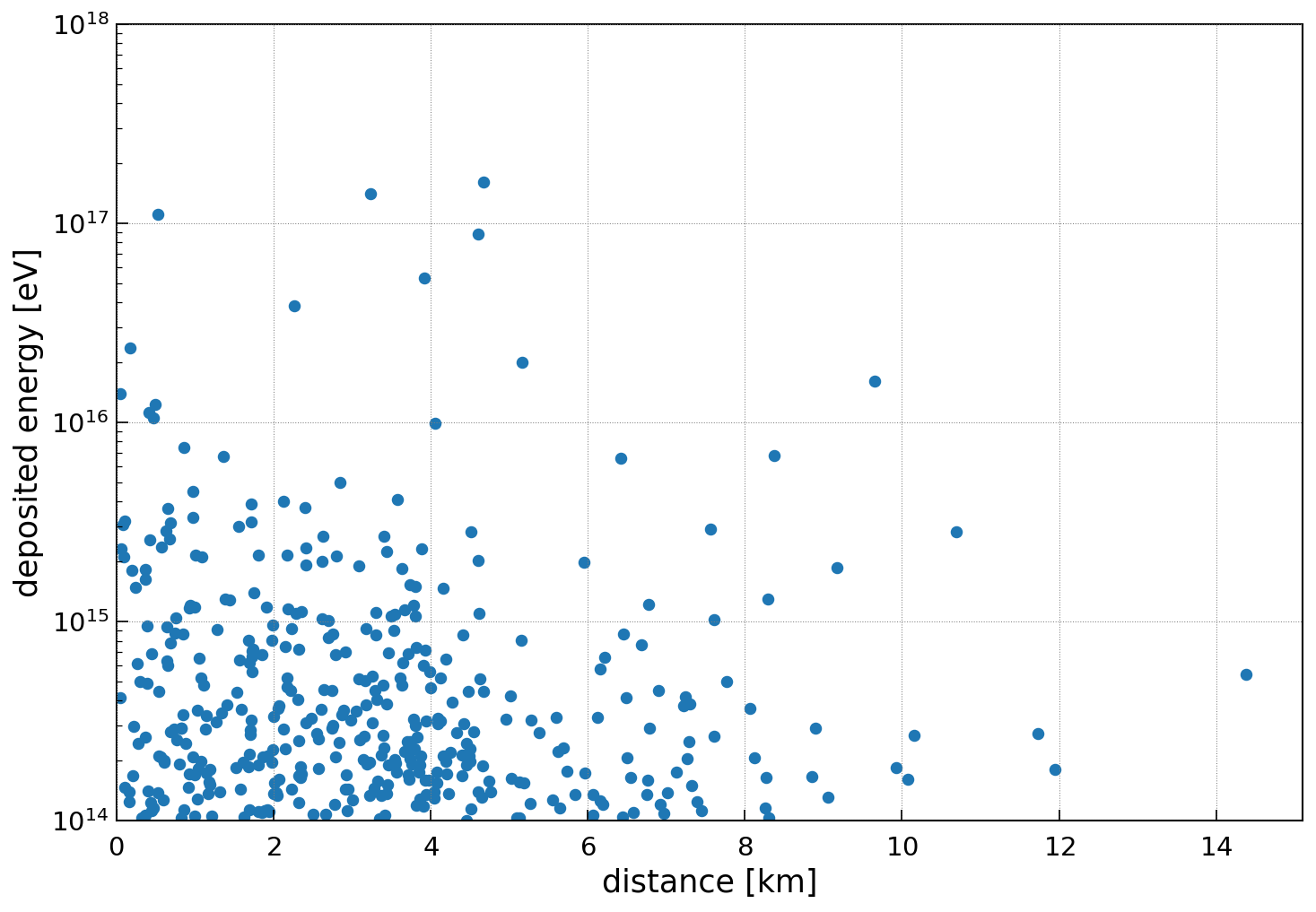}
    \caption{Stochastic energy losses above \SI{e14}{eV} of a \SI{e18}{eV} muon propagating through ice. Each energy deposit initiates a particle cascade that could produce detectable Askaryan radio emission. Energy losses were calculated using the PROPOSAL code \cite{proposal}. Figure from \cite{GlaserICRC2021Leptons}.}
    \label{fig:muon_dEdX}
\end{figure}

The probability that a muon decays within the ice sheet is negligible but muons, as well as taus, will lose energy while propagating through the ice which provides another chance of detection. Most of the energy losses are small and do not produce a large particle shower to be detectable via radio emission, but sometimes a significant amount of energy is deposited resulting in a measurable radio signal. A typical example of the energy losses of a high-energy muon that propagates almost horizontally through the ice is shown in Fig.~\ref{fig:muon_dEdX}. In this particular example, a \SI{e18}{eV} muon creates three showers with an energy of more than \SI{e17}{eV} and many showers above \SI{e16}{eV} where each of them produces a potentially detectable radio signal. Including the interactions of muons in the estimate of the sensitivity of a radio detector to muon neutrinos, increases the sensitivity by up to 50\% at around \SI{1}{EeV} compared to only considering hadronic shower of the initial $\nu_\mu$ interaction \cite{GarciaFernandez2020, GlaserICRC2021Leptons}. 

Also taus lose energy along the trajectory which adds to the decay channel. At low energies, most showers are produced through the tau decay but already at an energy of \SI{3e16}{eV} more showers produced via energy losses although at mostly lower energies. Therefore, up to an energy of \SI{7e17}{eV} the detected tau interactions are dominated by tau decays. The increase in sensitivity by including taus is largest at low energies (more than 50\% below \SI{e17}{eV}) where most taus decay within the observable volume. At energies of \SI{e18}{eV} and beyond the increase in sensitivity decreases to 25\% \cite{GarciaFernandez2020}.

\subsubsection{Topologies}
We can classify the interaction products into three different topologies: The first and simplest topology are neutral current interactions of any neutrino flavor that produces only a hadronic shower. The second topology is charged-current electron neutrino interactions which produce both a hadronic and electromagnetic shower at almost the same position in the ice. The third topology is charged-current interactions of muon and taus that produce an initial hadronic shower and additional spatially displaced showers through energy losses and decay in the case of the tau. These different topologies provide the potential to identify the neutrino flavor but studies on how well these signatures can be resolved are still in early stages \cite{Alvarez-Muniz1999,Gerhardt2010,GarciaFernandez2020}. A golden signature will be the detection of two showers in one or more radio detector stations. An early study found that a secondary shower from muon energy losses can be detected for more than 50\% of the muon neutrinos at energies beyond \SI{e19}{eV} where the details depend on the experimental setup in particular the separation between the radio detector stations \cite{GarciaFernandez2020, GlaserICRC2021Leptons}.

\subsubsection{Inelasticity}
\begin{figure}
    \centering
    \includegraphics[width=0.7\textwidth]{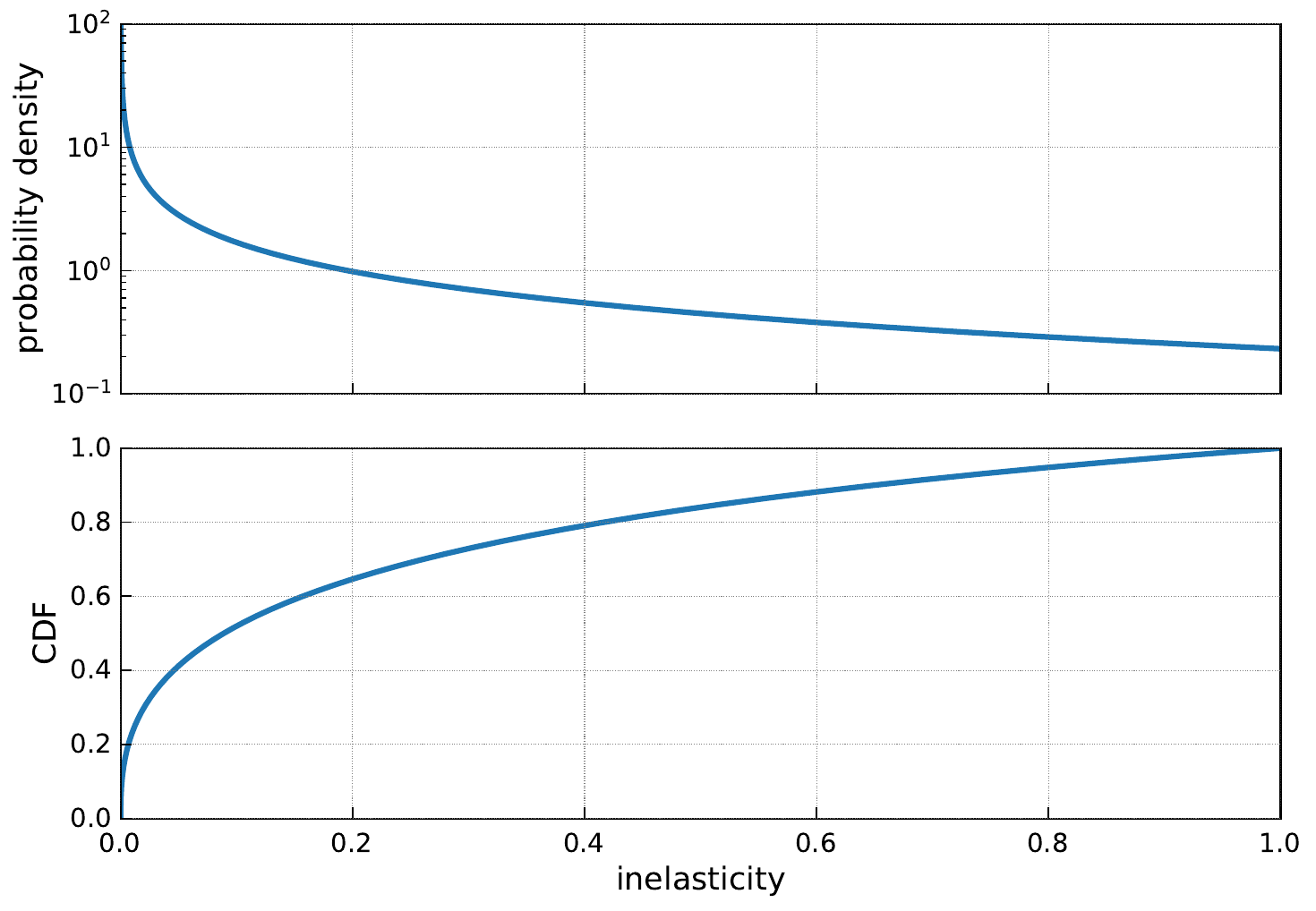}
    \caption{Inelasticity distribution of high-energy neutrino interactions. (Top) probability density, (bottom) Cumulative distribution function (CDF), adapted from \cite{PersichilliPhD}}
    \label{fig:inelasticity}
\end{figure}

The amount of energy that is transferred from the neutrino into the particle shower is a random process. Typically most energy stays with the neutrino (in the neutral current interaction) or remains with the lepton that the neutrino is converted to (in the charged current interaction, cf. Fig.~\ref{fig:feynman_diagram}). The fraction of energy that is transferred from the neutrino to the nucleon (and subsequent hadronic shower) via the $Z$ or $W$ Boson is called inelasticity and an approximation of the probability distribution is shown in Fig.~\ref{fig:inelasticity}. (For a more detailed presentation including the dependence on neutrino energy and interaction type the reader is referred to  \cite{Bertone:2018dse,Valera:2022ylt}). Half of the induced hadronic showers will have less than 10\% of the neutrino energy. 

Because of the strong peak at low inelasticities, the chance of measuring an electron neutrino undergoing a charged-current interaction is highest, as most energy remains with the electron which converts all its energy into a particle shower. 
The probability of a CC interaction is approx. 70\% with a weak energy dependence \cite{Bertone:2018dse,Valera:2022ylt}.

\section{Askaryan radio emission}
Radio emission is generated by the particle shower through the Askaryan effect which was postulated in 1962 by the Soviet-Armenian physicist Gurgen Askaryan \cite{Askaryan1962}. A particle shower that develops in a dielectric medium such as ice will develop a time-varying negative charge-excess which gives rise to coherent radio emission. This is primarily because electrons from the ice are dragged along with the shower front. 
The emission can be understood simplest in a macroscopic picture where the time-varying charge-excess induces a longitudinal current, and one can imagine the emission of electromagnetic radiation from a dipole source propagating along the shower path. 

\begin{figure}[t]
    \centering
    \includegraphics[width=\textwidth]{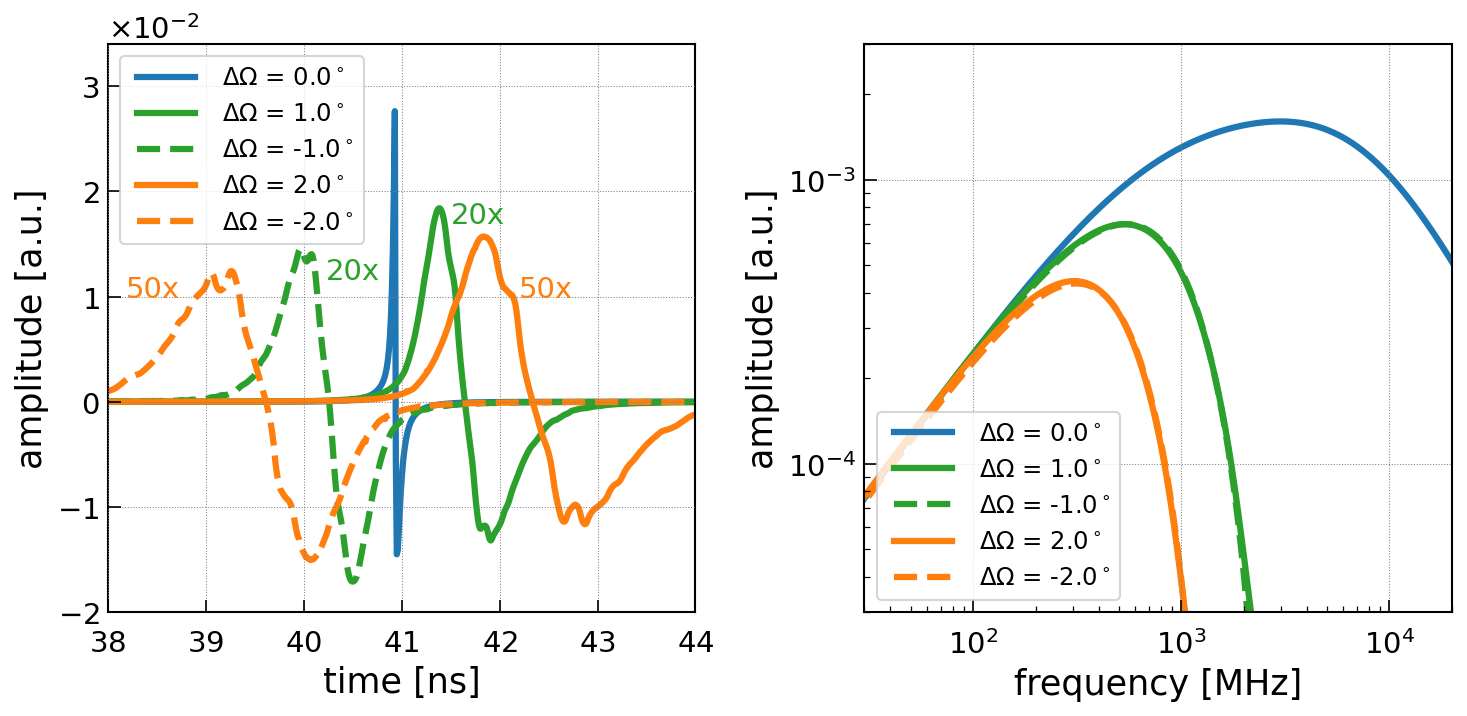}
    \caption{Askaryan signal calculated using the semi-analytical ARZ2020 model \cite{ARZ,ARZ2,NuRadioMC2019} for a hadronic shower with an energy of \SI{e17}{eV} at \SI{5}{km} distance. We show the electric field amplitude of the Askaryan signal for five different observation angles of the shower: 1) directly on the Cherenkov cone (blue
    ), 2) one degree away from the Cherenkov cone (green) and 3) two degrees away from the Cherenkov cone (orange). Solid curves show observation angles larger than the Cherenkov angle. Dashed curves show observation angles smaller than the Cherenkov angle. The left panel shows the signal in the time domain, and the right panel shows the corresponding frequency spectrum. In the time domain plot, the amplitude of the $\Delta \Omega = \pm1^\circ$ ($\pm2^\circ$) pulse is multiplied by a factor of 20 (50) for better visibility.}
    \label{fig:askaryan_pulse}
\end{figure}

As discussed in the introduction, the Askaryan radiation exhibits strong Cherenkov-like time compression effects because the shower propagates with the vacuum speed-of-light $c_0$ whereas the radio emission propagates with the slower speed-of-light of $c_0/n$ due to the index-of-refraction $n$ being larger than one. An observer far enough away from the shower will see all signals emitted along the shower path arriving at the same time when the shower is observed at the Cherenkov angle of $\arccos(1/n)$ which will lead to a maximum constructive interference. The signal amplitude increases linearly with frequency up to a characteristic cutoff frequency. The cutoff frequency is a consequence of the lateral extent of the shower front which limits the coherence at small wavelengths \cite{ARZ2}. A back-on-the-envelope estimate is given by the Moli\'{e}r radius in ice of \SI{10.55}{cm}, which specifies the transverse dimension of electromagnetic showers. Thus, the coherence for wavelength smaller than the Moli\'{e}r radius $R_M$ is limited. The corresponding estimate of the cutoff frequency is $c/R_M = \SI{2.8}{GHz}$.  An example is shown in Fig.~\ref{fig:askaryan_pulse}. The Askaryan pulse is bipolar with a slight asymmetry which is a consequence of the charge-excess distribution. The initial rise is steeper than the decay of the charge-excess (cf. Fig. ~\ref{fig:ceprofiles}).  

If the shower is observed from angular directions slightly away from the Cherenkov cone, the observer still sees the radiation from all points along the shower track but the signals do not arrive simultaneously anymore. As a consequence, the coherence is reduced depending on the observation angle and frequency. The high-frequency components lose coherence earlier than the low-frequency components. This leads to the remarkably simple feature that just the cutoff frequency reduces with increasing offset from the Cherenkov angle. In Fig.~\ref{fig:askaryan_pulse}, we show a detailed calculation of the electric-field amplitude of the Askaryan signal for a \SI{e17}{eV} hadronic shower at different observation angles for one of the charge-excess profiles from Fig.~\ref{fig:ceprofiles}. For observation angles larger the Cherenkov angle $\theta_C$, the observer sees first the electric field produced by the early stages of the shower. At observation angles smaller than $\theta_C$, the time order reverses and the electric field emitted by the later part of the shower is observed first \cite{ARZ2}. As a consequence, the Askaryan pulse at $\theta = \theta_C + \Delta\Omega$ is an asymmetric copy of the pulse at $\theta = \theta_C - \Delta\Omega$.

\subsection{Calculation of Askaryan signal}
\label{sec:AskaryanEmission}
The Askaryan radiation can be calculated theoretically from the particle shower with great accuracy because the emission is purely described by classical electrodynamics, a theory without any free parameters. In a microscopic picture, radiation is generated by the acceleration of charge, which is often referred to as Bremsstrahlung. A particle shower is nothing but a large number of charges that get accelerated. Using the fundamental equations of electrodynamics, the radiation that is generated by each particle trajectory can be calculated, and the sum of the contributions of all particles is the Askaryan radiation \cite{Endpoint2011}. These calculations can be performed with the ZHAireS and CoREAS Monte Carlo codes \cite{AlvarezMuniz:2011bs, CoREAS2013} that give a precise and accurate prediction of the Askaryan radiation. 

Microscopic calculations are very accurate but also impractical for many purposes because the calculation takes about 1 CPU day for a single shower and observer position and needs to be repeated for every observer position. Therefore, a set of microscopic MC simulations was used to parameterize the Askaryan signal based solely on the shower type, shower energy, and viewing angle \cite{Alvarez_box,Alvarez2012,Alvarez2009,AlvarezMuiz2000,AlvarezMuiz1998}. Please refer to the documentation of the NuRadioMC code for a detailed description, comparison and reference implementation of all available models \cite{NuRadioMC2019}. One of the first and simplest model \cite{AlvarezMuiz2000} has a remarkably simple form but already gives a good description of detailed microscopic calculations. Because of its simplicity, it provides qualitative and easily understandable, however, not necessarily precise insights into the main dependencies of the Askaryan signal. For pedagogical reasons, we explicitly provide the parameterization of this model here:

If the shower is observed on the Cherenkov angle, the electric field in the frequency domain (scaled to a distance of \SI{1}{m}) is given by
\begin{equation}
  \frac{\varepsilon_{c}^{1\mathrm{m}}}{\si{V/m/MHz}}(E_{sh},f) = 2.53 \times 10^{-7} \cdot \frac{E_{sh}}{\text{TeV}} \cdot \frac{f}{f_0} \cdot \frac{1}{1 + (\frac{f}{f_0})^{1.44}} \, ,
  \label{eqn:SelfMCEFieldAmp}
\end{equation}
with the shower energy $E_{sh}$, frequency $f$ and $f_0 = \
\SI{1.15}{GHz}$. Thus, the signal amplitude increases linearly with frequency up to the cutoff frequency $f_0$.

Signal amplitudes at angles away from the Cherenkov cone, $\varepsilon^{1\mathrm{m}}$, are modeled as a Gaussian profile according to 
\begin{equation}
  \varepsilon^{1\mathrm{m}}(E_{sh},f,\theta_v) =  \varepsilon_{c}^{1m}(E_{sh},f) \cdot \frac{\sin \theta_v}{\sin \theta_c} \cdot \exp\bigg[-\ln(2) \cdot \Big(\frac{\theta_v - \theta_c}{\sigma_\theta}\Big)^2\bigg]
  \label{eqn:ShelfMCConeAngle}
\end{equation}
with $\varepsilon_{c}^{1m}$ given in Eq.~\eqref{eqn:SelfMCEFieldAmp}, and where $\theta_v$ is the viewing angle relative to the shower axis. The angular width of the cone around the Cherenkov angle $\sigma_\theta$ is a function of both frequency and energy and separately parameterized for electromagnetic and hadronic showers. The higher the frequency, the smaller the width $\sigma_\theta$ of the Cherenkov cone. This corresponds to the decrease in cutoff frequency with increasing offset to the Cherenkov angle as seen in Fig.~\ref{fig:askaryan_pulse}.

Frequency domain parameterizations are extremely fast to evaluate and provide an accurate description of the magnitude of the frequency spectrum but lead to inaccuracies in the time-domain because no phase information is provided. Therefore, a semi-analytic model was developed, that allows calculating the time domain waveform directly from the charge-excess profile via convolution with a Form factor. It turns out that the Form factor is universal and only needs to be parameterized once for each shower type \cite{ARZ, ARZ2}. This model, called \emph{ARZ} in the following, is able to reproduce the results of a microscopic simulation within 3\% but at a fraction of the computing time \cite{ARZ2}. A modern implementation of this model is provided in NuRadioMC including an extensive library of charge-excess profiles \cite{NuRadioMC2019}. 
The combination of the \emph{ARZ} model with a shower library allows to precisely model the effects of LPM elongation  \cite{Landau:1953um,Migdal:1956tc} and the resulting shower-to-shower fluctuations on the Askaryan signal on a single event basis (cf. Fig.~\ref{fig:ceprofiles}), rather than describing an average behavior as done in improved frequency domain parameterizations \cite{Alvarez-Muniz2009}. The model also captures subtle features of the cascades like sub-showers and accounts for stochastic fluctuations in the shower development which can alter the Askaryan signal amplitudes significantly.

So far, the microscopic simulations were performed in a homogeneous medium with a constant index-of-refraction. This assumes that the signal generation can be decoupled from the signal propagation (discussed in the next chapter) which is expected to be a good approximation for most geometries \cite{NuRadioMC2019}. An area of current research is to verify these approximations by performing microscopic simulations in inhomogeneous media. These simulations will be enabled through the development of the CORSIKA8 code in near future \cite{Engel:2018akg}.

\subsection{Experimental evidence of Askaryan radiation}
Because of the low flux of high-energy neutrinos and the limited size of current radio arrays, no neutrino has been measured yet via Askaryan radiation. However, there is still substantial evidence for Askaryan radiation. On the one hand, Askaryan radiation was measured in a laboratory setup. On the other hand, Askaryan radiation is also produced in \emph{air showers} that are created when cosmic rays strike the Earth atmosphere which has been measured successfully with radio detector arrays for decades. 

\subsubsection{Lab measurements of Askaryan radiation}
Generating a particle shower with an energy of around \SI{e18}{eV} in the lab seems impossible at first glance, given that the most powerful particle accelerator on Earth, the LHC, can only accelerate protons up to a maximum energy of $\mathcal{O}(\SI{e13}{eV})$. The experimental trick is to not accelerate a single particle up to \SI{e18}{eV} but to accelerate a bunch of \num{e9} particles up to a moderate energy of a few \si{GeV}. Hence, a particle shower is generated that is already partly developed, but the energy of each electron is still significantly larger than the critical energy of \SI{80}{MeV}. 

This was done in a series of measurements performed at the Stanford Linear Accelerator (SLAC) that constitute the first direct experimental observation of Askaryan radiation \cite{Saltzberg2001,Miocinovic2006, Gorham2005, Gorham2007}. The experimental setup was the following: A beam of \SI{28.5}{GeV} electrons with up to \num{e10} electrons per bunch was converted into Bremsstrahlung photons. The photon beam was directed into a dense target material producing electromagnetic showers several meters long with shower energies ranging between $(0.06 - 1.1) \times 10^{19}~$eV. Initially, silica sand was used as the target material \cite{Saltzberg2001,Miocinovic2006}. Later, the experiment was repeated using synthetic rock salt \cite{Gorham2007} and ice \cite{Gorham2007}. Radio emission was detected by an array of antennas sensitive from a few \SI{100}{MHz} to several \si{GHz}.

The measurements probed all relevant properties of Askaryan radiation: Short broadband pulses were observed with an upper limit on the width of \SI{500}{ps} limited by the detector resolution. The amplitude was found to scale linearly with the shower energy (which is proportional to the number of shower particles which is proportional to the charge-excess). Thus, the observed radiation is coherent. The radiation was found to be 100\% linearly polarized in the plane containing the antenna and shower axis. Furthermore, the measurement is in agreement with theoretical predictions on an absolute scale. 

\begin{figure}[t]
    \centering
    \includegraphics[width=0.55\textwidth]{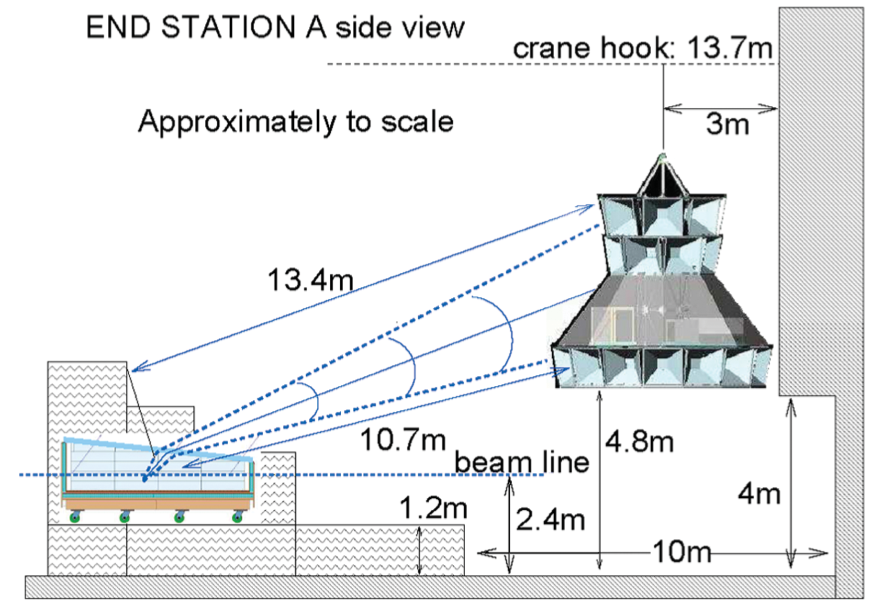}
    \includegraphics[width=0.44\textwidth]{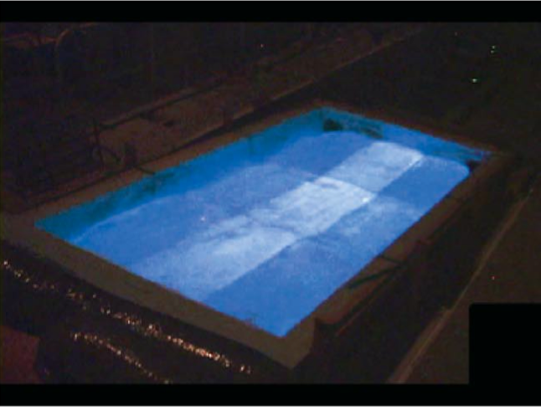}
    \caption{(left) Sketch of the measurement setup at SLAC to measure Askaryan radiation in ice. (right) Photo of the ice target illuminated from interior scattered optical Cherenkov radiation. Figures from reference \cite{Gorham2007}.}
    \label{fig:slac_sketch}
\end{figure}

\subsubsection{Measurement of Askaryan radiation from air showers}
Another direct evidence of Askaryan radiation is the measurement of radio emission from air showers. These particle cascades are initiated by ultra-high energy cosmic rays in the atmosphere and develop over several kilometers because of the much lower density of air compared to ice. As a consequence, the geomagnetic field has a significant effect and accelerates electrons and positrons into opposite directions leading to so-called geomagnetic radio emission. In a macroscopic picture, one can think of a current in the direction of the Lorentz force for the geomagnetic emission, and a longitudinal current (along the shower direction) for the Askaryan emission. 

\begin{figure}[t]
    \centering
    \includegraphics[width=0.7\textwidth]{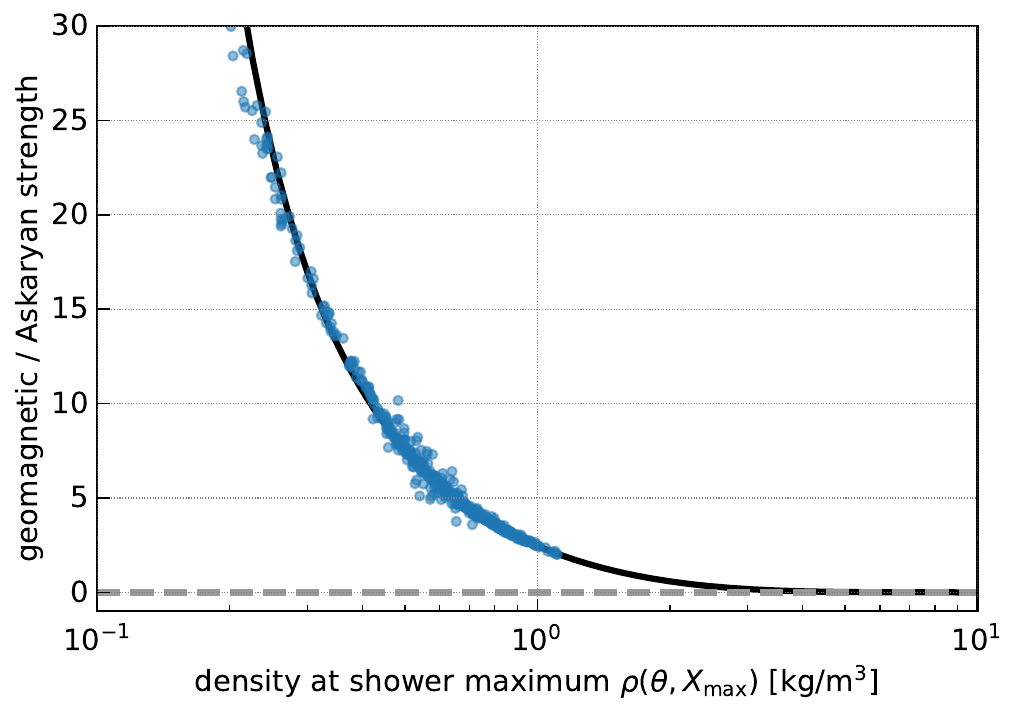}
    \caption{The relative strength of the geomagnetic emission over the Askaryan emission as a function of density at the shower maximum. Each data point corresponds to one microscopic air shower simulation using the CoREAS code \cite{CoREAS2013}. At large densities (such as ice with \SI{917}{kg/m^3}) the geomagnetic emission becomes negligible. Shown is the maximum geomagnetic emission corresponding to a shower propagation direction perpendicular to the magnetic field where the Lorentz force acting on the shower particles is maximum. The figure is adapted from  \cite{GlaserErad2016}.}
    \label{fig:charge_excess_strength}
\end{figure}

In air, the geomagnetic emission is the dominant effect. The exact strength depends strongly on the density in which the shower develops but is typically significantly larger than the Askaryan emission \cite{GlaserErad2016}.  As shown in Fig.~\ref{fig:charge_excess_strength}, the strength of the geomagnetic emission drops quickly with increasing density and is completely irrelevant for dense media such as ice. 

The geomagnetic and Askaryan emission can be distinguished by the polarization of the signal. Both emission processes produce linearly polarized signals. In the geomagnetic case, the polarization is in the direction of the Lorentz force whereas, in the Askaryan emission case, the polarization points towards the shower axis. This also leads to an interference pattern and to an asymmetric footprint which is yet another way of distinguishing the two emission mechanisms. The Askaryan emission was confirmed in several cosmic-ray radio detectors and is in excellent agreement with theoretical predictions \cite{AERAPolarization,LofarPolarization2014,Huege2016,Schroeder2016}.

Also, the lab measurements at SLAC were extended to probe the geomagnetic emission as well by applying a magnetic field to the target material where the magnetic field strength was scaled up to match the much shorter shower lengths \cite{SlacT510}. Also, this experiment confirmed the theoretical expectation. It is worth noticing that the radio signals from air showers provide a useful calibration signal for radio neutrino detectors. They have been used by the ARIANNA collaboration successfully to confirm detector operation \cite{Barwick2017-Airshowers} and because of the excellent understanding of the polarization of radio signals, air shower signals can even be used to probe the capabilities to reconstruct the signal polarization \cite{Arianna:2021lnr}.

\section{Propagation of radio signals in ice}
In this section, we describe the propagation of radio waves in ice. Four things need to be considered: First, the attenuation of radio signals in the ice. Second, changes of the index-of-refraction (as a consequence of changing density in the firn) that lead to a bending of signal trajectories. Third, birefringence that leads to differences in propagation time for different polarization components. Forth, density fluctuations around a smooth profile that lead to additional propagation modes. 

\subsection{Attenuation}
\label{sec:attenuation}
The attenuation length -- defined as the propagation distance after the signal amplitude reduces by a factor of $1/e$ -- is in the order of $\mathcal{O}$(\SI{1}{km}) but varies significantly between different sites. The main quantity that determines the transparency to radio waves is the temperature and the chemical composition of the ice \cite{Aguilar:2022kgi} which varies between the three considered sites: South Pole \cite{Barwick2005-PoleAtten}, Greenland \cite{Avva:2014ena,Aguilar:2022kgi}, and Moore's Bay \cite{BarwickBergBessonEtAl2014}. As the temperature typically increases with increasing depth, the attenuation length also decreases with depth. Furthermore, the attenuation length has a slight frequency dependence and is larger at lower frequencies. 

\begin{figure}[t]
    \centering
    \includegraphics[width=0.7\textwidth]{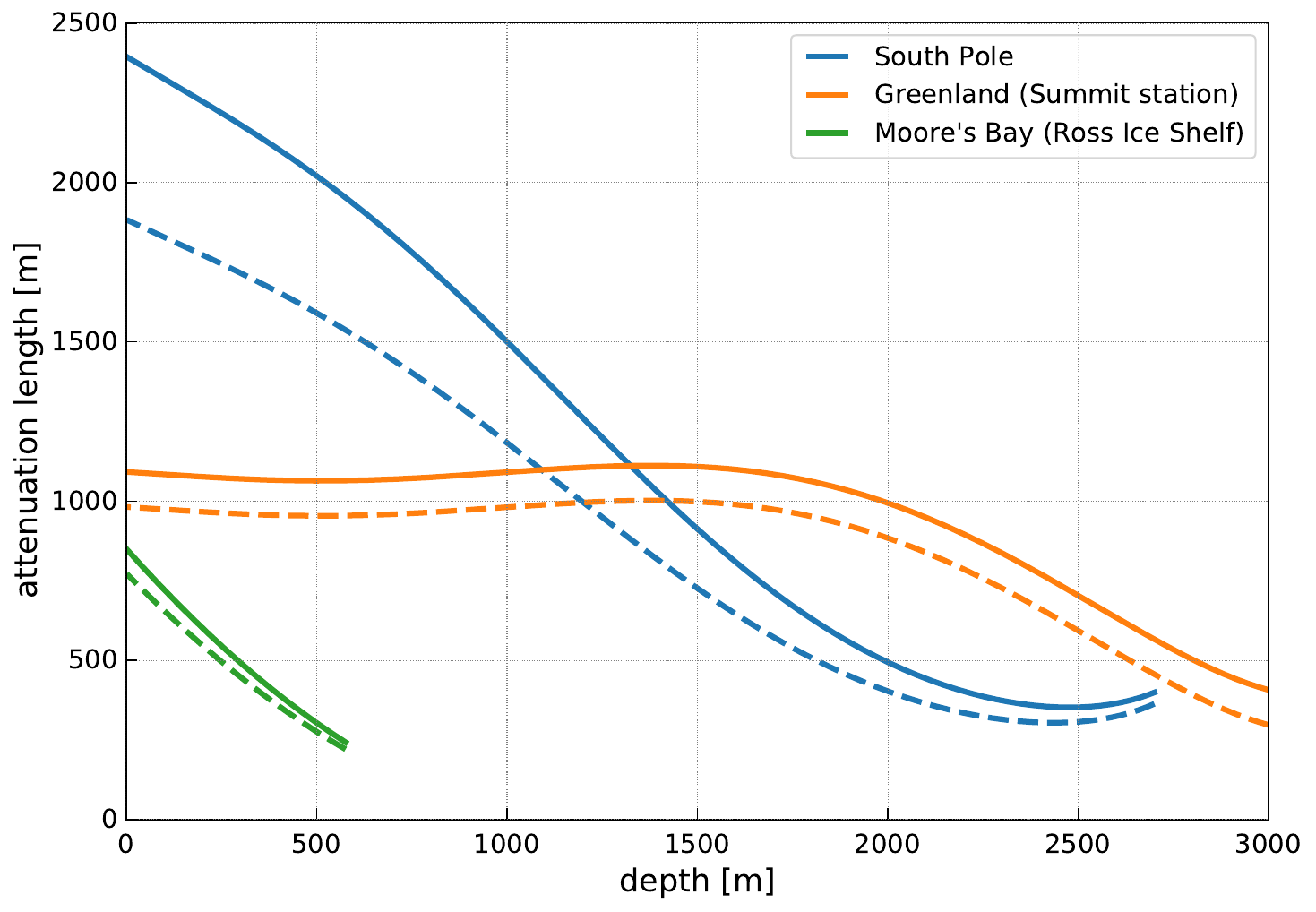}
    \caption{Attenuation length vs. depth for three different sites of a radio neutrino detector at South Pole, Moore's Bay on the Ross Ice Shelf and Greenland at Summit station as implemented in NuRadioMC \cite{NuRadioMC2019}. The solid lines show the attenuation length at \SI{200}{MHz} and the dashed lines show the attenuation length at \SI{400}{MHz}.}
    \label{fig:attenuation}
\end{figure}

An overview of the attenuation lengths of the three sites is shown in Fig.~\ref{fig:attenuation}. The site with the best ice it the South Pole which has the largest attenuation lengths in the more relevant upper \SI{1}{km} of the ice sheet. Remember that the antennas are placed at a maximum depth of \SI{200}{m}, thus, the ice quality close to the antennas matters most, especially at low neutrino energies where the neutrino interaction happens closer to the antennas. 
The site with the shortest attenuation lengths is Moore's Bay on the Ross Ice Shelf but because of the reflective properties of the bottom of the ice shelf and the resulting large sky coverage, the site provides important physics capabilities for a future detector.

\subsection{Index-of-refraction profile}
\begin{figure}[t]
    \centering
    \includegraphics[width=0.7\textwidth]{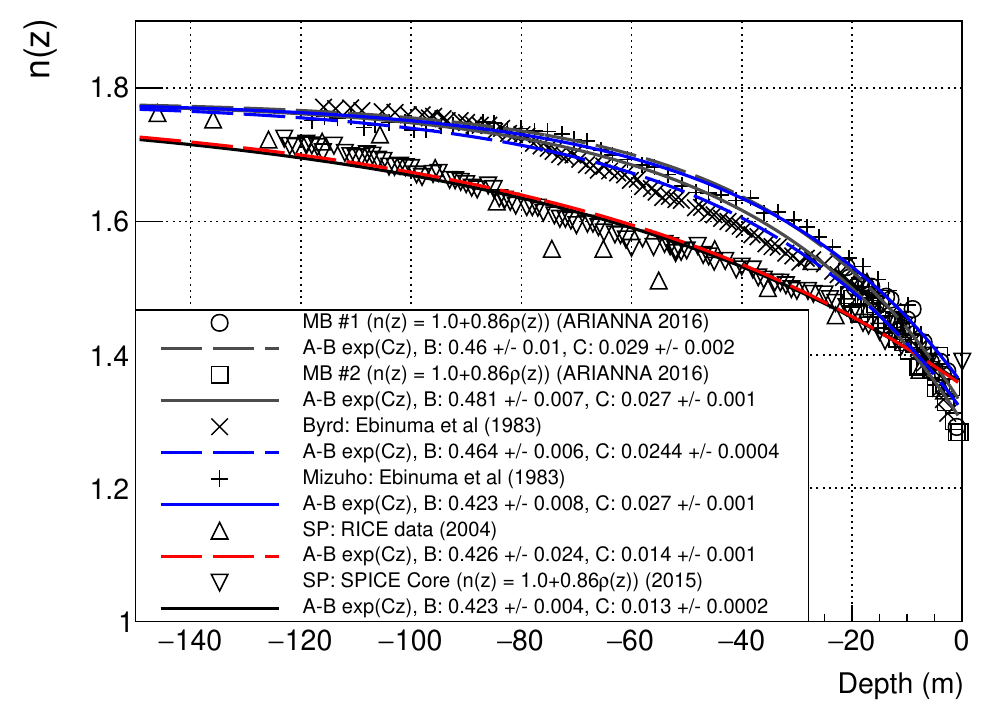}
    \caption{A compilation of various index-of-refraction measurements at Moore's Bay, Byrd station, Mizuho station and the South Pole. Also shown are a fit of an exponential function to the different data sets. Figure taken from \cite{Barwick2018}.}
    \label{fig:n_z_measurments}
\end{figure}

All sites have in common that the density changes in the upper $\mathcal{O}$(\SI{100}{m}). Due to the increasing gravitational pressure, the firn gets compressed into clear ice with increasing depth. To first order, the density at a given depth is directly dependent on the overburden of ice at that depth. Then, the density profile is continuous and follows an exponential \cite{Kravchenko2004}. An empirical depth–density relation has been given \cite{Schytt1958} as $\rho(z) = \rho_\mathrm{ice} - (\rho_\mathrm{ice} - \rho_\mathrm{surface}) \, \exp(-C\, z)$, where $\rho_\mathrm{ice}$ is the asymptotic density of polar ice (\SI{917}{kg/m^3}) and $\rho_\mathrm{surface}$ is the density at the surface. The index-of-refraction $n$ scales linearly with the density and is often parameterized as $n(z) = 1 + \SI{0.86}{cm^3/g}\ \times \rho(z)$ and thus follow the same shape as the density profile \cite{Kravchenko2004}. A compilation of measurements from different sites across Antarctica is shown in Fig.~\ref{fig:n_z_measurments} that can be described well with an exponential fit to the data points which is also shown in the figure. 

\begin{figure}[t]
    \centering
    \includegraphics[width=0.8\textwidth]{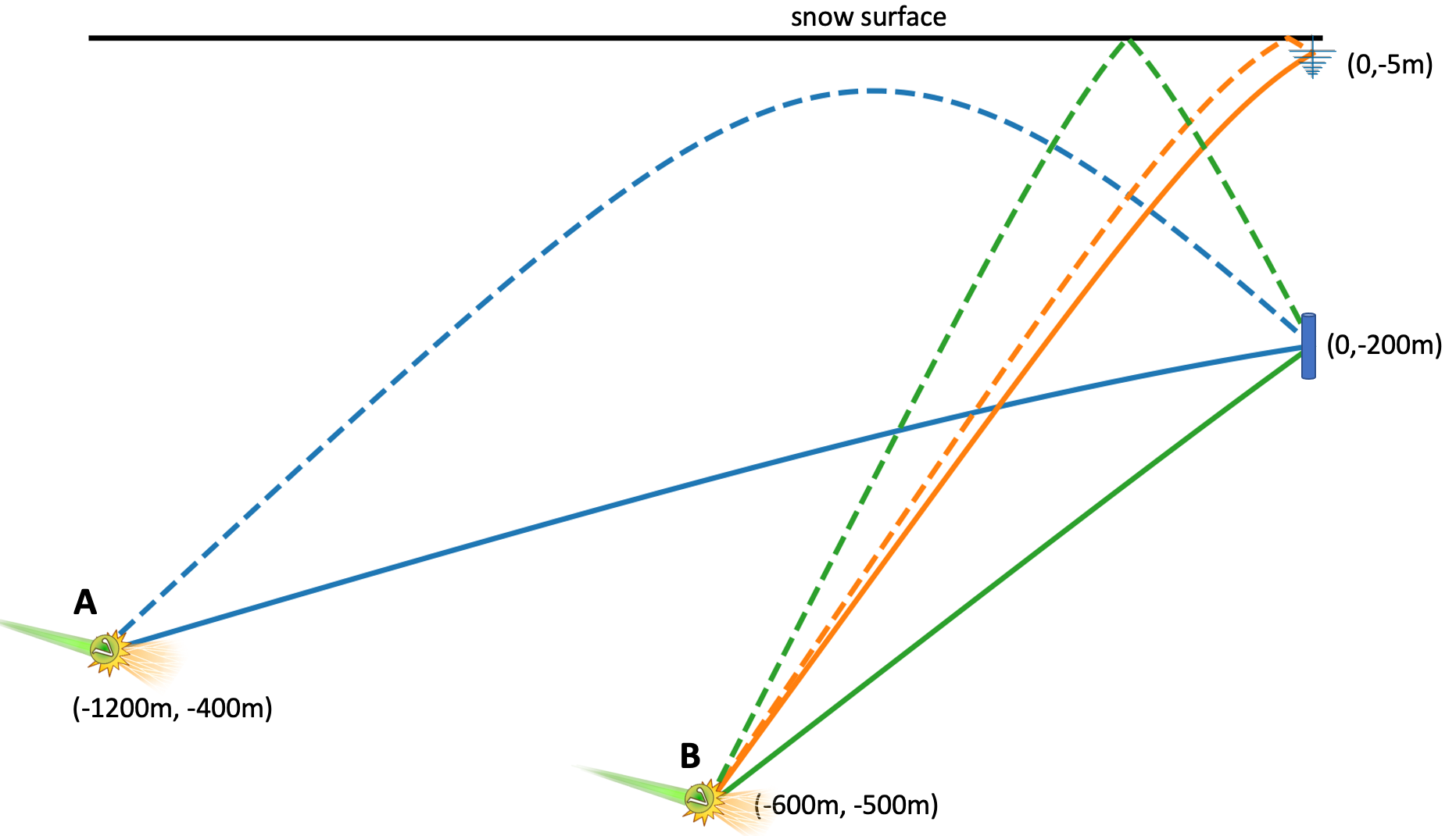}
    \caption{Example of signal trajectories for the index-of-refraction profile at the South Pole.}
    \label{fig:RT_example_SP}
\end{figure}

As a consequence of the changing index-of-refraction, radio waves propagating through the firn will bend downwards due to continuous Fresnel refraction which will result in curved signal trajectories. A signal can reach the receiver often via a direct path, and another path where the trajectory is bent downwards and reaches the receiver from above (refracted trajectory). We show a few typical trajectories in Fig.~\ref{fig:RT_example_SP}. The firn-air interface also reflects radio waves back into the firn, and for most expected geometries of neutrino radio signals, the condition of total-internal-reflection (TIR) is fulfilled which results in a lossless reflection of the signal leading to additional reflected signal trajectories. 
In general, between any two positions exist either zero or two signal trajectories, where the two signal trajectories are a combination of direct, refracted, and reflected trajectories. 

\begin{figure}[t]
    \centering
    \includegraphics[width=0.6\textwidth]{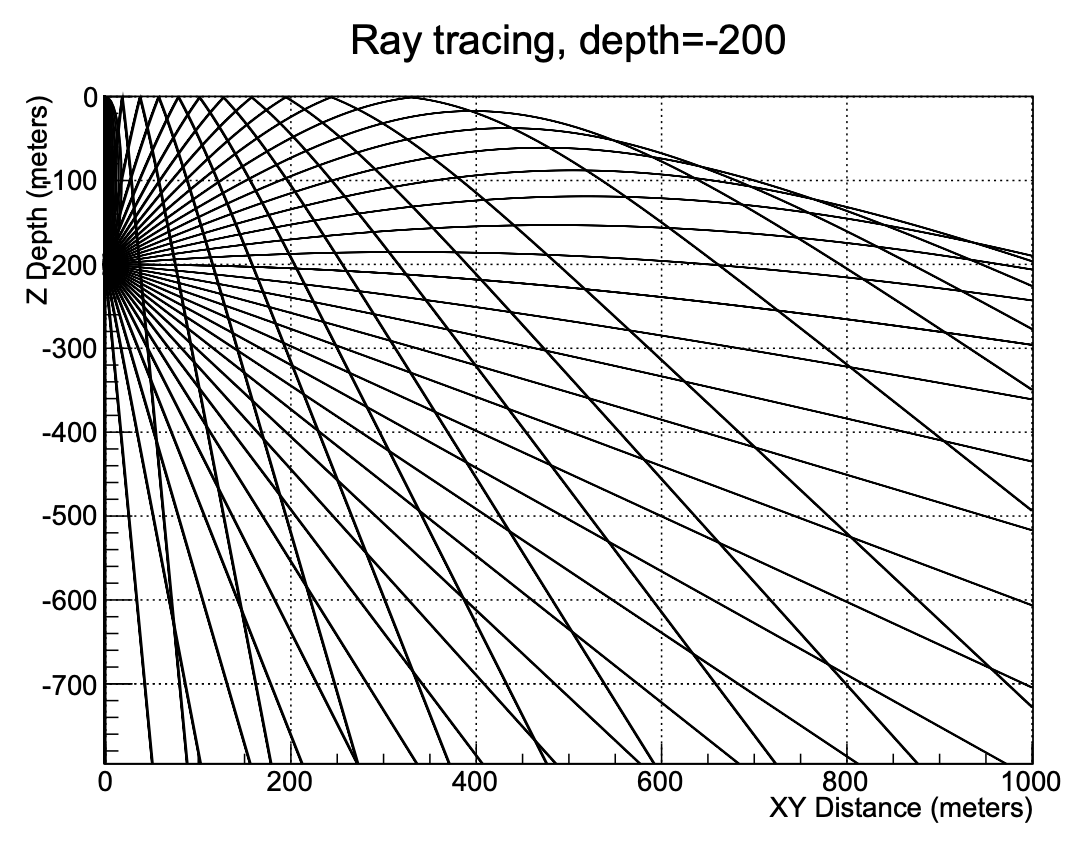}
    \caption{Visualization of the shadow zone effect. The black curves show the possible signal trajectories that end in a receiver at \SI{200}{m} depth. No signal trajectories from the white region in the upper right part to the receiver exist. This region is called the shadow zone. Figure from \cite{Barwick2018}.}
    \label{fig:shadow_zone_1}
\end{figure}

Another consequence of the bending of signal trajectories is that a receiver in the firn can only be reached from certain parts of the ice. This is illustrated in Fig.~\ref{fig:shadow_zone_1} that shows all possible signal trajectories from a \SI{200}{m} deep emitter. The region in the upper right of the plot can never be reached. This region is called the shadow zone. Another example was already visible in  Fig.~\ref{fig:RT_example_SP} where no signal path from position $A$ to the LPDA receiver close to the surface exists. The size of the shadow zone depends on the depth of the emitter. The deeper a receiver is placed, the more ice can be observed. 

\begin{figure}[t]
    \centering
    \includegraphics[width=0.6\textwidth]{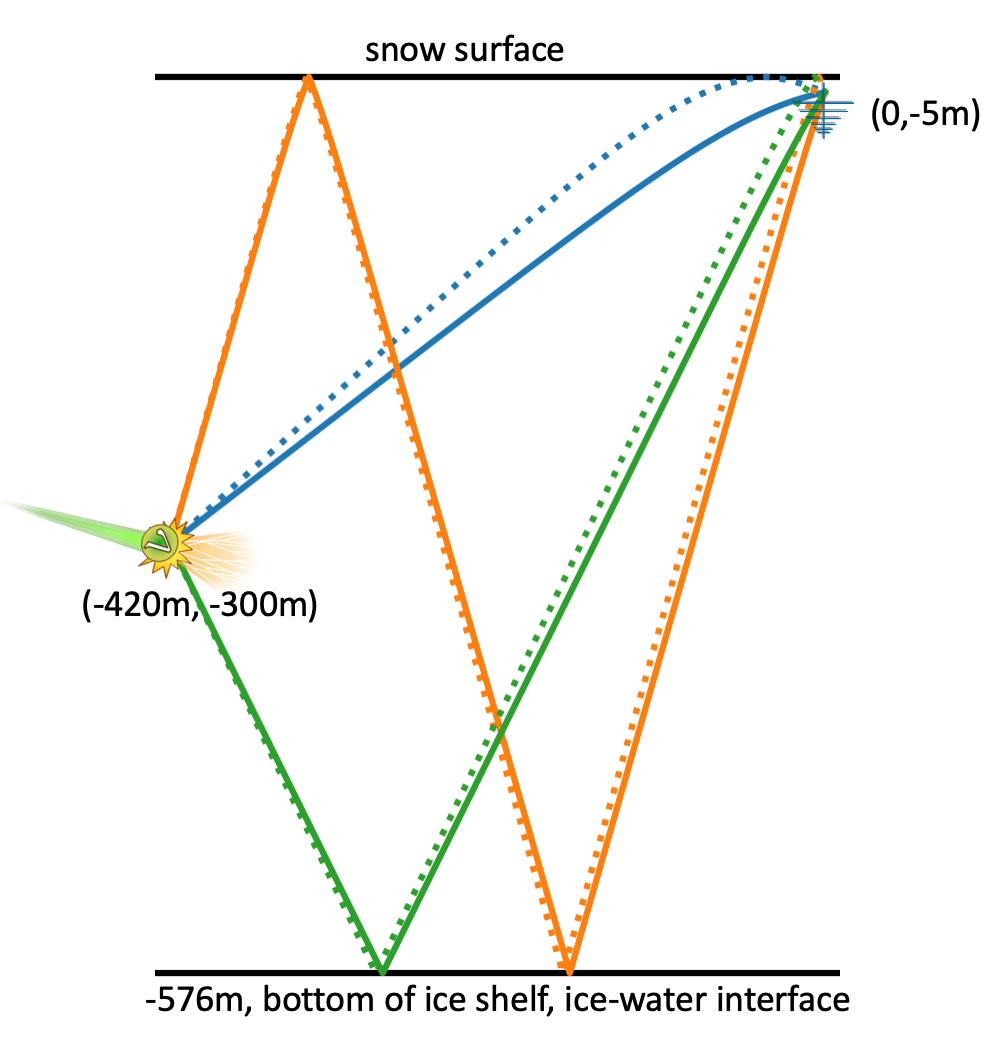}
    \caption{Example of signal trajectories for the index-of-refraction profile at Moore's Bay. The reflective layer at the bottom of the ice shelf leads to an infinite number of possible signal paths through multiple reflections between the surface and the bottom layer. Here, only signal trajectories with at most one reflection at the bottom layer are shown.}
    \label{fig:RT_example_MB}
\end{figure}

A special case is Moore's Bay where the ice-water interface of the bottom of the ice shelf reflects signals back upwards \cite{BarwickBergBessonEtAl2014} which is shown in Fig.~\ref{fig:RT_example_MB}. In principle, this would lead to an infinite number of possible signal paths, due to succeeding reflections at the surface and bottom of the ice shelf. However, due to the increasing attenuation of the signal, at most one bottom reflection is relevant. Even at very high neutrino energies (above \SI{e19}{eV}), only a small fraction of neutrinos are expected to be observed via signal trajectories with two bottom reflections. 

\subsection{Sky coverage}
\label{sec:skycoverage}
The sensitivity of a radio detector is largest to neutrinos with directions slightly above the local horizon. To both sides of the horizon, the sensitivity is limited: Neutrinos with directions below the local horizon get absorbed by the Earth as discussed in Sec.~\ref{sec:interaction} and shown in Fig.~\ref{fig:earth_absorption}. Towards the other side, i.e., for neutrino directions with a higher elevation, the shadowing effect limits the observable sky (except for Moore's Bay which we'll discuss later): As in almost all cases, the neutrino interaction takes place in the ice below the antennas and the radio signal needs to propagate upwards to reach the antenna. As the Askaryan signal is emitted on a cone with an opening angle of \SI{56}{\degree} around the neutrino direction, the elevation angle of the neutrino direction needs to be smaller than \SI{56}{\degree} for the signal to propagate upwards. But then, the signal trajectories are bend downwards due to the changing index-of-refraction. Thus, the Askaryan signal needs to be emitted at an even steeper angle to reach the receiver, resulting in smaller elevation angles of the neutrino direction. As for the shadow zone, this effect is stronger the shallower the receiver is. 

\begin{figure}[t]
    \centering
    \includegraphics[width=0.95\textwidth]{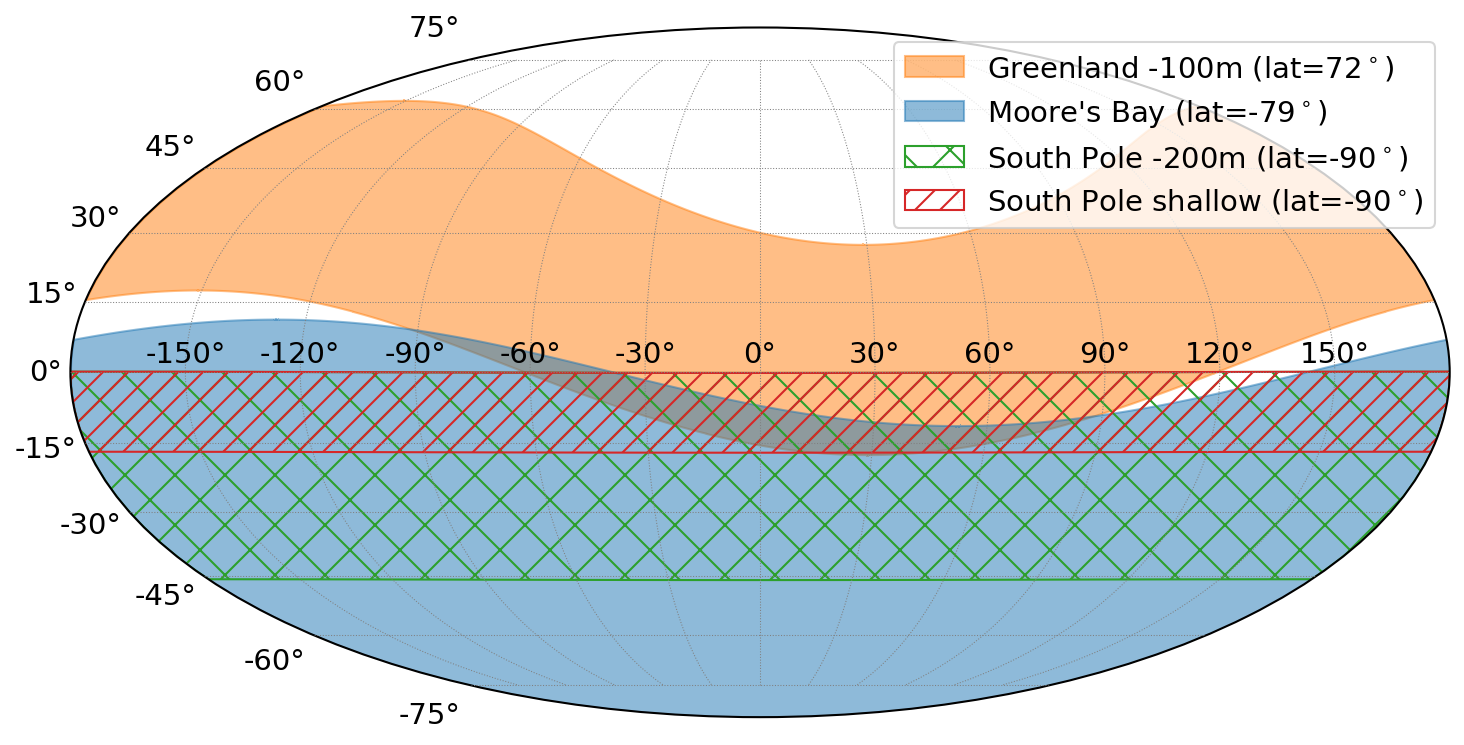}
    \caption{Sky coverage of Askaryan detectors for a neutrino energy of \SI{e18}{eV} at different sites at one particular time of the day in Equatorial coordinates. During the course of one day, the field of view will rotate around the vertical axis due to the rotation of the Earth. Only a detector at the South Pole observes always the same sky due to its special location. For Greenland \cite{Aguilar:2020RNOG} (latitude \SI{72}{\degree}), the antennas are placed at a depth of \SI{100}{m}. For the South Pole \cite{HallmannICRC2021} (latitude \SI{-90}{\degree}), two cases are shown, one with antennas at \SI{200}{m}, and one with shallow antennas just a few meters below the surface. For Moore's Bay \cite{ARIANNA200} (latitude \SI{-79}{\degree}), the depth of the antennas is irrelevant due to the reflective bottom layer of the ice shelf.}
    \label{fig:sky_coverage}
\end{figure}

A special case, again, is Moore's Bay on the Ross Ice shelf. Due to the high reflectivity of the bottom of the ice shelf, downward signal trajectories can still reach the antennas as the signals get reflected back upwards. Therefore, a neutrino detector at Moore's Bay can view the whole sky above the local horizon at any instant. 
Only at low neutrino energies of $\mathcal{O}$(\SI{e17}{eV}), the generated radio signals are too weak, and only neutrino interactions close to the antennas, where the signal reaches the antennas on a direct path, can be detected. At the target energies of \SI{e18}{eV} and beyond, the emitted radio signals are strong enough to still lead to a measurable signal in the antennas after propagating back up through the ice shelf. 

We show the field of view in Fig.~\ref{fig:sky_coverage}. A detector at Moore's Bay has the largest sky coverage. The complete southern hemisphere and an important part of the northern hemisphere is observed. Also, a detector at Greenland covers a significant part of the sky over the course of one day due to its comparably low latitude of \SI{72}{\degree}. A combination of detectors at Moore's Bay and Greenland will cover almost the entire sky except for a small region around the northern pole in Equatorial coordinates. A detector at the South Pole does not benefit from Earth rotation and observes the same sky at any instant of time. 

\subsection{Birefringence}
Ice is a birefringent crystal, which means that the index-of-refraction has a (small) dependence on the polarization and propagation direction of the radio signal. As a consequence, the Askaryan pulse splits up into two orthogonal polarization states with slightly different propagation speeds, and the time delay between the two components scales linearly with propagation distance \cite{Allison:2017jpy,Besson:2021wmj}. The size of this effect is depth-dependent and varies between detector sites.  For randomly oriented ice crystals, as deposited by snowfall or wind blown, the net birefringence of the ice fabric is expected to be very small.  Near the surface, the ice crystals retain their random orientation at deposition, but mechanical stresses from the pressure created by an ice overburden will induce a small alignment of the optical axis  (which are correlated with the anisotropy of mechanical stress properties of the ice crystal). Due to the limited overburden of ice at Moore's Bay and the melting of the bottom of the ice sheet as it slides into the water, no significant birefringence is expected.  However, at the South Pole, birefringence was measured with time delays of \SI{10}{ns} over a propagation distance of \SI{1}{km} \cite{Jordan2020}.

There is still a lot of work to do in understanding and parameterizing the birefringence of polar ice but a first fairly simple model of South Pole ice already shows good agreement with experimental data \cite{Jordan2020}: Birefringence depends on the bulk orientation of the ice crystals which is determined primarily by the vertical compression over overlying snow and ice and secondarily by the lateral flow of the ice, which tends to be predominantly in one direction due to the stability of gravity gradients over time.  Thus, the index-of-refraction tensor (or more generally the dielectric tensor) can be decomposed into the vertical direction and the two lateral directions parallel and perpendicular to the ice flow. 

The need to understand birefringence is clear.  Reliable modeling of birefringence is crucial to predict how Askaryan pulses will appear in the detector. Once birefringence effects are parameterized, the time delay between the two polarization components is a powerful estimator of the distance to the neutrino interaction,  which is needed to estimate the neutrino energy. But it may also serve to confuse the situation. In particular, because the polarization of the two birefringence eigenstates, i.e., the slower and faster propagating component, can change during propagation for certain geometries leading to interference effects \cite{Connolly:2021cum,Heyer:2022ttn}. This can alter the signal polarization and thereby effect the neutrino direction reconstruction or result in atypical waveforms that require special treatment. Recent work allows a detailed simulation of these effects \cite{Heyer:2022ttn} and work is ongoing to quantify its impact on neutrino detection. 

Perhaps most worrisome, at this point, there is no firm idea on the scope of the effort to fully measure and characterize the variation of birefringence for all relevant propagation paths and interaction depths at the various polar sites under development. It may require considerable effort (and related support) to reach the necessary level of precision.

\subsection{Second order propagation effects}

Additional propagation effects arise from density fluctuations around a smooth index-of-refraction profile that result in the existence of potentially detectable (though generally small) signals coming from regions where there is no ray-tracing solution \cite{Barwick2018}, diffraction and interference of the radio waves, and the presence of caustics, where the small electric field may be significantly amplified in some geometries \cite{Deaconu2018,RadarEchoTelescope:2020nhe}. Research is ongoing to study these effects through finite-difference time-domain (FDTD) simulations that evolve Maxwells equation in the time domain. However, their large computational cost limits their practical usability and faster surrogate models are needed \cite{RadarEchoTelescope:2020nhe}. Furthermore, layered impurities in the ice can reflect up to 1\% of downgoing signals back up \cite{Besson:2021wmj} (see also discussion of backgrounds below).

Apart from a better modelling in simulations, experimental data will be key to understand and quantify these effects. More precise measurements of the density and index-of-refraction profiles are needed to inform the simulations, as well as precise in-situ calibration measurements of radio propagation through polar ice to experimentally prove these second order effects. 

\section{Detector Designs and Performance}

The study of different detector designs requires a reliable and flexible simulation code. As discussed in the previous sections, the four components needed for an accurate MC simulation (neutrino interaction in ice, Askaryan emission from in-ice showers, propagation of radio signals, and modeling of the detector response) are well understood. The formation of particle showers upon a neutrino interaction and the Askaryan generation is theoretically well understood and can be calculated in great detail in microscopic MC simulations that were confirmed in accelerator measurements as well as air-shower measurements. Also the radio detector response can be modeled in great detail as demonstrated by radio detectors of air showers. Most propagation effects are also understood and experimentally verified including attenuation as a function of frequency, refraction in the firn, and reflections from ice-water and firn-air boundaries. Other effects like birefringence and radio propagation from the shadow zone need more work and may require extensive experimental campaigns to fully understand how to properly model this physics. 

The in-ice radio community developed the NuRadioMC simulation code \cite{NuRadioMC2019,NuRadioReco2019} that incorporates our current best knowledge of the physical processes in a flexible way so that different detector designs can be simulated quickly. It was shown that the results agree within a few percent with previous MC codes if the same physics settings were used. NuRadioMC allows to make reliable predictions on the sensitivity of a radio detector and can be used for a reasonable estimate of reconstruction performance. Current developments focus on a more detailed simulation of second-order propagation effects such as birefringence or modeling the effects of density fluctuations which are important for improving event reconstruction. 

The properties of relevance that each detector design needs to accomplish is primarily a sufficient sensitivity to high-energy neutrinos and secondarily the ability to reconstruct the neutrino properties of interest, i.e., the neutrino direction and energy, and ideally also the neutrino flavor. All of that should be achieved while keeping the costs, deployment efforts, and maintenance requirements low. In addition, the observable sky (which we already discussed in the previous section) and the ability for real-time processing of events to send out multi-messenger alerts play an important role. In this chapter, we discuss the sensitivity of a radio neutrino detector and its ability to reconstruct the neutrino properties of interest. The more practical aspects of deployment, maintenance, and data processing will be discussed in chapter \ref{sec:operations}. 

An in-ice radio detector consists of an array of largely independent radio detector stations with typical distances between stations of about one kilometer. This allows for cost-efficient instrumentation of large volumes but also means that each station needs to be capable of triggering on the Askaryan signal, of discriminating it against noise, and of reconstructing the neutrino properties from it. This is unlike the IceCube detector where a part of the ice is continuously instrumented with optical sensors that all work together, or air-shower radio arrays where a coincident observation of the radio signal in at least three detector stations is required. 

\subsection{Sensitivity to high-energy neutrinos}
The sensitivity to neutrinos is typically quantified via the effective volume $V_\mathrm{eff}$ which can be calculated from MC simulations by multiplying the simulation volume $V$ with the fraction of observable neutrino events where each event $i$ gets an additional weight $\omega_i$ corresponding to the probability of the neutrino to reach the simulation volume, i.e., to not get absorbed in the Earth (cf. Fig.~\ref{fig:earth_absorption})
\begin{equation}
    V_\mathrm{eff} = V/N \sum_{i \in \mathrm{triggered}} \omega_i \, .
\end{equation}
The effective volume can be converted into an effective area by dividing by the interaction length of neutrinos in ice. The expected sensitivity can be calculated directly from the effective area and is inversely proportional to it. The number of detectable events for a given flux is also linearly proportional to the effective area. 

The easiest way to increase the sensitivity -- but also the most expensive way -- is to build more radio detector stations and to place them with sufficient separation so that their active volume has little overlap. In this case, the effective volume scales (almost) linear with the number of stations. Because of logistical considerations, the detector stations can't be placed at arbitrary distances, which means that overlap can't be avoided completely, especially at high energies. For example, for detector stations with a \SI{1.5}{km} separation and antennas placed at \SI{200}{m} depth, 20\% of the neutrino interaction at \SI{e18}{eV} are observed in more than one station. For an order of magnitude lower energy of \SI{e17}{eV}, the coincidence fraction drops quickly to zero, and for an order of magnitude larger energy of \SI{e19}{eV}, about 60\% of the neutrinos are observed in more than one station. The optimal spacing will be a compromise between total sensitivity, logistical constraints, and the need for a golden event sub-sample of multi-station coincidences.

The better way to increase the overall sensitivity is to increase the sensitivity of each detector station. This can be achieved by lowering the trigger threshold. The challenge is that current Askaryan detectors already operate at such low trigger thresholds so that the vast majority of triggered events are just unavoidable thermal noise fluctuations. The trigger threshold is then set by the maximum data rate the detector can handle. To give an example, the thermal-noise trigger rate decreases by an order-of-magnitude every time the amplitude threshold is increased by 10\% - 15\% \cite{Glaser2020}. There are several approaches to overcome this problem. 

\begin{itemize}
    \item \textbf{More sensitive antennas:} One way to increase the sensitivity is to use better antennas. The best antennas are LPDA antennas which are essentially an array of dipole antennas of different lengths that are spaced at a distance to achieve constructive interference. This leads to enhanced sensitivity and broad frequency response. The downside is that these antennas occupy a relatively large area with a triangular shape of a height of \SI{1.4}{m} and maximum width of \SI{1.5}{m} \cite{ARIANNATimeDomain2015}. Therefore, these antennas can only be deployed close to the surface where a slot can be dug or melted. 
    \item \textbf{Interferometry of low gain antennas:} If antennas are deployed deeper in the ice, for example at \SI{200}{m}, the antennas must fit into a narrow borehole which limits the choice of antennas. To still increase the sensitivity, several low-gain antennas in close proximity can be combined digitally. In an ideal case, the signal amplitude increases with the number of antennas $N$ whereas the noise only increases with $\sqrt{N}$. Thus, the signal-to-noise ratio scales with $\sqrt{N}$. Such an interferometric phased array trigger system has been build and successfully tested at the South Pole \cite{ARA2019-PA} where 7 dipole antennas are placed vertically above each other with \SI{1}{m} separation. The vertical orientation results in an azimuthal symmetry, the close proximity results in similar Askaryan signals in the antennas, and sufficient \emph{beams} are formed (corresponding to different time delays between the channels to cover different incoming signal directions) so that the improvement in signal-to-noise ratio is close to the theoretical $\sqrt{N}$. 
    \item \textbf{Optimization of the trigger bandwidth:} The amplitude of the Askaryan signal increases linearly with frequency up to a cutoff frequency that depends on the viewing angle but extends up to \SI{1}{GHz} if the shower is observed close to the Cherenkov angle. However, it was found that the sensitivity to neutrinos can be increased by up to 50\% if the bandwidth is reduced to the lower end of the frequency band of around $80-200~\mathrm{MHz}$ \cite{Glaser2020}. This is a) because the noise RMS decreases with the square root of the bandwidth, b) because antennas are typically more sensitive at low frequencies (the relevant vector effective length is inverse proportional to the frequency and proportional to the square root of the gain), and c) because, for a given amplitude at the detector, the geometry of off-cone events is more favorable where the frequency cutoff quickly drops into the hundred MHz range.
\end{itemize}

\begin{figure}[t]
    \centering
    \includegraphics[width=0.7\textwidth]{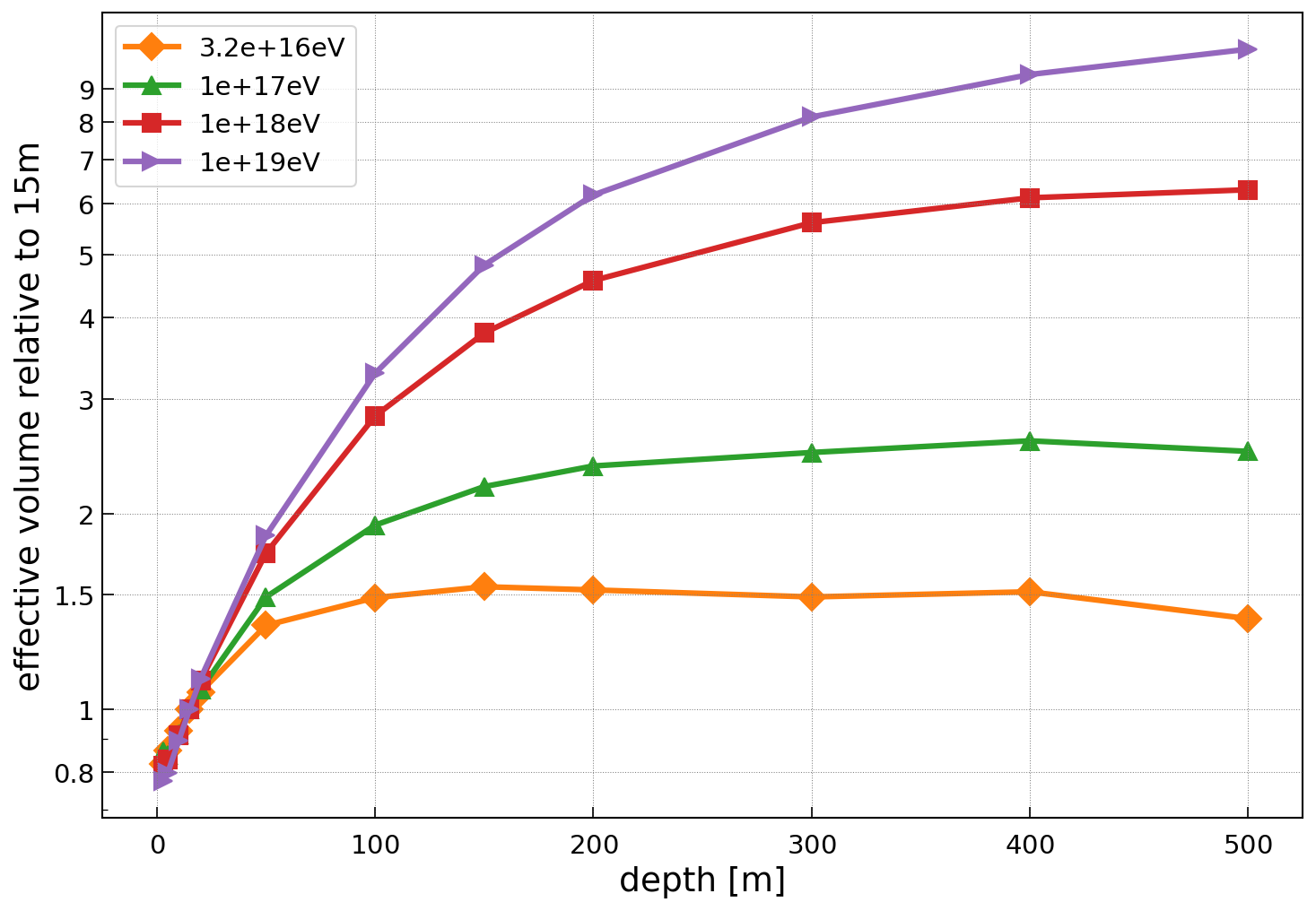}
    \caption{The dependence of the neutrino sensitivity (measured in effective volume) of a single radio detector station with depth of the triggering antennas normalized to the sensitivity at \SI{15}{m}. The simulation was done using NuRadioMC for the South Pole ice and using dipole antennas to measure and trigger the radio signal.}
    \label{fig:Veffdepthdependence}
\end{figure}

Another aspect that should be mentioned is the depth of the antennas. The sensitivity of a \emph{single} radio detector station can be increased by placing the antennas at a deeper depth which allows observing more of the surrounding ice. This is due to the bending of signal trajectories as discussed in the previous section. To get a feeling for this effect we show the scaling of the sensitivity with depth for the South Pole in Fig.~\ref{fig:Veffdepthdependence}. The improvement is strongest for larger neutrino energies. By going down to \SI{200}{m}, the sensitivity can be more than doubled at \SI{e17}{eV} and increased to almost a factor of five at \SI{e18}{eV}. However, also the cost of a radio detector station increases with depth which removes most of the apparent benefit. Also, logistical considerations play a role, e.g., it is considered cost-prohibitive to go below \SI{200}{m} with current drilling technology, and the larger observable ice volume leads to more overlap between radio detector stations which reduces the sensitivity of the array for the same station spacing. On the other hand, more shallow stations need to be deployed to achieve the same sensitivity. It turns out that the achievable sensitivity for the same total cost is similar between an array of shallow stations and an array of deep stations which can, for example, be seen by comparing the expected sensitivities of the RNO-G and ARIANNA-200 detector which we'll discuss in Sec.~\ref{sec:prototype_arrays}.
In the end, it is a complex optimization problem where not only the total sensitivity of an array of radio detector stations per unit cost is relevant but also the ability to reconstruct the neutrino direction, energy, and flavor, as well as the ability to efficiently reject all backgrounds which we discuss in Sec.~\ref{sec:Background}. 

\subsection{Sensitivity to neutrino direction and energy}
In this section, we discuss which quantities of the Askaryan signal need to be measured in order to reconstruct the direction and energy of the neutrino which is largely based on Ref.~\cite{GlaserICRC2019}. This understanding will be the basis of discussing different station designs at the end of this chapter. 

We start by presenting a simplified parameterization of the Askaryan signal at the antenna as a function of the neutrino parameters and event geometry. Although this is an overly simplified picture, all main dependencies are described. The electric field $\vec{\varepsilon}(f)$ as a function of frequency $f$ arriving at the antenna can be expressed as
\begin{equation}
    \vec{\varepsilon}(f) = \vec{e}_p(\vec{v}_\nu, \vec{r}, R) \times |\vec{\varepsilon}_0|(f) \times \frac{e^{-R/L(f)}}{R} \times \exp\left[\frac{-(\theta - \theta_C)^2}{2 \sigma_\theta(f)^2}\right] \, .
    \label{eq:all}
\end{equation}
The four parts of the equation are discussed in the following.

\begin{figure}[t]
    \centering
    \includegraphics[width=0.8\textwidth]{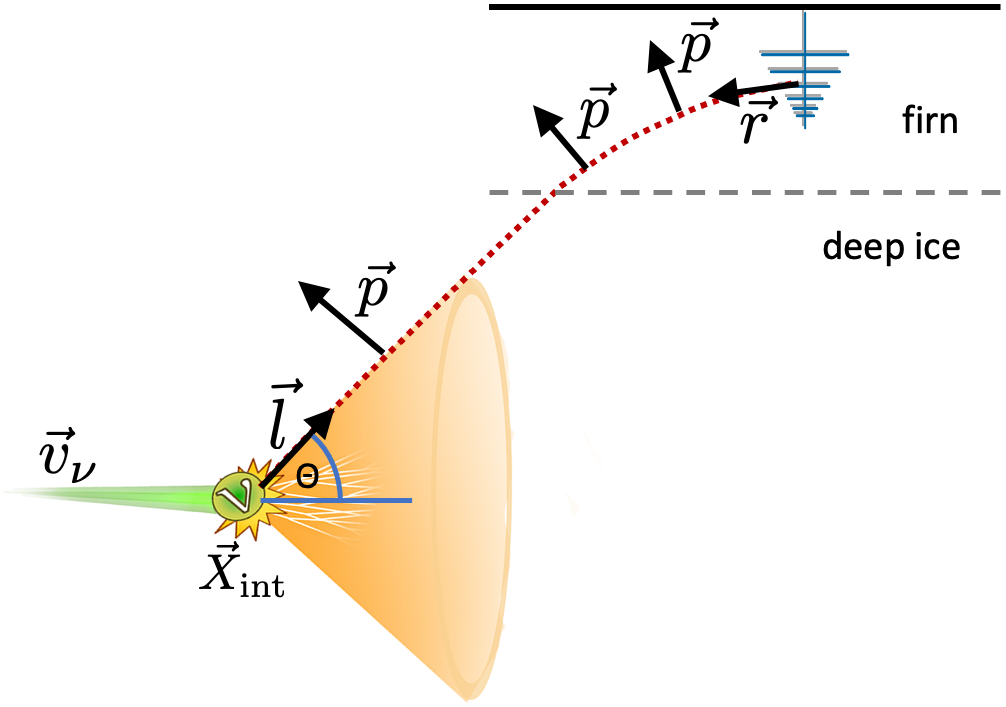}
    \caption{Sketch of the event geometry and how the polarization of the radio signal depends on the geometry. Please note that this is just a 2D projection of the problem.}
    \label{fig:directioreco_geometry}
\end{figure}

\subsubsection{Signal polarization}
The first part of the equation $\vec{e}_p$ describes the polarization of the signal (i.e. the orientation of the electric field vector). The polarization is approximately perpendicular to the Cherenkov cone (see Fig.~\ref{fig:directioreco_geometry} for an illustration). To be exact: the polarization of the Askaryan signal $\vec{p}$ at the point of emission is perpendicular to its direction of propagation (the launch vector $\vec{l}$) and the plane spanned by the neutrino direction $\vec{v}_\nu$ and the direction of signal propagation:
\begin{equation}
    \vec{p} = \vec{l} \times (\vec{v}_\nu \times \vec{l}) \, ,
    \label{eq:polarization}
\end{equation}

Because the signal direction can change due to the changing index-of-refraction in the firn,
the polarization can also change according to the bending of the signal path to remain perpendicular to it. To infer the launch vector from the measurable signal arrival direction at the antenna (the receive vector $r$ in Fig.~\ref{fig:directioreco_geometry}), the signal trajectory is back propagated through the ice using the knowledge of the index-of-refraction profile and using the measured distance to the neutrino interaction. For most cases though, one can safely assume that the neutrino interaction happened below the firn. 

\subsubsection{Inelasticity}
\label{sec:y}
The second term of Eq.~(\ref{eq:all}) $|\vec{\varepsilon}_0|(f)$ represents the dependence of the Askaryan signal amplitude on the neutrino energy. The signal amplitude does not depend directly on the neutrino energy but scales linearly with the shower energy. The fraction of energy transferred into the shower (the inelasticity) is a stochastic process that limits the achievable energy resolution. 

An exception is $\nu_e-CC$ (electron neutrino charged current) interactions where the complete neutrino energy is transferred into an electromagnetic and a hadronic shower (cf. Sec.~\ref{sec:interaction}). If both showers are measured, the sum of both shower energies gives the neutrino energy. Thus, no uncertainty from unknown inelasticity is present. At shower energies where the LPM effect is not yet important, the electromagnetic and hadronic shower develop in-phase and the two showers can be approximated with a single particle shower with an energy equivalent to the full neutrino energy. Here, the uncertainty from an unknown inelasticity is negligible. For larger energies where the LPM effect becomes relevant, the two showers do not develop in phase anymore and at even higher energies the electromagnetic shower on its own will consist of several spatially displaced sub showers. This complicates the measurement but it might still be possible to measure the different showers which reduces the uncertainty from inelasticity. We note that the signatures of $\nu_e-CC$ interactions allow to discriminate them from all other interaction channels which provides flavour sensitivity \cite{Stjarnholm:2021xpj}.

Another exception is charge-current interactions of muon or tau neutrinos where the initial neutrino interaction is observed as well as a secondary interaction of the muons or tau lepton. For these event topologies, the neutrino energy can likely be estimated more precisely but has so far not been quantified. 

Nevertheless, a good benchmark can be obtained by inspecting only hadronic showers, i.e., all interactions that are not $\nu_e-CC$, where one finds that the unknown inelasticity limits the achievable energy resolution to about a factor of two \cite{DnR2019}. This sets the scale for the required experimental precision: The uncertainties of other quantities impacting the energy reconstruction should be small enough to not significantly increase the energy uncertainty beyond the inelasticity limit but do not need to be much more precise either.

\subsubsection{Signal attenuation}
The third term of Eq.~(\ref{eq:all}) describes the attenuation of the radio signal during propagation. The signal amplitude decreases proportionally to the distance $R$ to the neutrino vertex. In addition, the signal is attenuated exponentially as $e^{-R/L(f)}$ where $L(f)$ is the frequency-dependent attenuation length with typical values ranging from \SI{500}{m} to \SI{2.5}{km} at the South Pole (cf. Sec~\ref{sec:attenuation}). 

\subsubsection{Viewing angle}
The last term of Eq.~(\ref{eq:all}) describes the dependence on the viewing angle $\theta$, i.e., the angle between the neutrino direction and the launch vector. As discussed in Sec.~\ref{sec:AskaryanEmission}, the signal amplitude is largest if the shower is observed at the Cherenkov angle $\theta_C = \cos^{-1}(1/n_\mathrm{ice}) \approx \SI{55.8}{\degree}$ in deep ice. If the viewing angle deviates from the Cherenkov angle, the signal amplitude drops quickly and can be modeled approximately with a Gaussian function where the width $\sigma_\theta$ depends on frequency. The width decreases with increasing frequency and typical values are a few degrees.  

To summarize: The quantities an in-ice radio detector needs to be sensitive to are the \textbf{incoming signal direction}, the \textbf{signal polarization}, the \textbf{distance to the neutrino vertex} and the \textbf{viewing angle}. Now that we obtained an understanding of how the radio signal depends on the event geometry and neutrino properties, we will discuss in more detail how the neutrino direction, energy and flavor can be measured.

\subsection{Measurement of neutrino direction}    
\label{sec:angularresolution}
To obtain the neutrino direction from a measurement, Eq.~(\ref{eq:polarization}) can be solved for the neutrino direction 
\begin{equation}
    \vec{\hat{v}}_\nu = \sin \theta \vec{\hat{p}} + \cos \theta \, \vec{\hat{l}} \, ,
\end{equation}
where the $\hat{}$ symbol indicates that the vectors have unit length, $\theta$ is the viewing angle, and $\vec{\hat{p}}$ is the signal polarization.
The launch vector $\vec{l}$ corresponds to the incoming signal direction after correcting for the bending in the firn. In practice, almost all neutrinos observed with an Askaryan detector will have the interaction vertex below the firn (see example 1 of \cite{NuRadioMC2019}). Thus, the exact distance to the interaction vertex is not needed to correct the incoming signal direction and polarization for the bending in the firn. Thus, a measurement of the incoming signal direction, polarization, and viewing angle is required to determine the neutrino direction.

\subsubsection{Incoming signal direction} 
The incoming signal direction can be reconstructed precisely from the pulse arrival times in multiple antennas. The exact resolution varies with the number of antennas that see a signal, the respective signal-to-noise ratios, and the distance between antennas but typically a resolution of (much) better than \SI{1}{\degree} is achieved (see e.g. \cite{ARIANNAPolarization2020}).  

\subsubsection{Signal polarization}
The precision of the polarization reconstruction depends strongly on the experimental setup and can range from a one to a few degrees to being largely unconstrained. The determination of the polarization requires a measurement of the radio signal in at least two antennas with orthogonal polarization response. Because of the transversality of electromagnetic waves, the radial polarization state is zero and only the two polarization states orthogonal to the propagation direction need to be determined with two orthogonal measurements. 

This can be achieved with two LPDA antennas that are rotated \SI{90}{\degree} with respect to each other which measure the two orthogonal horizontal polarization states. This approach is used in the ARIANNA experiment. The usage of the same antenna type with different orientations has the advantage that most systematic uncertainties cancel out in the polarization measurement. To improve the polarization resolution, a vertically oriented dipole antenna can be added to directly measure the vertical polarization state. Another option, that is used in deep radio detectors, is to combine vertically oriented dipoles with quad-slot antennas that are sensitive to the horizontal polarization component and still fit into a narrow borehole. The polarization measurement is more challenging for the latter setup as the quad-slot antennas have significantly smaller gain compared to the dipole antennas, and a good description of the antenna response is required as systematic uncertainties in the antenna response do not cancel out. 

An experimental test of the polarization reconstruction capabilities was performed by the ARIANNA collaboration with a shallow radio detector station at the South Pole \cite{ARIANNAPolarization2020}. The station comprises two pairs of orthogonal downward facing LPDA antennas and 4 vertically oriented dipole antennas. Short radio pulses of known characteristics were emitted from about a kilometer away from various depths by lowering an emitter into a \SI{2}{km} deep borehole. The measurement showed a polarization resolution of $\SI{2.7}{\degree}$. Most of the scatter originated from a variation of the measured polarization with the depth of the emitter which might be attributed to an emitter effect. The polarization resolution for any small depth interval was only \SI{1}{\degree}. Furthermore, radio signals from cosmic-ray-induced air-shower were used to measure the polarization resolution to $\mathcal{O}(\SI{1}{\degree})$ \cite{Arianna:2021lnr}.

\subsubsection{Viewing angle}
The viewing angle can be measured via two complementary techniques: First, via mapping of the Cherenkov cone via the measurement of the Askaryan signal in multiple antennas that observe the shower under different viewing angles. This requires a sufficient spatial separation between the antennas. The optimal spacing will depend on the vertex distance, hence, it will be difficult to optimize a detector layout equally well for all possible neutrino events.
For antennas close enough to the surface, already a single antenna can perform this measurement. A dipole antenna at a depth of \SI{10}{m} - \SI{15}{m} will observe one direct signal pulse and another pulse that is reflected off the snow surface \cite{DnR2019}. For most geometries, the reflection occurs under total-internal-reflection and a receiver depth of \SI{10}{m} is deep enough to have the two pulses be separated sufficiently in time to be resolved. 

Second, the viewing angle can be determined by measuring the frequency spectrum of the Askaryan signal \cite{Schoorlemmer:2015afa,Welling:2019scz,Aguilar:2021uzt}. The frequency spectrum of Askaryan pulses increases in amplitude with frequency up to a cutoff frequency. The cutoff frequency depends on the viewing angle. It is highest ($> \SI{1}{GHz}$) at the Cherenkov angle ($\theta_C$) and decreases with increasing deviation from the Cherenkov angle ($\Delta\Omega$, cf. Fig.~\ref{fig:askaryan_pulse}). Thus, the requirement for the detector to measure the viewing angle is a broad frequency response. Using only this method to measure the viewing angle often results in an ambiguity of being inside or outside of the Cherenkov cone, i.e., $\theta_C + \Delta\Omega$ and $\theta_C - \Delta\Omega$ result in almost the same frequency spectrum.

\subsubsection{Illustration of uncertainty contours of neutrino direction}
\begin{figure}[t]
    \centering
    \includegraphics[width=0.7\textwidth]{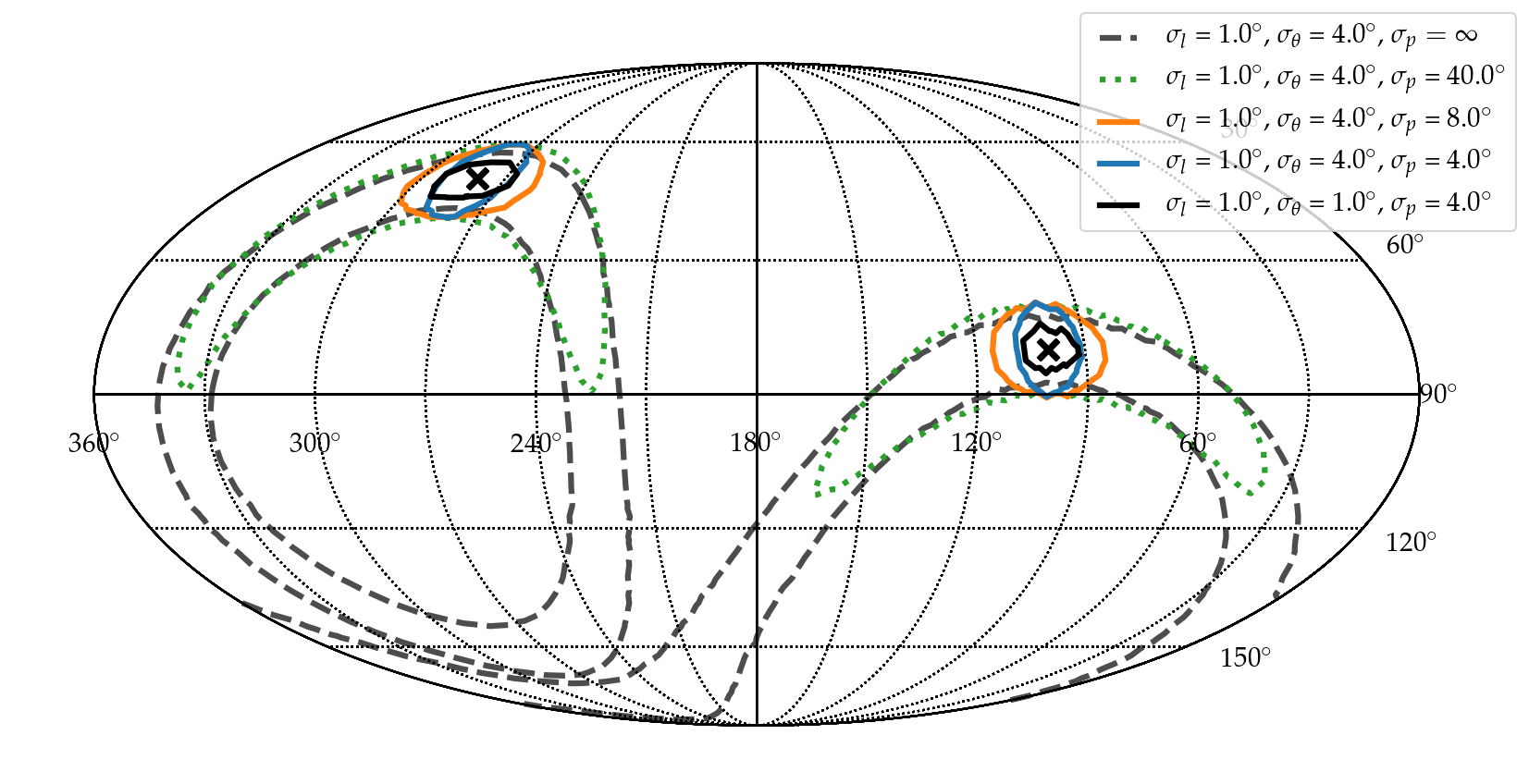}
    \caption{Skymap in local coordinates. The crosses indicate two hypothetical neutrino directions. The lines show the 90\% CL contours of the reconstructed neutrino direction for different assumptions on uncertainties. Figure from \cite{GlaserICRC2019}.}
    \label{fig:skymap_uncertainty}
\end{figure}

The complex dependence of the neutrino direction on the signal direction, polarization, and viewing angle results in non-Gaussian error contours. Especially for a large uncertainty in signal polarization
the uncertainty contour has a banana-like shape. In Fig.~\ref{fig:skymap_uncertainty}, the uncertainty contours of the neutrino direction for different assumptions on the uncertainties on signal direction, viewing angle, and polarization are shown. We quantify the resolution by calculating the solid angle covered by the 90\% CL contour and present the results in Tab.~\ref{tab:direction}. 
The banana-like shape of the contours is a consequence of the constraint of the polarization vector being perpendicular to the signal direction. 

\begin{table}[t]
    \label{tab:direction}
    \caption{Resolution of neutrino direction quantified as solid angle of the 90\% CL contour ($A_{90\%\, \mathrm{CL}}$) as a function of uncertainty of the signal direction $\sigma_l$, viewing angle $\sigma_\theta$ and polarization $\sigma_p$. For small uncertainties where the resulting neutrino direction uncertainty can be approximated with a 2d Gaussian distribution, we also specify the $\sigma$ parameter.}
    \centering
    \begin{tabular}{c c c c c}
    \hline \hline 
        $\sigma_l$ & $\sigma_\theta$ & $\sigma_p$ & $\sigma_{68\%}$ & $A_{90\%\, \mathrm{CL}}$  \\ \hline
        \SI{0.2}{\degree}& \SI{1}{\degree} & \SI{2}{\degree} & \SI{2.1}{\degree} & \SI{0.01}{sr} \\
         \SI{1}{\degree}& \SI{1}{\degree} & \SI{4}{\degree} & \SI{3.9}{\degree} & \SI{0.02}{sr} \\
         \SI{1}{\degree}& \SI{4}{\degree} & \SI{4}{\degree} & \SI{5.7}{\degree} &\SI{0.06}{sr} \\
         \SI{1}{\degree}& \SI{4}{\degree} & \SI{8}{\degree} & \SI{8.3}{\degree} & \SI{0.12}{sr} \\
         \SI{1}{\degree}& \SI{4}{\degree} & \SI{40}{\degree} & - & \SI{0.60}{sr} \\
         \SI{1}{\degree}& \SI{4}{\degree} & unconstrained & - & \SI{1.23}{sr} \\ \hline \hline 
    \end{tabular}
\end{table}

\subsection{Measurement of the neutrino energy}
To measure the neutrino energy, also the distance to the neutrino interaction vertex needs to be determined to correct for the signal attenuation, and the incident electric field needs to be reconstructed. These two measurements, together with the viewing angle measurement discussed above, will allow determining the energy of the particle shower. Translating the shower energy into the most probable neutrino energy adds some intrinsic uncertainty due to the unknown inelasticity. As discussed above, an exception is electron neutrino charged-current interactions where all neutrino energy is transferred into particle showers, or muon/tau neutrino charged-current interactions where the produced muon/tau is measured as well.

\subsubsection{Measurement of distance to neutrino interaction vertex}
The measurement of the vertex distance relies on a precise measurement of the signal arrival time. The distance can either be reconstructed using a 3D array of antennas \cite{Aguilar:2021uzt}, or if the receiver is close enough to the surface, through the D'n'R technique \cite{ARIANNA:2019scz, DnR2019,Allison:2017jpy}. If an antenna is placed $\sim\SI{15}{m}$ below the surface it will observe two Askaryan pulses for most neutrino events  \cite{NuRadioMC2019,DnR2019}, one signal from a direct path to the antenna, and a second delayed signal that is reflected off the ice surface. For most geometries, we get total-internal-reflection at the ice-air boundary leading to two pulses with comparable amplitude. The vertex distance is a function of the time delay between the two pulses $\Delta t$ and the incoming signal direction. At deeper depths though, the efficiency to detect both pulses reduces quickly  \cite{NuRadioMC2019}. Therefore, a detector station installed deeper in the ice requires more antennas that are spatially separated to obtain sensitivity to the vertex distance.   

One advantage of the D'n'R technique is that the time delay $\Delta t$ can be measured very precisely. An in-situ measurement with the ARIANNA detector showed a time resolution of \SI{80}{ps}, and a simulation study showed that the vertex distance can be determined with a precision of around 10\% \cite{DnR2019} by a geometrically compact surface design. A similar precision is expected for a deep 3D array of antennas \cite{Aguilar:2021uzt}.

\subsubsection{Reconstruction of the electric-field pulse}
Similar to the polarization reconstruction, the determination of the three-dimensional electric field pulse requires the measurement of the signal pulse in at least two antennas with orthogonal polarization response, as well as knowledge of the incident signal direction as the antenna response depends on it. The polarization angle is actually just the angle between the amplitudes of the two orthogonal electric-field polarization states. 

Several methods have been developed. The simplest way that works for any type of signal is to solve the following system of equations that relates the electric field $\mathcal{E}^{\theta, \phi}$ to the measured voltages $\mathcal{V}_i$ in the antennas \cite{NuRadioReco2019}:
\begin{equation}
    \begin{pmatrix} \mathcal{V}_1(f) \\ \mathcal{V}_2(f) \\ ...\\ \mathcal{V}_n(f)\end{pmatrix} = 
    \begin{pmatrix} \mathcal{H}_1^\theta (f)& \mathcal{H}_1^\phi (f)\\ \mathcal{H}_2^\theta (f) & \mathcal{H}_2^\phi (f)\\ ... \\ \mathcal{H}_n^\theta (f)& \mathcal{H}_n^\phi (f)\end{pmatrix} 
    \begin{pmatrix} \mathcal{E}^\theta(f) \\ \mathcal{E}^\phi(f)\end{pmatrix} \, ,
    \label{eq:H_full}
\end{equation}
where $\mathcal{V}_i$ is the Fourier transform of the measured voltage trace of antenna $i$, $\mathcal{H}_i^{\theta, \phi}$ represents the response of antenna $i$ to the $\phi$ and $\theta$ polarization of the electric field $\mathcal{E}^{\theta, \phi}$ from the direction $(\varphi_0,\vartheta_0)$. 
The disadvantage of this method is that it does not perform well for small signal-to-noise ratios and the direct unfolding of the antenna response often amplifies noise. Therefore two other techniques have been developed.

The \emph{analytic forward folding} technique uses an analytic model for the electric-field pulse. It turns out that the pulse can be adequately described with just three free parameters for neutrinos as well as for cosmic rays \cite{NuRadioReco2019, GGaswintPhD}. The electric field is then folded with the antenna response yielding a prediction of the measured voltages. The predicted voltages are compared with the measured ones, and the model parameters are varied to get the best match between measured and predicted voltages. This method outperforms the unfolding method significantly at low signal-to-noise ratios but requires a signal model. In the case of neutrinos, the model parameters are directly the neutrino direction and the shower energy that allows predicting the electric field at each antenna if an estimate of the vertex position exists  \cite{GGaswintPhD}.

Another technique uses information field theory to determine the electric field. The advantage is that this method also performs well at low signal-to-noise ratios and makes fewer model assumptions \cite{WellingIFT2021}. The disadvantage is that it (so-far) assumes the same signal in each antenna which only allows using the signal from nearby antennas. 

\section{Backgrounds and rejection techniques}
\label{sec:Background}

Due to the much higher energy threshold, radio arrays do not suffer from the same intrinsic backgrounds that are dominant in neutrino detectors based on optical techniques. In particular, the vast majority of downward traveling atmospheric muons are too low of energy to generate a detectable radio pulse. In addition, the flux of atmospheric neutrinos at energies above \SI{e17}{eV} is too small \cite{Gaisser2019-AtmNu} to be of concern for detectors with sensitivities near $E_\nu^2\Phi\le$ \SI{4e-10}{GeV cm^{-2}s^{-1}sr^{-1}}. So what, then, is responsible for most of the background in high-energy radio-based neutrino detectors?  We identify sources associated with human activity (anthropogenic noise), environmental, and physical backgrounds due to similar signals being generated by non-neutrino particles. 

The most common source of background signal is due to radio emission from thermal agitation of the ice medium. The amplitude of the fluctuations scale as the square root of the bandwidth of the detector and ambient temperature of the ice.  This is not the only contributor to thermal fluctuations though.  High-gain amplifiers and connector cables will produce spurious signals.   Fortunately, thermally-induced waveforms are typically very short in duration (a few ns), which are readily distinguished from the characteristics of neutrino signals produced in LPDA receivers. The differences are sufficiently distinct that the ARIANNA collaboration plans to include a real-time filter based on deep learning techniques \cite{Arianna:2021vcx, ICRC2021Anker} to identify and de-prioritize thermally induced events.  The goal is to increase the trigger rate of the system by four orders of magnitude while keeping the same transfer rate of high-priority events over the communication channels. If proven to be successful in field tests, then the increased trigger rate implies a reduced trigger threshold, and consequent increase in sensitivity.

Radio receivers must contend with anthropogenic sources of radio noise such as communication transmitters, spark plugs, static discharge, etc. Radio frequency emission from radio stations and communication transmitters can be observed at great distances.  Due to the relative scarcity of radio transmitters and human activities in the polar environment, anthropogenic noise is quite low at locations that are more than \SI{100}{km} from research stations. Obviously, anthropogenic sources increase in the vicinity of research stations, but the prototype radio arrays have shown that these events can be rejected at reasonable efficiency \cite{ARA2020-limit,ARA:2022rwq}. In general, waveforms from mono-frequency continuous-wave emitters are readily identified and rejected in neutrino analysis, but can be a nuisance at the trigger level if the collective contributions are strong enough.  The added power may increase the trigger rates, typically based on requiring several antennas to observe similar signals within the maximum time to travel from the nearest to the furthest antenna in the station, to unacceptably high levels that exceed the maximum collection rate of the data acquisition system.  

In addition to radio noise created by human activity, a significant increase in trigger rates was observed during periods of high winds \cite{Aguilar:2021voo}.  The physical mechanism of this radio emission is not known but thought to be related to the triboelectric effect, i.e., due to to static discharge from snow particles tumbling across the snow surface.  Unfortunately, due to the apparent stochastic nature of the waveforms, it is possible to mimic the signals expected from neutrinos. The amplitude, shape, and duration of events induced by high winds are quite variable, but rarely exhibit all the features expected from neutrino interactions. The ARIANNA collaboration reported that wind-induced events can be identified reliably by deep learning tools \cite{ICRC2021Anker}.  

Both pilot arrays ARA (with \SI{200}{m} deep receivers) and ARIANNA (with shallow receivers) showed that backgrounds could be rejected in analysis while maintaining a large fraction of captured neutrino events  \cite{Anker:ARIANNAlimit2019,PersichilliPhD,ARA2020-limit,ARAICRC2021,ARA:2022rwq,leshan_zhao_2022_6785194}. This is very encouraging, but the challenges increase for larger arrays that envision improving the sensitivity by four orders of magnitude (or more!) relative to current capabilities. 

To give one concrete example, although the ARIANNA collaboration achieved background rejection at high signal efficiency using only the information contained in the waveforms collected by the downward-facing LPDA, the research team concluded that the most severe backgrounds are detected during periods of high winds.  Therefore, the next-generation surface station includes additional safety measures to reject the rarest wind-generated signals that closely align with the expected shape in the downward-facing LPDA.  In addition to downward-facing LPDA, the next-generation station includes upward-facing LPDA and a dipole located at a depth at approximately \SI{10}{m}. These two changes in station design provide an independent assessment of event identification. Neutrino signals, since they arrive from below the surface of the ice will produce a much stronger signal in the downward-facing LPDA and a comparatively weak signal in the upward oriented LPDA. Background signals generated at or above the snow surface will not follow this expectation. In addition, most upward traveling neutrino signals will reflect off the snow surface, providing a nearly identical delayed copy of the neutrino signal to compliment the direct signal observed by the dipole antenna. Downward traveling background signals will not produce a direct-reflected pair of signals. Since the same antenna measures both the directed and reflected pulse, many of the systematic errors associated with antenna response and relative time uncertainties are reduced to sub-tenths of nanoseconds.  The dipole provides a third independent tool to identify and tag background generated at or above the surface.  With the additional handles provided by the next-generation station design, there is optimism that the required rejection will be possible, perhaps with even greater efficiency to keep neutrino events. 

Cosmic rays that interact in the atmosphere provide both an opportunity and a background concern on multiple fronts.  The air showers produced by cosmic-ray interactions generate a large flux of short duration pulses of radio signals that serve as a natural, and far more frequent, source of signal for radio neutrino detectors than neutrinos themselves.   Since these signals travel downward in a compact conical beam  \cite{Huege2016} only a few degrees wide, the asymmetry in the measured signals by the upward and downward oriented LPDA provides reliable identification.  This is true even for cosmic rays arriving from the horizon because to reach the LPDAs, which are buried below the surface snow, the direction of the signals change direction by refraction to about \SI{45}{\degree} with respect to the normal of the snow surface.

\begin{figure}[t]
    \centering
    \includegraphics[width=0.99\textwidth]{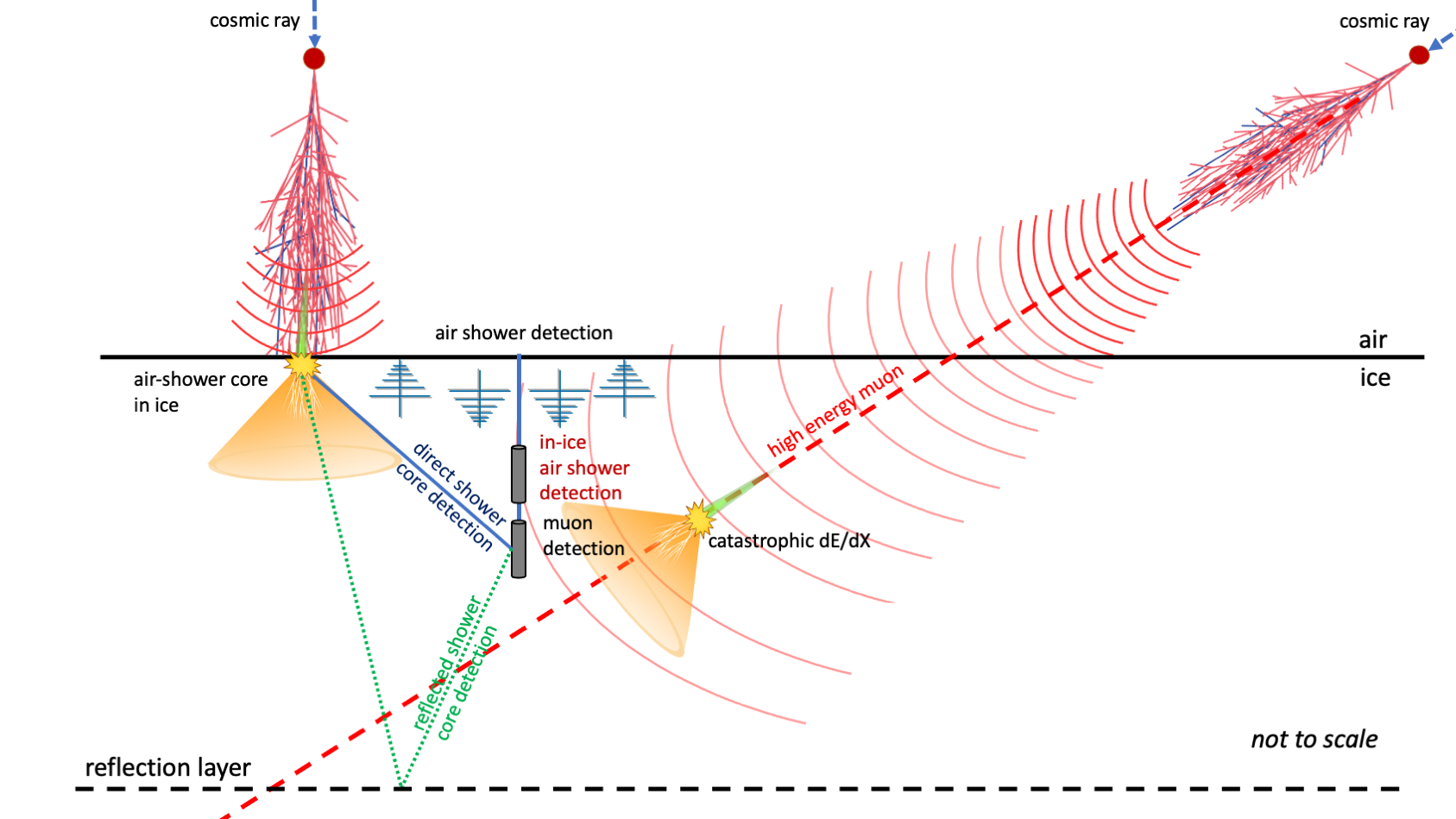}
    \caption{Illustration of backgrounds from cosmic-ray interactions in the atmosphere. Show the direct detection of the air-shower radio signal by upward facing LPDAs, the detection of the same signal in buried dipole antennas; the radio emission generated after a catastrophic energy loss of a high-energy muon; in-ice showers generated by undeveloped air showers when they hit ground and where the resulting radio signal can reach the detector via a direct trajectory and a reflection off a deep reflection layer.}
    \label{fig:background}
\end{figure}

Nevertheless, due to the high flux and strong radio emission of air showers, the rate of cosmic-ray events will dominate over the expected neutrino rates.  Thus, all physical mechanisms associated with cosmic-ray collisions are being investigated as possible sources of background. We illustrate the different background channels in Fig.~\ref{fig:background}.  One particularly insidious background is created by deeply penetrating high-energy atmospheric muons which then occasionally radiate a photon with energies comparable to the muon itself  \cite{GarciaFernandez2020,GlaserICRC2021Leptons}. 
The photon initiates a high-energy particle shower that generates a detectable radio pulse. This background is dangerous because the signal vertex can be (and usually is) below the detector station. Since the generation of the radio pulse is produced by the same Askaryan mechanism as neutrino interactions, the distribution of muon directions is constrained to the zenith angle interval as expected for detectable neutrinos. The upward traveling signal from this background process is indistinguishable in shape, frequency content or direction from neutrino signals. However, due to the steeply falling flux of cosmic rays primaries, the even steeper flux of high-energy muons produced in cosmic-ray interactions, and the exponential suppression of photons with a substantial fraction of the muon energy, lead to a suppression of this background with increasing shower energy. Though the largest background rates are expected to occur at the lowest relevant energies, we want to emphasize that the calculation of the flux of this background is highly uncertain due to imprecise knowledge of the elemental composition of cosmic rays in the energy region just above the knee, i.e., above several PeV, and the uncertainties of high-energy muon production in air showers mostly from uncertainties in meson production at large fractions of the incident nucleon energy. Estimates of the rates of muon-induced cascades currently span four orders of magnitude. The strong energy dependence of this background can be exploited in an analysis strategy but requires that the energy resolution of the detector be as small as possible to avoid spillover into higher energy events.  The potential seriousness of this background suggests that radio detector designs must confidently address the most pessimistically large predictions to avoid nagging concerns, i.e., to be on the safe side. One strategy focuses on the detection of the progenitor cosmic ray that initiated the atmospheric air shower to tag an event that triggers the detector station.  Of course, this is precisely the job of the upward-facing LPDA in a surface station.  The only question is the areal density to ensure the detection of nearly 100\% of the cosmic rays that initiate this background.  Fortunately, the cosmic rays that generate this background tend to arrive from inclined trajectories that produce large radio footprints on the snow surface, are detectable down to energies of $3-5\times10^{17} eV$.  The precise requirement for the detection fraction depends on the actual rate of this uncertain background. 

There is perhaps an even more worrisome background which we discuss next. As mentioned, the ice matrix in polar regions is inhomogeneous and anisotropic. The ice within approx.~\SI{150}{m} of the surface (known as firn) is variable density (increasing with depth from \SI{0.3}{g/cm^3} to \SI{0.92}{g/cm^3}) and striated due to seasonal accumulation.  Below the firn, the striated layers compress to a more uniform ice matrix, punctuated at specific depths with layers of volcanic ash, dust, and acidic aerosols that can act as reflective, absorptive or scattering layers. Ash layers and ice density perturbations may reflect as much as 1\% of the radio power at normal incidence angles \cite{Fujita-reflection,Besson2010-biref-reflect,Barwick2005-PoleAtten,Besson:2021wmj}, but grazing angles can reflect much more power.  At the South Pole, several layers have been identified \cite{Bay:ReflectionLayers,Besson:2021wmj} beginning at a depth of \SI{300}{m}.  Reflected background events are troublesome for the same reasons given for radiated photons from high-energy muons. That is,  the reflected signal will be traveling in the upward direction, which is the same direction as neutrino signals. 

Is there a mechanism that can produce non-normal incidence angles at a reflective boundary within the ice?  Possibly. The developing cascade of shower particles induced by cosmic rays in the atmosphere grows to a maximum and then fades out. If the shower crosses the snow surface, the shower will continue to develop in the denser medium.  For sea-level detectors, the energy of the cosmic ray must be very high to intersect the snow surface before it reaches the maximum particle content, but at higher elevations like the South Pole at almost \SI{3}{km} above sea level, relatively low energy showers of \SI{e17}{eV} can reach the snow surface before they reach maximum particle content. Interestingly, the core of the shower remains sufficiently compact at surface elevations of a few kilometers to generate Askaryan signals in the firn ice near the surface \cite{DeKockere:2022bto}. The radio signal is expected to be similar to the Askaryan signals generated by neutrino interactions in the ice though with less high frequency power. The radiated power forms a much broader cone compared the narrow atmospheric emission. Thus, the angle of incidence with the reflective layer can reach large values which might lead to a stronger reflection.

Cosmic rays that strike the surface with large residual core energies arrive predominantly from directions close to vertical.  Simulations show that the upward facing antennas in neutrino stations do not detect these cosmic rays with much efficiency due to the smaller footprint and for higher elevation sites and the relatively short distance to the atmospheric location that generates most of the radio emission. The required density of detector stations to observe $>90\%$ of the cosmic rays for zenith angles less than 30 degrees is cost prohibitive. One possibility to increase the detection efficiency of near vertical cosmic rays is to exploit the phenomena of shadow zone propagation.

Also, the radio signal emitted by the air shower in the air itself can propagate into the ice and can be reflected back up. Though refraction into the firn limits the incidence angle to $\sim$~\SI{45}{\degree} and the bending of signal trajectories in the firn will further bend the signal downward.
The rate of this process will depend on many details, some of which are poorly understood, but it is conceivable that this background could dominate at larger shower energies. The background mitigation is similar to the strategy employed for photon emission by high energy muons that penetrate into the ice;  tag the initiating cosmic-ray event with upward-facing LPDA in the surface stations. This process may require an even higher surface density than required to mitigate the background generated by high energy muons because the most dangerous events correspond to the case where more of the shower develops in the firn, and comparatively less is generated in the atmosphere. This reduces the in-air radio emission and favors vertical showers that have a smaller footprint on the surface than the horizontal air showers relevant for the high-energy muon background. 

The last two cases are examples of physical backgrounds that occur at low rates, too low to manifest themselves in small-scale prototype instruments. Theoretical calculations are plagued by considerable uncertainty. These types of backgrounds are illustrative of the challenges when scaling a detector system by four orders of magnitude in sensitivity.  It is imperative to control rare, but non-negligible, backgrounds.  Due to the considerable uncertainties, and lack of experimental normalization, future detectors must employ mitigation strategies to the fullest extent possible.  In our view, background rejection must be implemented with redundancy and with sufficient safety margin to handle surprises.

\section{Operational Considerations}
\label{sec:operations}

In this section, we describe the considerations that influence the design of a large area radio-based neutrino detector.

Several authors (e.g. see \cite{Price:RadioComparison}) have shown that the radio technique is superior to optical techniques for neutrino energies in excess of \SI{e16}{eV} (UHE neutrinos).  Experimental upper limits on the flux of neutrinos above this energy require that a radio-based detector has a volume on the order of 1000 cubic kilometers to have a realistic chance of detecting UHE neutrinos. Constructing a fully instrumented detector of this volume is not economically feasible. Instead, radio-based detector designs utilize a dense dielectric medium with kilometer-long attenuation lengths for radio propagation, such as cold Antarctic ice. This allows the UHE neutrino detector to be comprised of many relatively compact, independent stations that are separated by \SI{\sim1}{\km} while still achieving the cubic kilometer effective volume per detector station. Therefore, many neutrino detectors have focused their efforts on the polar regions of Earth (i.e. in Antarctica and Greenland). Further, due to relative lack of human activities, polar regions provide a pristine radio-quiet environment which allows a radio detector to run without interference (of course, the detector might be near a populated research station, but even then the radio interference is far below levels found over most of the planet). For example, the ARIANNA site at Moore's Bay has been measured to have the lowest anthropogenic noise level on Earth.  At this site, the primary source of noise is due to the unavoidable thermal motion of the electrons in the ice itself (and instrumental contributions). One consequence of this characteristic of the Antarctic environment is that detectors can operate with very low threshold requirements - typically only a factor 4 larger than the root mean square (RMS) noise value. As mentioned, there is another source of RF noise, which arises during the relatively infrequent episodes of high winds. At wind speeds in excess of \SI{10}{m/s}, it is thought that blowing snow particles will charge up surface snow and local metal towers until static discharge, creating impulsive radio noise \cite{Aguilar:2021voo}. 

Given the relative youth of the radio-neutrino technique and the absence of astrophysical signals at the relevant energies, there are good reasons to incorporate independent experimental architectures in the next generation radio detector.  In the energy regime targeted by in-ice radio detectors, two techniques have been developed over the past decade.  The ARIANNA collaboration pioneered the development of a near-surface station that relies on directional LPDA antennas. Another technique installs antennas between \SI{100}{\meter} and \SI{200}{\meter} beneath the snow surface.  The deep ice technique provides more volume per station (cf. Fig.~\ref{fig:Veffdepthdependence}), but more limited options for antenna receivers because the antennas are constrained to fit in a relatively narrow hole. There is a higher cost per station due to drilling, more antennas per station, longer cables and downhole electronics. The deep ice relies on a volumetric reconstruction that differs from the ARIANNA technique that relies on more planar reconstruction using directional LPDA antennas, augmented by a dipole antenna located at a depth of approximately 10 meters.  The differences in technique may be useful in identifying and removing rare background processes.  

At lower energies, it is conceivable that radio techniques will overlap with optical neutrino detectors, especially if the next generation of larger optical Cherenkov detectors are constructed. Not only would a detailed comparison between the two detector architectures be immensely valuable in boosting confidence in the radio results, but radio could complement the highest energy observations by IceCube by providing independent angular resolution for sources in the southern sky for a detector in Antarctica (for example). 

It is notoriously difficult to predict the antenna response over all angles and frequencies when it is embedded in a non-uniform medium.  For the South Pole location, the upper 200 meters of ice varies in density, with the largest gradients near the surface. For the Moore's Bay sea-level location, the variation occurs down to about 100 meters. In addition, surface roughness in the air-snow boundary will create challenges to modeling the antenna response. The ARIANNA station was designed to minimize many systematic issues associated with antenna responses by installing the receivers in a symmetric geometric layout.  Many systematic errors associated with antenna responses are eliminated in the measurement of time delays between parallel antennas using a cross-correlation technique that requires that both waveforms show similar shapes. The antenna layout has an important impact on the resolution of measurable quantities such as angular direction of the radio pulse, the vertex resolution or neutrino angular resolution. The same argument holds for the rest of the front-end signal chain such as amplifiers, cable effects, and readout electronics. The 4 fold symmetry provides two parallel measurements for two orthogonal polarization directions, minimizing the effects of systematic uncertainties in the response functions. Though the angular direction of the radio pulse can be triangulated from just three antennas, the antenna effects in mixed media near a discrete boundary must be understood with excellent precision. 

The combination of orthogonal LPDA and dipole antennas in the surface station measures three polarization components and provides the ability to over-constrain the reconstruction of the electric field. The vertically orientated dipole antenna, located at a depth of around \SI{10}{\m},  can also provide an important measurement of the direct and surface reflected signal within the same time record. The time delay between the direct and reflected signal is related to the path length distance to the event vertex. The timing resolution between direct and reflected signals was measured to be \SI{80}{picoseconds} \cite{DnR2019}.  This high level of precision is a consequence of the design:  the effects of the antenna model, channel-to-channel time delay offsets, and amplifier response are mitigated when both the direct and reflected pulse is measured by a common antenna.

As mentioned, the density of the upper \SI{200}{\meter} of snow surface increases with depth. To a very good approximation, the index of refraction for radio frequency electromagnetic (EM) waves increases in direct proportion to density. EM waves originating from within the bulk ice will refract to more horizontal trajectories. Due to the gradient in the index of refraction, EM waves will begin to travel downwards even though the initial direction is upward toward the surface.  It is not hard to see that this phenomenon will restrict the geometric locations that can produce signals that reach the surface.  We use the term "shadow zone" to describe the region where signals cannot propagate to the surface. This limitation can be compensated by installing the detector on an ice shelf. A radio detector installed on the surface of an ice shelf will observe signals reflecting off of the bottom ice/water interface, which acts as a near-perfect reflector. This increases the effective volume that a detector can achieve, and allows for the detection of downward traveling neutrinos.  Due to the unique feature of the ice shelf (which is a block of ice floating on water, like a very large iceberg), an ARIANNA detector can view UHE neutrinos from the entire Southern Sky.  In contrast, a detector without the water-ice reflector is limited to viewing a more restricted fraction of the sky as shown in section~\ref{sec:skycoverage}.  For surface detectors at the South Pole, for example, the range of zenith directions is within 17 degrees of the horizon.  For dipole antenna located below the firn layer (\SI{100}{m} in Greenland, \SI{200}{m} at the South Pole), the range of observable zenith angles extends to 45 degrees above the local horizon.  A deep station design profits from having the antennas embedded in a near uniform medium. However, experience has shown that a thorough in-situ calibration of the antenna response is required. For example, the reconstruction of the neutrino direction requires a precise determination of the relative response of the vertically polarized dipole antennas to the horizontally polarized quad-slot antennas to determine the signal polarization.

We next address a few more technical considerations associated with the construction and operation of a large array for radio-based neutrino detectors.  The prototype detectors mentioned below are more completely described in the next section. 

The ARIANNA collaboration has reported highly reliable wireless data transfer using the Iridium satellite network over 5 years of operation, an important milestone. The frequency of transmission is above the relevant radio frequencies observed by neutrino stations (between 80-500~MHz), so there is no direct in-band interference.  However the data transfer rate is very slow, limiting the event transfer rate to about one per hour for low dead-time operation (less than 5\% deadtime). It can be boosted by a factor of 20 by modifying the data acquisition system to be able to collect new data while simultaneously transferring high priority data. Since the vast majority of triggered events are created by thermal fluctuations, most events can be quickly discarded by on-board real-time filters.  Recent work has focused on implementing deep learning techniques based on convolutional neural networks to retain less than 1 thermal trigger per 100,000, yet passes 95\% of all neutrino signals \cite{Arianna:2021vcx}. This filter allows high rate triggering (which lowers the threshold and increases sensitivity by up to a factor of 1.8 at low energies) while keeping the transfer rate of high priority events within the limitations imposed by the Iridium network. This scheme benefits from the relatively few antenna channels (8) and relatively compact waveforms (256 samples) of the surface station architecture. In the future, the next generation of Iridium satellites, Iridium Next, are expected to increase data throughput substantially so event rates, channel count and/or waveform duration may be optimized, but costs remain a large unknown.

By placing a detector near an existing research base, data transmission rates can be increased dramatically by connecting all neutrino stations to a cable grid with a centralized storage facility.  The ARA collaboration has implemented a point to point scheme at the South Pole, using the infrastructure of the South Pole Station to store data and transfer using high speed satellite connections. The cost to install a buried grid of cables over areas of \SI{500}{km^2} using traditional techniques is prohibitive, so less costly cable-laying strategies are under investigation.  The IceCube neutrino detector installed data/power cables to a distance of \SI{0.5}{km} or so from the central operations hub, providing insight on technologies for much larger scale installations. For the next generation of large scale neutrino detectors, point to point connections will likely need to be replaced by a fault-tolerant communication networks that takes into account the unique challenges of working in cold, isolated environments.  For example, repair and maintenance will be very difficult during the Arctic or Antarctic winters.  There are open questions on the environmental impact and potential removal of a large grid of cables.

Local wireless networks, such as LTE (which is a standard for wireless broadband communication by mobile devices), offer another avenue to increase the data transfer rate relative to Iridium.   Obviously, with wireless networks, there is no need to bury cable over large areas of the polar plateau, but a set of dedicated towers with LTE base stations must be erected to communicate with the neutrino stations.  The long term durability of LTE wireless networks have not been tested in the polar environment.  Earlier tests using the WiFi technology revealed unacceptable levels of operational failure. Part of the reason was the lack of ground and the need for continuous high power transmission to a remote tower.  Beginning the summer of 2021, the RNO-G collaboration tests LTE data transmission network to evaluate the suitability for future radio neutrino detectors, where LTE is used as the primary communication method to connect the radio detector stations to a base station installed at the Summit research station in Greenland.  

What provides power to a large area, sparse array of detector stations?  Since there is sufficient continuous sunlight during the polar summer, solar panels are used.  They are robust, low cost, and can generate several hundred watts of power.  This is more than enough: the surface station technology requires 5-10 watts and recent designs for deep string stations are expected to run on about 20 watts of power. This technology has been extensively tested by ARIANNA and other polar experiments.  The risks are low and costs are well known.  

Operation during the dark winter months provides many more challenges, and the risks associated with various technology options are hard to assess due to insufficient information. The ARA collaboration laid electrical cable to provide power, but as previously discussed, this method is difficult to scale up to cover the surface area required by the next generation designs.  As with communication networks, a power network must be designed with low losses and robust operation to accommodate the unique environmental requirements of the high polar plateau.  Alternatively, wind turbines may be able to supply sufficient power to an individual station if the power requirements are low, but questions remain on their mechanical durability. The ARIANNA collaboration has tested a relatively low cost, light weight wind turbine that generated the required 5 watts for about 30\% of the winter. Commercial wind turbines were not robust, so a special-purpose device was designed and constructed by researchers at Uppsala University in Sweden.  For \SI{20}{W} power consumption expected for deep ice stations, the wind turbine must be larger and more costly. There is an added difficulty for high elevation sites: the air density is lower so the available energy to extract from the wind is less (for a given wind speed) compared to sea level sites. Moreover, wind does not blow continuously, so large battery systems are required to store power. Unfortunately, inexpensive lead-acid batteries are heavy and, in general, work poorly at cold temperatures but they have a proven track record in Arctic environments if discharge is controlled to be small relative to total capacity \cite{Aguilar:2020RNOG}.  Lithium ion batteries with Iron Phosphate chemistry require less weight to be transported to the remote geographical location and the ARIANNA collaboration reported good reliability and storage capacity at -30C, but their costs are 10-15 times greater than lead acid batteries. An alternative - that is currently under investigation - are special purpose Nickel-Cadmium (NiCd) batteries where the chemistry is adjusted to allow low temperature operation.

Winds can be light for arbitrarily long periods of time, but this situation is infrequent. The ARIANNA station operated for 4 days on battery power, which provided stability during sun-setting periods, and some level of continued operation during the winter when wind speeds are low. This experience provides input to design larger wind turbines and battery storage systems. Simulations indicate that an uptime of 80\% during the winter month can be reached with a reasonable upscaling of the current wind generator design and battery system. Prototypes of the upscaled wind generators are under construction and are planned to be tested at the South Pole and Greenland.

The last technical consideration involves initial deployment, and in the case of deep string technologies, drilling. Though a hot water drill was the appropriate tool for the IceCube Project, the preparation and transportation of the requisite equipment proved to be very cumbersome for radio-based high-energy neutrino stations, where the drill has to be re-positioned over much larger areas. Rather than using a hot water to drill the holes in the Greenland project, RNO-G employ a novel mechanical drill designed by researchers at the British Antarctic Survey. It is hoped that this drill will reliably and quickly (on the time scale of 8 hours) create an open hole to a depth of \SI{200}{m}, with the goal of reducing the operational complexity (and cost) of drilling operations. There are several open questions associated with an open hole (in other words, the cored hole is not filled with a fluid). First, do open holes eventually close back and possibly alter the antenna response?  If so, on what time scale? Recently, rain was observed to fall at Summit Station in Greenland.  What complications might rain introduce?  

In terms of deployment, there are complementary advantages of surface designs and deep-string technologies. The ARIANNA project has demonstrated that an 8-channel surface station can be deployed by a team of 5 people in one work day.  The time and personnel required to prepare and deploy a 24-channel deep-string station has not yet been established, though it is certainly longer than a surface station. Establishing an accurate time scale to deploy a deep string station is an important goal of the RNO-G project.  One thing to keep in mind, fewer deep string stations are required to reach the same effective volume as the surface station array.

\section{Prototype arrays dedicated to neutrino detection - proof of principle}
\label{sec:prototype_arrays}

Due to the extremely low neutrino flux at energies above \SI{10}{PeV}, no neutrino has yet been detected using the radio technique. However, several experiments have shown the feasibility of this detection method and its potential, which build on the experience of previous radio neutrinos detectors, like the pioneering RICE \cite{RICE2003-Performance,RICE2003-limits} and ANITA \cite{ANITA2008-Design,ANITA-2ndFlight} experiments, which informed the development of prototype in-ice arrays such as ARIANNA \cite{Barwick:ARIANNA2014,Design:ARIANNA2014} and ARA \cite{ARA2011,ARA2020-limit} experiments.  These efforts were critical to the demonstration of technologically important aspects of operating in remote locations in harsh polar conditions.

The first experience with in-ice radio detectors was gained with the Radio Ice Cherenkov Experiment (RICE \cite{RICE2003-Performance}) at the South Pole, which operated from 1999 until 2010. RICE provided the first neutrino limits \cite{RICE2003-limits} from radio detectors and valuable experience in deploying and operating radio detectors at depths of down to \SI{200}{m}.

The Antarctic Impulsive Transient Antenna (ANITA) experiment is a balloon-borne radio receiver array that scans the ice surface from afar for upcoming neutrino signals generated below the ice surface. It has flown four separate missions over Antarctica. Several components of the ANITA hardware have been incorporated into the ARA and ARIANNA designs \cite{ANITA2008-Design,ANITA-2ndFlight,ANITA-4thFlight}. The dual-polarization receiver antennas provide an identical response to orthogonal polarization components, which reduces the systematic errors associated with absolute antenna response.  This design element was incorporated into the ARIANNA surface design.  If non-identical antenna designs are required to determine the polarization, then the absolute response for each antenna type must be determined, which poses a significant challenge for antennas embedded in a non-uniform medium (such as air-ice for dipoles deployed in a borehole). A data acquisition system with high timing accuracy and thorough calibration is needed to reliably reconstruct neutrino or cosmic-ray signals. ANITA was the first radio-neutrino experiment to report the detection of air shower signals \cite{ANITA2010-CR}, which helped to verify the simulation chain and the understanding of the energy calibration \cite{ANITA2016-CR}. The ANITA collaboration observed several events which, if neutrinos, would seem to be in tension with Standard-Model cross-sections \cite{ANITA2016-4up, ANITA2019-upward, ANITA2019-noTau} and experimental constraints from IceCube \cite{IceCubeICRC2019-ANITA,IceCube2020-ANITA}. Several authors have suggested that they may be the result of potential ice effects \cite{deVries2019,Shoemaker2019}.

\begin{figure}[tp]
    \centering
    \includegraphics[width=0.49\textwidth,valign=t]{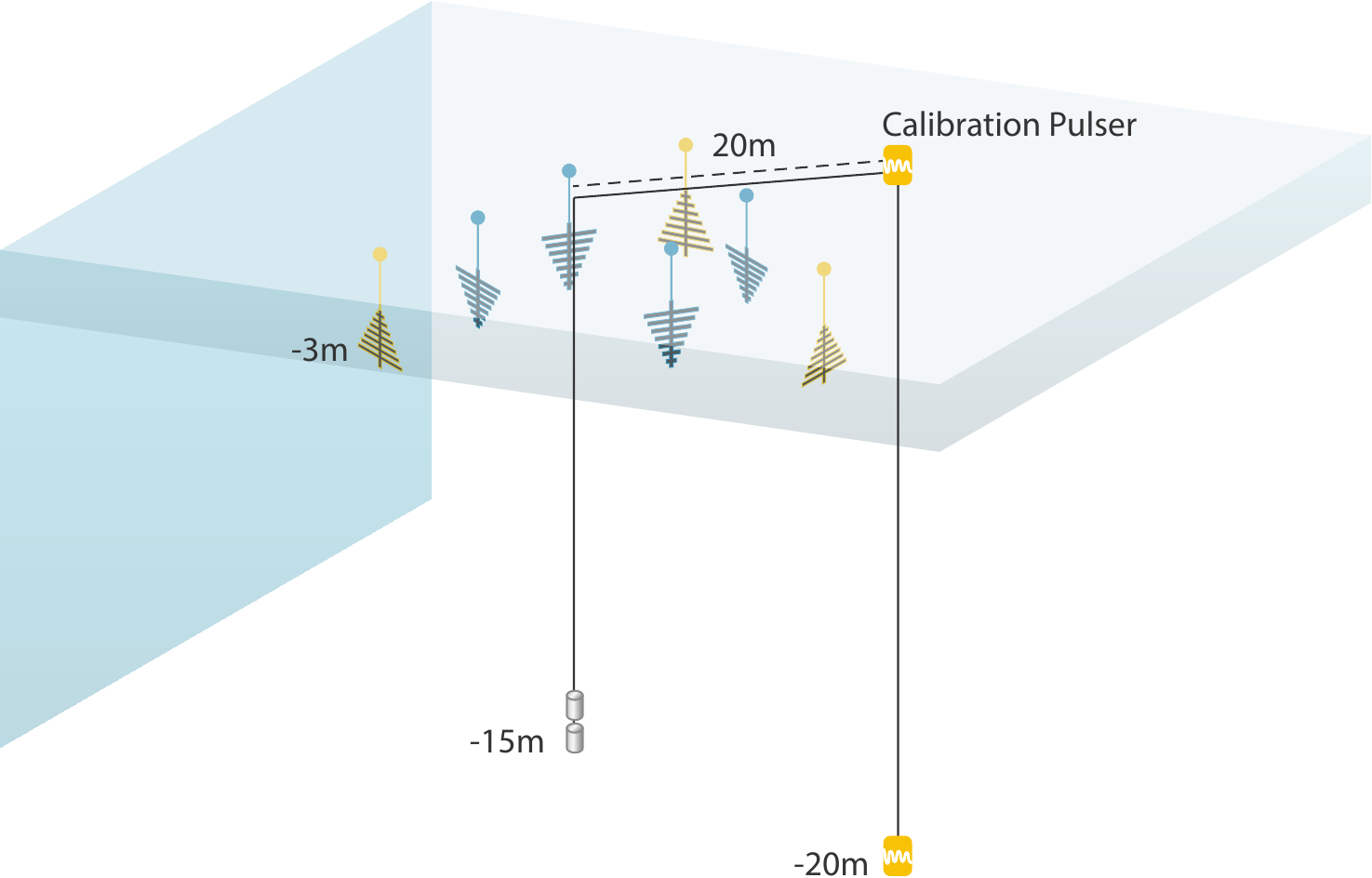}
    \includegraphics[width=0.49\textwidth,valign=t]{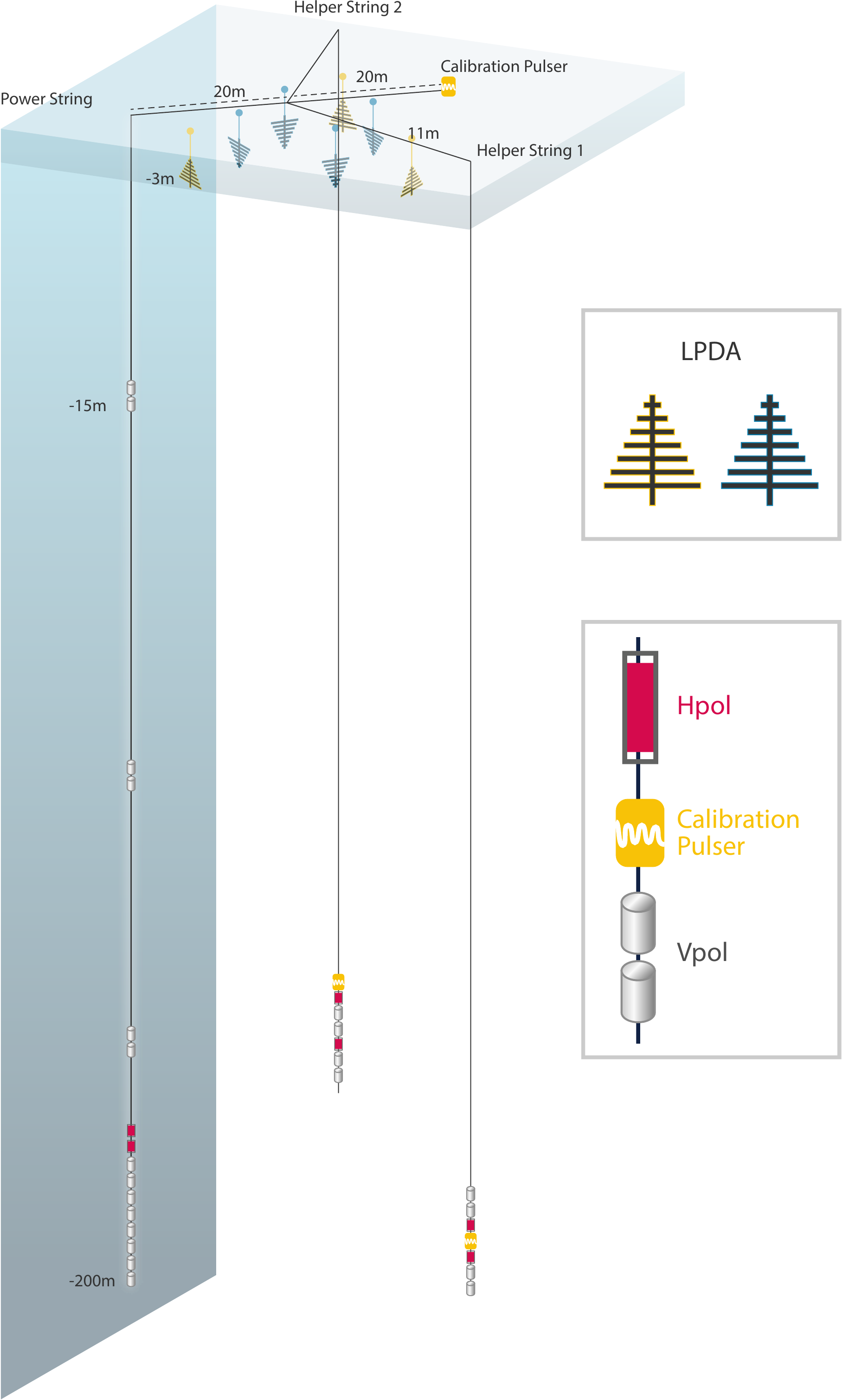}
    \caption{Illustration of the shallow (left) and hybrid (right) station design as foreseen for IceCube-Gen2. Figure from \cite{HallmannICRC2021}. The shallow design consists of only 8 channels with 4 downward facing LPDAs to trigger on the neutrino radio signal with the additional \SI{15}{m} deep dipole to aid event reconstruction, and 3 upward facing LPDA antennas for air-shower measurements. The hybrid design features the same shallow component and has additional antennas installed at \SI{200}{m} depth on three strings with a phased array trigger system on one of them. The proposed station design for ARIANNA-200 equals the shallow station of Gen2. The station design of RNO-G is similar to the hybrid design shown here but with the deepest antenna at \SI{100}{m} and additional surface LPDAs.}
    \label{fig:stationdesign}
\end{figure}

The Antarctic Ross Ice-Shelf ANtenna Neutrino Array (ARIANNA \cite{Barwick:ARIANNA2014,Design:ARIANNA2014}) began construction at the Ross Ice-Shelf in 2010, with a first hexagonal radio array being completed in 2015. The ARIANNA concept is based around surface stations, i.e., the antennas are deployed just underneath the snow surface. High-gain log-periodic dipole antennas (LPDAs) are deployed in shallow slots in the snow, where they are not restricted by the borehole geometry and exhibit broadband characteristics and dedicated polarization sensitivity, particularly to horizontally polarized signals. By placing the antennas at Moore’s Bay on the Ross Ice-Shelf, the neutrino-detection strategy utilizes the reflective surface at the bottom of the ice at the water interface, which reflects downward going neutrino signals back to the stations. Without external infrastructure, ARIANNA pioneered autonomous low-power stations ($\sim$5 Watts), based on renewable energy sources, operated via wireless communications. More recently, wind turbines were added to the solar panel power system \cite{Nelles:ICRC2019-WindTurbine}.

ARIANNA has successfully detected the radio signal of air showers as calibration and verification signals \cite{Barwick2017-Airshowers,ARIANNAICRC2021CosmicRays} and published limits on the UHE neutrino flux \cite{Anker:ARIANNAlimit2019}. The collaboration also published the effectiveness of recording signals reflected from the surface by monitoring snow accumulation \cite{DnR2019}. Two ARIANNA stations have also been deployed at South Pole to test the robustness of the hardware under environmental circumstances differing from the Ross Ice Shelf. The same calibration source as used for ARA from the SPICE borehole was then also used to verify the reconstruction capabilities of the ARIANNA experiment with respect to arrival direction and polarization \cite{ARIANNAPolarization2020,ARIANNAICRC2021Direction}.

The Askaryan Radio Array (ARA \cite{ARA2011}) has operated at South Pole since 2010 and is a direct successor to RICE. While the RICE antennas were co-located with the AMANDA and IceCube experiments at South Pole, all five ARA stations operate in dedicated dry holes of depths \SIrange{50}{200}{m}. While different hardware has been deployed in different ARA stations, the station layout is mostly uniform. Every station consists of four receiver strings down to \SI{200}{m}. Each string is equipped with two vertically-polarized dipole antennas (VPol) and two ferrite-loaded slot antennas (Hpol) to reconstruct the radio signals. In addition, one or two calibration strings, as well as surface antennas (on the earlier stations), are deployed. As the narrow cylindrical borehole geometry limits the intrinsic antenna gain, ARA pioneered the phased-array technique for radio detection of neutrinos at the most recently completed station \cite{ARA2019-PA}. To date, the ARA collaboration has published constraints on the diffuse ultra-high energy (UHE) neutrino flux \cite{ARA2020-limit} based on 2 of the 5 deployed stations, as well as the phased array system \cite{ARA:2022rwq}. The performance of the instrument has been verified using transmitters lowered into the SPICE borehole \cite{SPICEhole}, which also allowed for the measurement of glaciological properties of the ice  \cite{Allison:2017jpy,ARA-LongBaseline2020}.

The planned RNO-G project \cite{Aguilar:2020RNOG}, based at Summit Station in Greenland, weds the surface design of ARIANNA with the deep hole array of ARA into a single station. Due to the dominant cost and power demands of the station infrastructure for the deep hole design, adding near-surface receivers provides additional surface stations to augment the required number of dedicated surface stations at marginal cost.  Many technological improvements are envisioned, especially for the deep hole instruments compared to ARA.  The power requirements of the down-hole electronics have been reduced from 100 Watts to about 20 Watts 
(about a factor of 2-4 more power than consumed by a dedicated surface station) 
by a redesign of the data acquisition system and fiber optic readout from the deep antennas. Like with ARIANNA, power will be supplied by renewable resources, rather than a cabled power grid to a centralized power generator, reducing fuel requirements. Initially, the only power source will be solar panels, but plan to include wind turbines in later phases of the installation.  The holes will be drilled by a mechanical drill developed by the British Antarctic Survey (BAS) rather than the more cumbersome hot water drill utilized by the ARA project. The drill is relatively portable and designed to create a \SI{200}{m} hole, \SI{28}{cm} in diameter, in less than 24 hours. The first results from this crucial element of the deployment strategy will become known in 2021.  The primary trigger will be a 4-element phased array, which is superior to a simple majority logic trigger, to achieve a trigger threshold below $3 \times V_{rms}$, where $V_{rms}$ is the root mean square of the thermal noise fluctuations.  The noise figure of the amplifier chain is much lower, and the gain-bandwidth is optimized to the response of the antenna \cite{Glaser2020}.  The deep hole geometry has changed from a rectangular-symmetric arrangement of dipole receivers in ARA to a triangular geometry with a primary string uniformly instrumented between the surface and maximum depth. The primary string is augmented with 2 other strings, known as helper strings, consisting of relatively few antennas. Their purpose is to support and improve the identification and reconstruction capabilities of the station.  The station has 3 calibration transmitters, two around \SI{100}{m} depth on the helper strings, and one near the surface to provide the DnR calibration of surface location.  The data will be sent wirelessly with standard LTE cellphone technology. This innovation will be utilized for the first time in a high-altitude, dry, polar environment. The first RNO-G stations were installed in July 2021.

\begin{figure}[tp]
    \centering
    \includegraphics[width=0.7\textwidth]{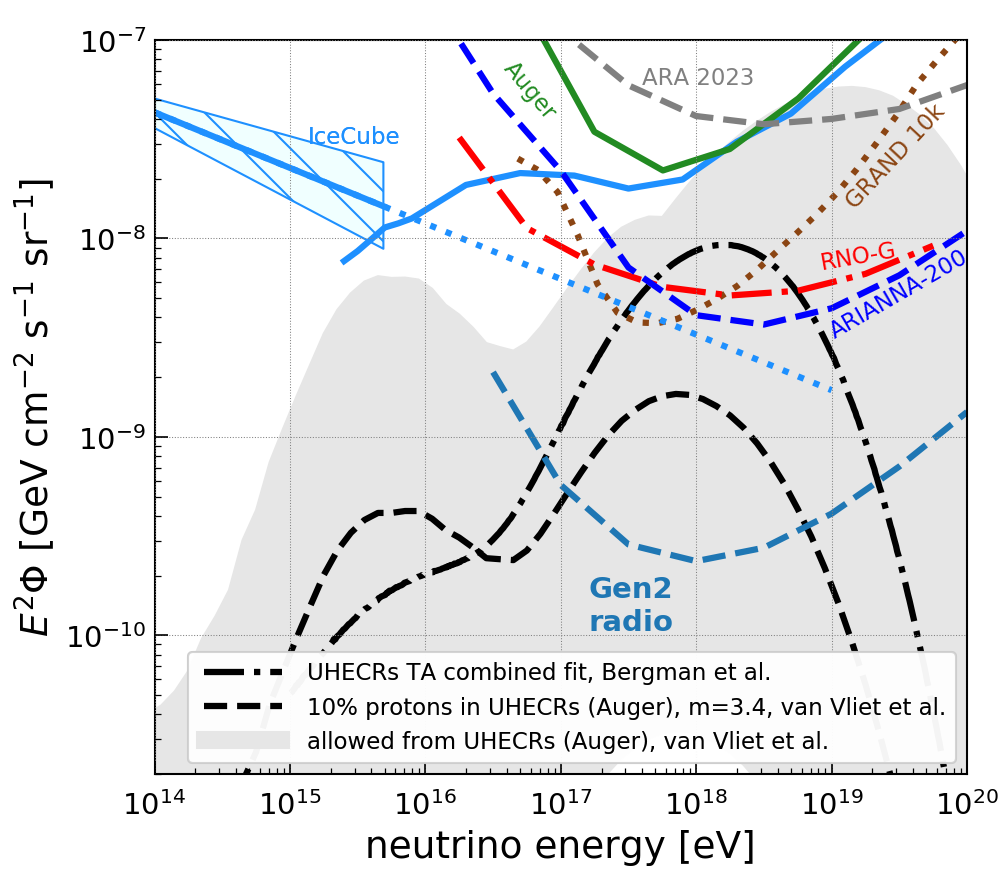}
    \caption{Overview of the sensitivity of current and future experiments to the diffuse neutrino flux. Shown is the measured astrophysical neutrino flux by IceCube \cite{icecube_spectrum} and its extrapolation to higher energies. Also shown are the 90\% CL limits on the neutrino flux set by Auger \cite{AugerNuLimit2019} and IceCube \cite{IceCubeFlux2018} as solid lines. The limits set by the prototype radio neutrinos detectors are outside the y-axis range. The expected 90\% CL sensitivity of ARA by the end of 2023 is shown as gray dashed curve  \cite{ARAICRC2021}. The colored dashed, dotted and dash-dotted lines show the expected all-flavor 90\% CL sensitivity for future arrays for 10 years at trigger level assuming zero background for RNO-G \cite{Aguilar:2020RNOG}, ARIANNA-200 \cite{ARIANNA200}, GRAND-10k  \cite{GrandWhitePaper2018} and IceCube-Gen2 \cite{HallmannICRC2021}. Also shown are predictions of the GZK neutrino flux derived from measurements of the Pierre Auger Observatory \cite{Vliet2019} and the Telescope Array \cite{BergmanICRC2021}.}
    \label{fig:expected_sensitivity}
\end{figure}

In addition to RNO-G, a successor of the ANITA ballooning efforts -- PUEO -- is developed \cite{PUEOICRC2021}, a science-capable expansion of the ARIANNA array at Moore’s Bay \cite{ARIANNA200} is discussed and proposed, and an order-of-magnitude larger radio detector with a footprint of \SI{500}{km^2} with $\mathcal{O}$(300) stations as part of IceCube-Gen2 extension is planned  \cite{IceCubeGen2-2020,HallmannICRC2021}. An overview of the expected sensitivities of these future arrays as well as the current experimental landscape and theory predictions is shown in Fig.~\ref{fig:expected_sensitivity}.

\section{Alternative detection techniques}
In-ice radio detectors are the most efficient way to observe neutrinos in the energy range between \SI{e17}{eV} and a few \SI{e19}{eV} because it allows to instrument large amount of matter (the ice) with a moderate amount of detector units (the antennas). However, there are also alternative detection techniques. The most relevant ones for the energy range of \SI{e17}{eV} to a few \SI{e19}{eV} are air-shower detectors. They are not only sensitive to cosmic rays but also to tau neutrinos. A tau lepton has just the right properties to propagate sufficiently long distances through dense media without interacting and to decay after an average of \SI{3e-13}{s}. An Earth-skimming tau neutrino has a large probability to interact in the Earth's crust. If the interaction happens close enough to the surface, the newly created tau can escape the Earth and decay in air producing an air shower. More sensitivity can be achieved by installing the air-shower detector in a mountainous region where tau neutrinos can interact in the surrounding mountains, generating taus flying towards the detector and decaying in air. 

Through this channel, the Pierre Auger Observatory -- the largest cosmic ray detector in the world with 1660 particle detectors spread over an area of \SI{3000}{km^2} located next to the Andes -- was able to set strong limits on the ultra-high-energy neutrino flux \cite{AugerNuLimit2019}. A more cost-efficient way to build a large air-shower detector is to also use the radio technique \cite{Huege2016, Schroeder2016}. This idea to build a giant air-shower radio detector is explored by several groups (\cite{Wang:2020bqr,Liu:2020uyd,beacon2020, TAROGE:2022soh, GrandWhitePaper2018}), the most ambitious one being GRAND with plans to instrument \SI{200,000}{km^2} with 200,000 tripol antennas\footnote{A tripole antenna refers to three antennas with perpendicular polarization response combined in one mechanical structure. When compared to the in-ice arrays, one tripole antenna would be counted as three independent antennas.} separated by \SI{1}{km} in clusters of 10,000 antennas \cite{GrandWhitePaper2018}. The initial plan is to install an array of 10,000 tripol antennas in a remote, mountainous area in  Western China. Apart from logistical considerations, the largest challenge of this technique is the identification of neutrinos against a transient background (with a rate of noise pulses of tens of Hz per antenna even in a remote, radio-quiet area as proposed for the GRAND site) and the background of cosmic-ray induced air showers which are several orders of magnitude more abundant. So far it has not been shown that background can be rejected sufficiently well although promising results on the self-triggering and identification of air showers exist \cite{ARDOUIN2011717, Barwick2017-Airshowers,Monroe2019}. 

It is interesting to compare the amount of instrumentation required with the air shower detector compared to the in-ice technique to reach the same sensitivity to UHE neutrinos. The proposed GRAND10k array (i.e. 10,000 tripole antennas spread over an area of \SI{10,000}{km^2}) achieves a similar sensitivity as the RNO-G \cite{Aguilar:2020RNOG} and the proposed ARIANNA-200 \cite{ARIANNA200} detector. RNO-G comprised 35 detector stations spread over an area of \SI{35}{km^2} with 24 antennas per station which amounts to a total of 840 antennas. The ARIANNA-200 array comprises 200 detector stations spread over an area of \SI{300}{km^2} with 8 antennas per station which amounts to a total of 1600 antennas. In contrast, GRAND10k needs 30,000 antennas (or 10,000 tripole antennas) to achieve the same sensitivity. 
We can also compare the air-shower vs. in-ice detectors at an even larger size: The proposed GRAND200k array has a similar sensitivity to UHE neutrinos as the proposed IceCube-Gen2 radio detector \cite{IceCubeGen2-2020,HallmannICRC2021}. IceCube-Gen2 will likely consist of a mix of ARIANNA-like shallow detector stations and RNO-G like stations that combine a deep and shallow part with a total of around 300 detector stations with close to 9,000 antennas spread over an area of \SI{500}{km^2}. In contrast, GRAND200k needs 600,000 antennas (or 200,000 tripole antennas) that are spread out over an area of \SI{200,000}{km^2} to achieve the same sensitivity. Independent of the exact choice of metric (instrumented surface area, number of antennas, number of detector stations), the in-ice technique is 1-2 order of magnitude more sensitive with the same level of instrumentation. 
Comparing the instrumentation costs yield a similar result as GRAND10k has a similar budget as the IceCube-Gen2 radio detector but IceCube-Gen2 has a ten times higher sensitivity. 

There are several reasons why an air-shower detector is less efficient in measuring UHE neutrinos: First, an air-shower detector is only sensitive to tau neutrinos that interact via a charge-current (CC) interaction which make up only 24\% of all possible interactions. Second, the field-of-view is restricted to just \SI{+3}{\degree}/\SI{-2}{\degree} above/below the horizon \cite{GrandWhitePaper2018} slightly varying with the exact site location, mountain heights, etc. Third, the neutrino must interact relatively close to the edge of the mountain (or the surface of the Earth) for the tau to be able to escape into the air. In contrast, an in-ice radio detector is sensitive to all flavors and interaction channels, and the neutrino can interact anywhere in the ice volume as the ice is transparent to the generated radio pulses. However, the details of the Askaryan emission (strong emission only on a narrow cone), and the downward bending of signal trajectories in the firn also limit the observable volume for in-ice detectors but the instantaneous field-of-view is still substantially larger with a few degrees below and around \SI{45}{\degree} above the horizon (cf. Sec.~\ref{sec:skycoverage}). 

A large improvement in detection efficiency can be achieved by using interferometric techniques: The sensitivity per number of antennas of an air-shower detector can be improved by roughly a factor of 10 using a beamformer array where the beams are pointed towards the horizon. The signals of tens of antennas are combined in a dense cluster, increasing the signal-to-noise ratio by up to the square root of the number of antennas under ideal conditions. This concept is explored with BEACON \cite{beacon2020} and TAROGE \cite{TAROGE:2022soh}.

The air shower detectors have one significant advantage though. The neutrino direction can be reconstructed with sub-degree resolution if the different detector stations can be synchronized to sub-nanosecond precision. For an in-ice detector it will be challenging to go below a degree for most of the events (cf. Sec.~\ref{sec:angularresolution}). An exception here are muon and tau neutrino CC interactions where the initial interaction and a secondary shower induced by the muon or tau is observed. This gives a large lever arm and will allow constraining the neutrino direction much better. 

Another advantage of multiple UHE neutrino detectors at different locations is of course the increased sky coverage. The instantaneous field of view of an air-shower detector is small, but if installed at intermediate latitudes, a large portion of the sky will be covered within 24 hours. 

Sensitivity to even larger neutrino energies beyond \SI{e19}{eV} can be achieved with balloon or space-based experiments. The ANITA experiment \cite{ANITA-4thFlight} and its successor PUEO \cite{PUEOICRC2021} fly an array of radio antennas around Antarctica to measure tau neutrinos. The detection principle is the following. An Earth skimming neutrino interacts in the ice producing a particle shower (just as in the in-ice detector case), but then part of the radio signal propagates upwards to the surface and continues propagating through the atmosphere to be eventually detected by the balloon. The high altitude of the balloon allows to observe a huge part of the ice at once, but also requires high signal strengths and thus high neutrino energies. 

An alternative at neutrino energies below \SI{e17}{eV} is the RADAR technique. This technology might be able to better bridge the sensitivity of an optical in-ice detector and an Askaryan in-ice detector. The detection principle is to continuously transmit a radio wave into the ice and to search for reflections from the ionization deposit left in the wake of the cascade. This technique is in its early proof-of-concept phase but a first important milestone was achieved by measuring radar echoes from high-energy particle cascades in a lab measurement \cite{Prohira2020}. At the moment a pathfinder experiment to measure the more abundant cosmic rays is pursued \cite{Prohira2021}.

\section{Future}
So what is next for neutrino telescopes based on the detection of radio pulses? Any view into the future is highly speculative, representing the best guesses of the authors. To help inform our predictions,  the science community has identified two goals for the next generation of high-energy neutrino telescopes. Despite the impressive array of science missions discussed in other chapters of this book, the fact remains that there has been no observation of astrophysical neutrinos with energies above \SI{e16}{eV}.  So one thrust by the community involves the construction of an array that dramatically increases the effective volume with the goal of reaching a sensitivity below $E_\nu^2\Phi\le$ \SI{4e-10}{GeV cm^{-2}s^{-1}sr^{-1}}, for neutrino energies near \SI{e18}{eV}. To reach this goal, the array is planned to be installed at the South Pole which combines very cold ice \SI{2.7}{km} thick with excellent logistical support provided by Amundsen-Scott South Pole Station.  The grand scope of the array, encompassing an area of \SI{500}{km^2}, requires an unprecedented construction effort in the polar environment. The close proximity of the IceCube neutrino telescope provides a powerful rationale to combine the efforts of the mature optical Cherenkov community with the fledgling radio community. The design, construction, safety, and operation of the radio array on such an unprecedented scale at the South Pole will benefit greatly from the collective expertise of the IceCube collaboration.

There are many reasons to consider a hybrid design that includes design elements from both surface and deep string stations.  For example, the rejection of atmospheric muon-induced showers can be tagged more effectively by a higher areal density of surface stations (with dedicated upward-facing LPDA to tag cosmic-ray induced air showers).  The optimal separation distance is lower for surface stations since the station overlap is smaller due to the smaller effective volume per station, and equally important, a surface station has a lower cost than deep station designs, so it can be installed over the same total area at higher areal density.  Another feature of a hybrid concept: The deep and shallow detector components will allow triggering, identification, and reconstruction of UHE neutrinos with complementary systematics. Furthermore, a subset of golden events will be observed by both surface and deep components. Background rejection and event reconstruction can be cross-checked by independent systems, improving the integrity of the hybrid detector. 

Though the South Pole site has many advantages, it has one important limitation, at least for radio-based neutrino telescopes: The fraction of the sky which can be viewed is rather limited, perhaps about 30\% of the full sky. The reason was spelled out in Section \ref{sec:skycoverage}. At the relevant energies, neutrinos that propagate through the earth are greatly attenuated.  Without the benefit of a reflecting bottom surface, the view of an array at the South Pole is constrained by the geometry of the Cherenkov cone and the relatively shallow depth of the radio receivers. Interaction vertices occur mostly below the antennas, so signals must propagate upward.  For horizontally traveling neutrinos, approximately half the Cherenkov cone can trigger a detector, but as the zenith angle is reduced (and the arrival direction of the neutrino is slanted downward in local coordinates), less of the Cherenkov cone will propagate upwards, reducing the relative detection efficiency.  For zenith angles less than 50 degrees in local coordinates, the shower cannot send a ray on the Cherenkov cone upwards toward the radio receivers.  Therefore, a radio array located at the South Pole can only view an annular patch on the sky with declination between 0 and -45 degrees.  For near-surface stations, the range of declination that can be viewed is reduced.  To address this limitation, arrays in polar regions can be augmented with an additional array of high-energy neutrino stations at Moore's Bay, Antarctica to provide a nearly complete view of the sky. An array located in Greenland will scan most of the northern sky for sources that emit neutrinos for one day or longer. The combined approach satisfies the high priority objective to search for explosive or variable sources of high energy neutrinos over most of the sky. 

Looking further into the future, reductions in power consumption and improvements in battery technology may lead to the development of low cost, easily deployed, surface stations that could blanket large fractions of Antarctica and Greenland, taking maximal advantage of this ice resource. 

The final idea of this incomplete look into the future involves an intriguing proposal \cite{Miller:PRIDE} to send a satellite to one or more of the ice-covered moons in the solar system, such as Europa or Enceladus. There are several reasons to undertake such a project. First, the ice temperature for moons of the gas giant planets is significantly colder than Antarctic ice, with potentially a commensurate increase in attenuation length. The average surface temperature of Europa is -200C, which is 160 degrees colder than ice at the South Pole. However, the attenuation length also depends on the ice purity, and the salt composition of the ice sheets on these moons is not well known.  But if the ice purity is eventually determined to be high, radio attenuation lengths may be a large as \SI{100}{km}.  Another feature, the ice sheet may be much thicker than several kilometers, with some estimates as large as \SI{200}{km} thick for the ice sheet covering Europa. The final reason to consider sending a satellite to such a distant location is the variety of the diameters of the moons.  Recall that at the relevant energies, the earth will attenuate the flux of the vast majority of neutrinos, so only downward propagating neutrinos can be detected.  This limits the view of the sky and reduces the probability to view signals close to the Cherenkov cone, which contains the most power. The attenuation is greatly reduced for small diameter moons. For example, the diameter of Enceladus is only \SI{483}{km}.  Since it is possible for neutrinos to travel through the moon before interacting in the ice sheet, they will generate upward on-cone radio pulses.  The on-cone radio signals can reach orbiting receivers from the entire cone, which has two desirable effects. On-cone signals can be detected more readily from high altitudes above the ice sheet (say \SI{500}{km}) where a satellite must orbit. On-cone events exhibit the highest frequency content and therefore can be detected with smaller antennas. 
 
A radio neutrino detector placed in low orbit around the moon would be able to view about half the sky at any given instant and observe most of the sky every revolution.  Of course, the nearby planet would obscure a fraction of the sky. For example, from the perspective of Europa, Jupiter obscures a circular patch on the sky that is about \SI{6}{\degree} in radius.

\section{Summary}

Radio emission is generated in ice by particle showers through the Askaryan effect. If the shower occurs in a dielectric medium, such as ice, the shower develops a time-varying negative charge-excess in the shower front which is primarily due to a collection of electrons from the surrounding medium. The resulting radio emission can be calculated precisely using classical electrodynamics by tracking the movement of the individual particles using the well-tested Monte Carlo codes. The code incorporates important phenomena such as the LPM effect that affects the emission for $\nu_e$ charged current interactions. In ice, the electric field increases linearly from MHz frequencies up to a characteristic cutoff of a few GHz. 
Due to coherence effects, the emission is only strong at angles close to the Cherenkov angle, and linearly polarized in the plane defined by the shower direction and propagation direction of the radio signal. Because of this feature, the neutrino direction can be obtained from both the signal arrival direction as well as the polarization. The observed frequency range is strongest between \SI{100}{MHz} and \SI{1}{GHz} due to properties of the emission and ice attenuation.

In this brief overview of the field, we mentioned that the theoretical calculation of the Askaryan emission has been confirmed in accelerator experiments, and in particular for showers developing in ice. All measurements are consistent with the theoretical prediction within experimental uncertainties. Furthermore, the Askaryan effect has been observed in cosmic ray induced air showers where the Askaryan radiation is subdominant to radio signals emitted by the geomagnetic effect because of the much lower density of air compared to ice. As it was the case for accelerator measurements, the measurement of the Askaryan radiation from air showers is in agreement with theoretical calculations.

Measurements at Moore's Bay, Antarctica, the South Pole and Summit Station in Greenland reveal that radio signals propagate with little attenuation through ice, allowing an observation of large volumes with a sparse array of detector stations. In addition, studies at Moore’s Bay confirmed expectation of excellent reflection from the water-ice interface over the area of the hexagonal array or station. In-situ radio pulsers at Moore’s Bay and South Pole were used to construct the arrival direction and polarization from typical propagation paths originating from neutrino interactions. This proves that the refractive effects on radio signals is understood in bulk ice and through the upper firn.  Also the received signal amplitudes are consistent with expectation after correcting for attenuation from the propagation through the ice. 
In addition, the ARIANNA collaboration verified the polarization response by measuring cosmic rays which generate radio pulses in the atmosphere with well-characterized polarization,  providing an independent check on the predicted angular resolution.

Within the past few years, experience from the pilot programs have led to triggering innovations that produce dramatic improvements in sensitivity.  On-board filters based on deep learning techniques can increase the trigger rates by five orders of magnitude. In addition, recent work has shown that optimizing the bandwidth prior to the discriminator and transitioning to low noise amplifiers produces additional improvements in capabilities.  Furthermore,  a phased-array trigger has been developed to reduce the trigger threshold by a factor 2 compared to previously employed majority logic triggers.

Perhaps the most significant looming issue involves the rate and other physical characteristics of rare, or difficult to calculate, or unanticipated background processes.  In particular, reflected signals generated by cosmic ray cores or direct signals by secondary muons need further study.  At Moore's Bay, the specular water-ice reflecting surface at the bottom of the ice sheet is well studied, but at other site locations, the strength and reflective properties of internal scattering layers require more work to fully assess their impact on background processes. 

The field is poised to continue to make rapid progress.  With the construction of the Radio Neutrino Observatory in Greenland (RNO-G), an in-ice radio detector is being constructed with enough sensitivity to potentially measure the first ultra-high-energy neutrino. A surface detector, such as ARIANNA-200 on the Ross Ice Shelf,  might complement RNO-G with similar sensitivity and enhanced pointing resolution to achieve almost full sky coverage if funding can be secured. These projects make a leap forward in technology development and will evaluate the long term reliability, detector sensitivity, and costing of their technology choices.  Most importantly, both program will assess, and if necessary, develop the appropriate procedures to mitigate all backgrounds, including rare and difficult to calculate phenomena. This paves the way for an order-of-magnitude more sensitive radio detector as part of IceCube-Gen2 which will open a new window to the universe through UHE neutrino astronomy.

\bibliographystyle{JHEP}
\bibliography{references}

\providecommand{\href}[2]{#2}\begingroup\raggedright\begin{thebibliography}{100}

\bibitem{PhysRevLett.116.061102}
{\scshape LIGO Scientific Collaboration and Virgo Collaboration} collaboration,
  B.~P. Abbott et~al., \emph{Observation of gravitational waves from a binary
  black hole merger},
  \href{http://dx.doi.org/10.1103/PhysRevLett.116.061102}{\emph{Phys. Rev.
  Lett.} {\bfseries 116} (2016) 061102}.

\bibitem{IceCube2013-PeVNu}
M.~G. Aartsen et~al., \emph{First observation of {PeV}-energy neutrinos with
  {IceCube}},
  \href{http://dx.doi.org/10.1103/physrevlett.111.021103}{\emph{Phys. Rev.
  Lett.} {\bfseries 111} (2013) 021103}.

\bibitem{IceCube2016}
M.~G. Aartsen et~al., \emph{The {IceCube Neutrino Observatory}: Instrumentation
  and online systems},
  \href{http://dx.doi.org/10.1088/1748-0221/12/03/P03012}{\emph{JINST}
  {\bfseries 12} (2017) P03012},
  [\href{https://arxiv.org/abs/1612.05093}{{\ttfamily 1612.05093}}].

\bibitem{Askaryan1962}
G.~A. {Askar'yan}, \emph{{Coherent Radio Emission from Cosmic Showers in Air
  and in Dense Media}}, {\emph{Soviet Journal of Experimental and Theoretical
  Physics} {\bfseries 21} (1965) 658}.

\bibitem{Barwick2005-PoleAtten}
S.~Barwick, D.~Besson, P.~Gorham and D.~Saltzberg, \emph{{South Polar in situ
  radio-frequency ice attenuation}},
  \href{http://dx.doi.org/10.3189/172756505781829467}{\emph{J. Glaciol.}
  {\bfseries 51} (2005) 231--238}.

\bibitem{ARA-LongBaseline2020}
P.~Allison et~al., \emph{{Long-baseline horizontal radio-frequency transmission
  through polar ice}},
  \href{http://dx.doi.org/10.1088/1475-7516/2020/12/009}{\emph{JCAP} {\bfseries
  12} (2020) 009}, [\href{https://arxiv.org/abs/1908.10689}{{\ttfamily
  1908.10689}}].

\bibitem{IceCube-IceAttenuation}
{\scshape IceCube} collaboration, M.~G. Aartsen et~al., \emph{{Measurement of
  South Pole ice transparency with the IceCube LED calibration system}},
  \href{http://dx.doi.org/10.1016/j.nima.2013.01.054}{\emph{Nucl. Instrum.
  Meth. A} {\bfseries 711} (2013) 73--89},
  [\href{https://arxiv.org/abs/1301.5361}{{\ttfamily 1301.5361}}].

\bibitem{Barwick2018}
S.~W. Barwick, E.~C. Berg, D.~Z. Besson, G.~Gaswint, C.~Glaser, A.~Hallgren
  et~al., \emph{Observation of classically `forbidden' electromagnetic wave
  propagation and implications for neutrino detection},
  \href{http://dx.doi.org/10.1088/1475-7516/2018/07/055}{\emph{JCAP} {\bfseries
  07} (2018) 055}, [\href{https://arxiv.org/abs/1804.10430}{{\ttfamily
  1804.10430}}].

\bibitem{Deaconu2018}
C.~Deaconu, A.~Vieregg, S.~Wissel, J.~Bowen, S.~Chipman, A.~Gupta et~al.,
  \emph{Measurements and modeling of near-surface radio propagation in glacial
  ice and implications for neutrino experiments},
  \href{http://dx.doi.org/10.1103/physrevd.98.043010}{\emph{Physical Review D}
  {\bfseries 98} (2018) 043010},
  [\href{https://arxiv.org/abs/1805.12576}{{\ttfamily 1805.12576}}].

\bibitem{RICE2003-Performance}
{\scshape RICE} collaboration, I.~Kravchenko et~al., \emph{{Performance and
  simulation of the RICE detector}},
  \href{http://dx.doi.org/10.1016/S0927-6505(02)00194-9}{\emph{Astropart.
  Phys.} {\bfseries 19} (2003) 15--36},
  [\href{https://arxiv.org/abs/astro-ph/0112372}{{\ttfamily
  astro-ph/0112372}}].

\bibitem{SummitGreenland}
\emph{{Summit Station, Greenland}}, {\emph{\url{https://www.geosummit.org/}} }.

\bibitem{ARIANNA-Atten}
J.~Hanson et~al., \emph{{Radar Absorption, Basal Reflection, Thickness, and
  Polarization Measurements from the Ross Ice Shelf}},
  \href{http://dx.doi.org/10.3189/2015JoG14J214}{\emph{J. Glaciology}
  {\bfseries 61} (2015) 438--446},
  [\href{https://arxiv.org/abs/astro-ph/1410.7134}{{\ttfamily
  astro-ph/1410.7134}}].

\bibitem{Gandhi1998}
R.~Gandhi, C.~Quigg, M.~H. Reno and I.~Sarcevic, \emph{Neutrino interactions at
  ultrahigh energies},
  \href{http://dx.doi.org/10.1103/PhysRevD.58.093009}{\emph{Physical Review D}
  {\bfseries 58} (1998) 093009},
  [\href{https://arxiv.org/abs/hep-ph/9807264}{{\ttfamily hep-ph/9807264}}].

\bibitem{CooperSarkar2011}
A.~Cooper-Sarkar, P.~Mertsch and S.~Sarkar, \emph{The high energy neutrino
  cross-section in the standard model and its uncertainty},
  \href{http://dx.doi.org/10.1007/JHEP08(2011)042}{\emph{JHEP} {\bfseries 08}
  (2011) 042}, [\href{https://arxiv.org/abs/1106.3723v2}{{\ttfamily
  1106.3723v2}}].

\bibitem{Connolly2011}
A.~Connolly, R.~S. Thorne and D.~Waters, \emph{{Calculation of high energy
  neutrino-nucleon cross sections and uncertainties using the
  Martin-Stirling-Thorne-Watt parton distribution functions and implications
  for future experiments}},
  \href{http://dx.doi.org/10.1103/PhysRevD.83.113009}{\emph{Phys. Rev. D}
  {\bfseries 83} (2011) 113009},
  [\href{https://arxiv.org/abs/1102.0691}{{\ttfamily 1102.0691}}].

\bibitem{Bertone:2018dse}
V.~Bertone, R.~Gauld and J.~Rojo, \emph{{Neutrino Telescopes as QCD
  Microscopes}}, \href{http://dx.doi.org/10.1007/JHEP01(2019)217}{\emph{JHEP}
  {\bfseries 01} (2019) 217},
  [\href{https://arxiv.org/abs/1808.02034}{{\ttfamily 1808.02034}}].

\bibitem{NuRadioMC2019}
C.~Glaser et~al., \emph{{NuRadioMC}: simulating the radio emission of neutrinos
  from interaction to detector},
  \href{http://dx.doi.org/10.1140/epjc/s10052-020-7612-8}{\emph{The European
  Physical Journal C} {\bfseries 80} (2020) 77},
  [\href{https://arxiv.org/abs/1906.01670}{{\ttfamily 1906.01670}}].

\bibitem{Valera:2022ylt}
V.~B. Valera, M.~Bustamante and C.~Glaser, \emph{{The ultra-high-energy
  neutrino-nucleon cross section: measurement forecasts for an era of cosmic
  EeV-neutrino discovery}},
  \href{http://dx.doi.org/10.1007/JHEP06(2022)105}{\emph{JHEP} {\bfseries 06}
  (2022) 105}, [\href{https://arxiv.org/abs/2204.04237}{{\ttfamily
  2204.04237}}].

\bibitem{Esteban:2022uuw}
I.~Esteban, S.~Prohira and J.~F. Beacom, \emph{{Detector requirements for
  model-independent measurements of ultrahigh energy neutrino cross sections}},
  \href{http://dx.doi.org/10.1103/PhysRevD.106.023021}{\emph{Phys. Rev. D}
  {\bfseries 106} (2022) 023021},
  [\href{https://arxiv.org/abs/2205.09763}{{\ttfamily 2205.09763}}].

\bibitem{HalzenSaltzberg-TauRegen}
F.~Halzen and D.~Saltzberg, \emph{{Tau-neutrino appearance with a 1000
  megaparsec baseline}},
  \href{http://dx.doi.org/10.1103/PhysRevLett.81.4305}{\emph{Phys. Rev. Lett.}
  {\bfseries 81} (1998) 4305--4308},
  [\href{https://arxiv.org/abs/hep-ph/9804354}{{\ttfamily hep-ph/9804354}}].

\bibitem{Verhoeven2005}
O.~Verhoeven, A.~Rivoldini, P.~Vacher, A.~Mocquet, G.~Choblet, M.~Menvielle
  et~al., \emph{Preliminary reference earth model}, {\emph{Physics of the Earth
  and Planetary Interiors} {\bfseries 110} (2005) 297--356}.

\bibitem{Heitler}
W.~Heitler, \emph{{Quantum Theory of Radiation}}.
\newblock Oxford Univ. Press., 2nd~ed., 1944.

\bibitem{CarlsonOppenheimer1937}
J.~F. Carlson and J.~R. Oppenheimer, \emph{{On Multiplicative Showers}},
  \href{http://dx.doi.org/10.1103/physrev.51.220}{\emph{Phys. Rev.} {\bfseries
  51} (1937) 220--231}.

\bibitem{Matthews2005387}
J.~Matthews, \emph{{A Heitler model of extensive air showers}},
  \href{http://dx.doi.org/10.1016/j.astropartphys.2004.09.003}{\emph{Astropart.
  Phys.} {\bfseries 22} (2005) 387--397}.

\bibitem{Alvarez-Muniz1998}
J.~Alvarez-Mu{\~{n}}iz and E.~Zas, \emph{{The LPM effect for EeV hadronic
  showers in ice: Implications for radio detection of neutrinos}},
  \href{http://dx.doi.org/10.1016/S0370-2693(98)00905-8}{\emph{Phys. Lett.}
  {\bfseries 434} (1998) 396--406},
  [\href{https://arxiv.org/abs/astro-ph/9806098}{{\ttfamily
  astro-ph/9806098}}].

\bibitem{Landau:1953um}
L.~D. Landau and I.~Pomeranchuk, \emph{{Limits of applicability of the theory
  of bremsstrahlung electrons and pair production at high-energies}},
  {\emph{Dokl. Akad. Nauk Ser. Fiz.} {\bfseries 92} (1953) 535--536}.

\bibitem{Migdal:1956tc}
A.~B. Migdal, \emph{{Bremsstrahlung and pair production in condensed media at
  high-energies}},
  \href{http://dx.doi.org/10.1103/PhysRev.103.1811}{\emph{Phys. Rev.}
  {\bfseries 103} (1956) 1811--1820}.

\bibitem{Konishi1991}
E.~Konishi, A.~Adachi, N.~Takahashi and A.~Misaki, \emph{On the characteristics
  of individual cascade showers with the lpm effect at extremely high
  energies}, {\emph{Journal of Physics G: Nuclear and Particle Physics}
  {\bfseries 17} (1991) 719}.

\bibitem{Stanev1982}
T.~Stanev, C.~Vankov, R.~Streitmatter, R.~Ellsworth and T.~Bowen,
  \emph{Development of ultrahigh-energy electromagnetic cascades in water and
  lead including the landau-pomeranchuk-migdal effect}, {\emph{Physical Review
  D} {\bfseries 25} (1982) 1291}.

\bibitem{Alvarez-Muniz1997b}
J.~Alvarez-Mu{\~{n}}iz and E.~Zas, \emph{{Cherenkov radio pulses from EeV
  neutrino interactions: the LPM effect}},
  \href{http://dx.doi.org/10.1016/S0370-2693(97)01009-5}{\emph{Phys. Lett. B}
  {\bfseries 411} (1997) 218--224},
  [\href{https://arxiv.org/abs/9706064}{{\ttfamily 9706064}}].

\bibitem{SpencerLPMReview1999}
S.~Klein, \emph{Suppression of bremsstrahlung and pair production due to
  environmental factors},
  \href{http://dx.doi.org/10.1103/RevModPhys.71.1501}{\emph{Rev. Mod. Phys.}
  {\bfseries 71} (1999) 1501--1538}.

\bibitem{herwig}
J.~Bellm et~al., \emph{{Herwig 7.1 Release Note}},
  \href{https://arxiv.org/abs/1705.06919}{{\ttfamily 1705.06919}}.

\bibitem{AlvarezMuniz:2011bs}
J.~Alvarez-Mu{\~n}iz, W.~R. Carvalho, Jr. and E.~Zas, \emph{{Monte Carlo
  simulations of radio pulses in atmospheric showers using ZHAireS}},
  \href{http://dx.doi.org/10.1016/j.astropartphys.2011.10.005}{\emph{Astropart.
  Phys.} {\bfseries 35} (2012) 325--341},
  [\href{https://arxiv.org/abs/1107.1189}{{\ttfamily 1107.1189}}].

\bibitem{GarciaFernandez2020}
D.~García-Fernández, C.~Glaser and A.~Nelles, \emph{The signatures of
  secondary leptons in radio-neutrino detectors in ice},
  \href{http://dx.doi.org/10.1103/PhysRevD.102.083011}{\emph{Phys. Rev. D}
  {\bfseries 102} (2020) 083011},
  [\href{https://arxiv.org/abs/2003.13442}{{\ttfamily 2003.13442}}].

\bibitem{GlaserICRC2021Leptons}
C.~Glaser, D.~Garc\'{\i}a-Fern\'{a}ndez and A.~Nelles, \emph{Prospects for
  neutrino-flavor physics with in-ice radio detectors},
  \href{http://dx.doi.org/10.22323/1.395.1231}{\emph{PoS(ICRC2021)1231} }.

\bibitem{proposal}
J.-H. Koehne et~al., \emph{{PROPOSAL: A tool for propagation of charged
  leptons}},
  \href{http://dx.doi.org/https://doi.org/10.1016/j.cpc.2013.04.001}{\emph{Computer
  Physics Communications} {\bfseries 184} (2013) 2070 -- 2090}.

\bibitem{Alvarez-Muniz1999}
J.~Alvarez-Mu{\~{n}}iz, R.~A. V{\'{a}}zquez and E.~Zas, \emph{{Characterization
  of neutrino signals with radiopulses in dense media through the
  Landau-Pomeranchuk-Migdal effect}},
  \href{http://dx.doi.org/10.1103/PhysRevD.61.023001}{\emph{Phys. Rev. D}
  {\bfseries 61} (1999) 023001}.

\bibitem{Gerhardt2010}
L.~Gerhardt and S.~R. Klein, \emph{Electron and photon interactions in the
  regime of strong landau-pomeranchuk-migdal suppression},
  \href{http://dx.doi.org/10.1103/physrevd.82.074017}{\emph{Physical Review D}
  {\bfseries 82} (2010) 074017}.

\bibitem{PersichilliPhD}
C.~R. Persichilli, \emph{Performance and Simulation of the ARIANNA Pilot Array,
  with Implications for Future Ultra-high Energy Neutrino Astronomy}.
\newblock PhD thesis, 2018.

\bibitem{ARZ2}
J.~Alvarez-Mu\~niz, A.~Romero-Wolf and E.~Zas, \emph{\ifmmode \check{C}\else
  \v{C}\fi{}erenkov radio pulses from electromagnetic showers in the time
  domain}, \href{http://dx.doi.org/10.1103/PhysRevD.81.123009}{\emph{Phys. Rev.
  D} {\bfseries 81} (2010) 123009},
  [\href{https://arxiv.org/abs/1002.3873}{{\ttfamily 1002.3873}}].

\bibitem{ARZ}
J.~Alvarez-Mu\~niz, A.~Romero-Wolf and E.~Zas, \emph{{Practical and accurate
  calculations of Askaryan radiation}},
  \href{http://dx.doi.org/10.1103/PhysRevD.84.103003}{\emph{Phys. Rev.}
  {\bfseries D84} (2011) 103003},
  [\href{https://arxiv.org/abs/1106.6283}{{\ttfamily 1106.6283}}].

\bibitem{Endpoint2011}
C.~W. James, H.~Falcke, T.~Huege and M.~Ludwig, \emph{{General description of
  electromagnetic radiation processes based on instantaneous charge
  acceleration in endpoints}},
  \href{http://dx.doi.org/10.1103/physreve.84.056602}{\emph{Phys. Rev. E}
  {\bfseries 84} (2011) 56602}.

\bibitem{CoREAS2013}
T.~Huege, M.~Ludwig and C.~W. James, \emph{{Simulating radio emission from air
  showers with {CoREAS}}}, \href{http://dx.doi.org/10.1063/1.4807534}{\emph{AIP
  Conf. Proc.} {\bfseries 1535} (2013) 128--132}.

\bibitem{Alvarez_box}
J.~Alvarez-Mu\~niz, E.~Marqu\'es, R.~A. V\'azquez and E.~Zas, \emph{{Coherent
  radio pulses from showers in different media: a unified parameterization}},
  \href{http://dx.doi.org/10.1103/PhysRevD.74.023007}{\emph{Phys. Rev.}
  {\bfseries D74} (2006) 023007},
  [\href{https://arxiv.org/abs/astro-ph/0512337}{{\ttfamily
  astro-ph/0512337}}].

\bibitem{Alvarez2012}
J.~Alvarez-Mu\~niz, W.~R. Carvalho, M.~Tueros and E.~Zas, \emph{{Coherent
  Cherenkov radio pulses from hadronic showers up to {EeV} energies}},
  \href{http://dx.doi.org/https://doi.org/10.1016/j.astropartphys.2011.10.002}{\emph{Astroparticle
  Physics} {\bfseries 35} (2012) 287 -- 299},
  [\href{https://arxiv.org/abs/1005.0552}{{\ttfamily 1005.0552}}].

\bibitem{Alvarez2009}
J.~Alvarez-Mu\~niz, C.~James, R.~Protheroe and E.~Zas, \emph{Thinned
  simulations of extremely energetic showers in dense media for radio
  applications},
  \href{http://dx.doi.org/https://doi.org/10.1016/j.astropartphys.2009.06.005}{\emph{Astroparticle
  Physics} {\bfseries 32} (2009) 100 -- 111}.

\bibitem{AlvarezMuiz2000}
J.~Alvarez-Mu{\~{n}}iz, R.~A. V{\'{a}}zquez and E.~Zas, \emph{Calculation
  methods for radio pulses from high energy showers},
  \href{http://dx.doi.org/10.1103/physrevd.62.063001}{\emph{Physical Review D}
  {\bfseries 62} (2000) 063001}.

\bibitem{AlvarezMuiz1998}
J.~Alvarez-Mu{\~{n}}iz and E.~Zas, \emph{The {LPM} effect for {EeV} hadronic
  showers in ice: implications for radio detection of neutrinos},
  \href{http://dx.doi.org/10.1016/s0370-2693(98)00905-8}{\emph{Physics Letters
  B} {\bfseries 434} (1998) 396--406}.

\bibitem{Alvarez-Muniz2009}
J.~Alvarez-Mu{\~{n}}iz, C.~W. James, R.~J. Protheroe and E.~Zas, \emph{{Thinned
  simulations of extremely energetic showers in dense media for radio
  applications}},
  \href{http://dx.doi.org/10.1016/j.astropartphys.2009.06.005}{\emph{Astropart.
  Phys.} {\bfseries 32} (2009) 100--111}.

\bibitem{Engel:2018akg}
R.~Engel, D.~Heck, T.~Huege, T.~Pierog, M.~Reininghaus, F.~Riehn et~al.,
  \emph{{Towards a Next Generation of CORSIKA: A Framework for the Simulation
  of Particle Cascades in Astroparticle Physics}},
  \href{http://dx.doi.org/10.1007/s41781-018-0013-0}{\emph{Comput. Softw. Big
  Sci.} {\bfseries 3} (2019) 2},
  [\href{https://arxiv.org/abs/1808.08226}{{\ttfamily 1808.08226}}].

\bibitem{Saltzberg2001}
D.~Saltzberg, P.~Gorham, D.~Walz, C.~Field, R.~Iverson, A.~Odian et~al.,
  \emph{Observation of the askaryan effect: Coherent microwave cherenkov
  emission from charge asymmetry in high-energy particle cascades},
  \href{http://dx.doi.org/10.1103/physrevlett.86.2802}{\emph{Physical Review
  Letters} {\bfseries 86} (2001) 2802}.

\bibitem{Miocinovic2006}
P.~Mio{\v{c}}inovi{\'{c}}, R.~C. Field, P.~W. Gorham, E.~Guillian,
  R.~Milin{\v{c}}i{\'{c}}, D.~Saltzberg et~al., \emph{Time-domain measurement
  of broadband coherent cherenkov radiation},
  \href{http://dx.doi.org/10.1103/physrevd.74.043002}{\emph{Physical Review D}
  {\bfseries 74} (2006) 043002}.

\bibitem{Gorham2005}
P.~W. Gorham, D.~Saltzberg, R.~C. Field, E.~Guillian, R.~Milin{\v{c}}i{\'{c}},
  P.~Mio{\v{c}}inovi{\'{c}} et~al., \emph{Accelerator measurements of the
  askaryan effect in rock salt: A roadmap toward teraton underground neutrino
  detectors},
  \href{http://dx.doi.org/10.1103/physrevd.72.023002}{\emph{Physical Review D}
  {\bfseries 72} (2005) 023002}.

\bibitem{Gorham2007}
P.~W. Gorham, S.~W. Barwick, J.~J. Beatty, D.~Z. Besson, W.~R. Binns, C.~Chen
  et~al., \emph{Observations of the askaryan effect in ice},
  \href{http://dx.doi.org/10.1103/physrevlett.99.171101}{\emph{Physical Review
  Letters} {\bfseries 99} (2007) 171101}.

\bibitem{GlaserErad2016}
C.~Glaser, M.~Erdmann, J.~R. H{\"{o}}randel, T.~Huege and J.~Schulz,
  \emph{{Simulation of Radiation Energy Release in Air Showers}},
  \href{http://dx.doi.org/10.1088/1475-7516/2016/09/024}{\emph{J. Cosmol.
  Astropart. Phys.} {\bfseries 09} (2016) 24}.

\bibitem{AERAPolarization}
A.~Aab et~al., \emph{{Probing the radio emission from air showers with
  polarization measurements}},
  \href{http://dx.doi.org/10.1103/PhysRevD.89.052002}{\emph{Phys. Rev. D}
  {\bfseries 89} (2014) 52002}.

\bibitem{LofarPolarization2014}
P.~Schellart et~al., \emph{{Polarized radio emission from extensive air showers
  measured with LOFAR}},
  \href{http://dx.doi.org/10.1088/1475-7516/2014/10/014}{\emph{J. Cosmol.
  Astropart. Phys.} {\bfseries 10} (2014) 014},
  [\href{https://arxiv.org/abs/1406.1355}{{\ttfamily 1406.1355}}].

\bibitem{Huege2016}
T.~Huege, \emph{{Radio detection of cosmic ray air showers in the digital
  era}}, \href{http://dx.doi.org/10.1016/j.physrep.2016.02.001}{\emph{Phys.
  Rep.} {\bfseries 620} (2016) 1--52}.

\bibitem{Schroeder2016}
F.~G. Schr{\"{o}}der, \emph{{Radio detection of cosmic-ray air showers and
  high-energy neutrinos}},
  \href{http://dx.doi.org/10.1016/j.ppnp.2016.12.002}{\emph{Prog. Part. Nucl.
  Phys.} {\bfseries 93} (2017) 1--68}.

\bibitem{SlacT510}
K.~Belov, K.~Mulrey, A.~Romero-Wolf, S.~Wissel, A.~Zilles, K.~Bechtol et~al.,
  \emph{{Accelerator Measurements of Magnetically Induced Radio Emission from
  Particle Cascades with Applications to Cosmic-Ray Air Showers}},
  \href{http://dx.doi.org/10.1103/physrevlett.116.141103}{\emph{Phys. Rev.
  Lett.} {\bfseries 116} (2016) 141103}.

\bibitem{Barwick2017-Airshowers}
S.~W. Barwick, D.~Z. Besson, A.~Burgman, E.~Chiem, A.~Hallgren, J.~C. Hanson
  et~al., \emph{{Radio detection of air showers with the ARIANNA experiment on
  the Ross Ice Shelf}},
  \href{http://dx.doi.org/10.1016/j.astropartphys.2017.02.003}{\emph{Astropart.
  Phys.} {\bfseries 90} (2017) 50--68}.

\bibitem{Arianna:2021lnr}
{\scshape Arianna} collaboration, A.~Anker et~al., \emph{{Measuring the
  polarization reconstruction resolution of the ARIANNA neutrino detector with
  cosmic rays}},
  \href{http://dx.doi.org/10.1088/1475-7516/2022/04/022}{\emph{JCAP} {\bfseries
  04} (2022) 022}, [\href{https://arxiv.org/abs/2112.01501}{{\ttfamily
  2112.01501}}].

\bibitem{Aguilar:2022kgi}
J.~A. Aguilar et~al., \emph{{In situ, broadband measurement of the radio
  frequency attenuation length at Summit Station, Greenland}},
  \href{http://dx.doi.org/10.1017/jog.2022.40}{\emph{Journal of Glaciology}
  (2022) 1--9}, [\href{https://arxiv.org/abs/2201.07846}{{\ttfamily
  2201.07846}}].

\bibitem{Avva:2014ena}
J.~Avva, J.~M. Kovac, C.~Miki, D.~Saltzberg and A.~G. Vieregg, \emph{{An in
  situ measurement of the radio-frequency attenuation in ice at Summit Station,
  Greenland}}, \href{http://dx.doi.org/10.3189/2015JoG15J057}{\emph{J.
  Glaciol.} {\bfseries 61} (2015) 1005--1011},
  [\href{https://arxiv.org/abs/1409.5413}{{\ttfamily 1409.5413}}].

\bibitem{BarwickBergBessonEtAl2014}
S.~W. Barwick, E.~C. Berg, D.~Besson, T.~Duffin, J.~C. Hanson, S.~R. Klein
  et~al., \emph{{Radar Absorption, Basal Reflection, Thickness, and
  Polarization Measurements from the Ross Ice Shelf}},
  \href{http://dx.doi.org/10.3189/2015JoG14J214}{\emph{J. Glaciol.} {\bfseries
  61} (2015) 438--446}.

\bibitem{Kravchenko2004}
I.~Kravchenko, D.~Besson and J.~Meyers, \emph{In situ index-of-refraction
  measurements of the south polar firn with the {RICE} detector},
  \href{http://dx.doi.org/10.3189/172756504781829800}{\emph{Journal of
  Glaciology} {\bfseries 50} (2004) 522--532}.

\bibitem{Schytt1958}
V.~Schytt, \emph{{Snow studies at Maudheim. Snowstudies inland. The inner
  structure of the ice shelf at Maudheim asshown by core drilling,
  Norwegian-British Swedish Antarctic Expedition. 1949-52}}, {\emph{Sci.
  Results.} {\bfseries 4} (1958) 1–64}.

\bibitem{Aguilar:2020RNOG}
{\scshape RNO-G} collaboration, J.~A. Aguilar et~al., \emph{{Design and
  Sensitivity of the Radio Neutrino Observatory in Greenland (RNO-G)}},
  \href{http://dx.doi.org/10.1088/1748-0221/16/03/P03025}{\emph{JINST}
  {\bfseries 16} (2021) P03025},
  [\href{https://arxiv.org/abs/2010.12279}{{\ttfamily 2010.12279}}].

\bibitem{HallmannICRC2021}
{S. Hallmann, B. Clark, C. Glaser and D. Smith for the IceCube-Gen2
  collaboration}, \emph{Sensitivity studies for the {IceCube-Gen2} radio
  array}, \href{http://dx.doi.org/10.22323/1.395.1183}{\emph{PoS(ICRC2021)1183}
  }.

\bibitem{ARIANNA200}
A.~Anker, P.~Baldi, S.~W. Barwick, D.~Bergman, H.~Bernhoff, D.~Z. Besson
  et~al., \emph{White paper: Arianna-200 high energy neutrino telescope},
  {\emph{arXiv:2004.09841} },
  [\href{https://arxiv.org/abs/2004.09841}{{\ttfamily 2004.09841}}].

\bibitem{Allison:2017jpy}
P.~Allison et~al., \emph{{Measurement of the real dielectric permittivity
  $\epsilon_r$ of glacial ice}},
  \href{http://dx.doi.org/10.1016/j.astropartphys.2019.01.004}{\emph{Astropart.
  Phys.} {\bfseries 108} (2019) 63--73},
  [\href{https://arxiv.org/abs/1712.03301}{{\ttfamily 1712.03301}}].

\bibitem{Besson:2021wmj}
D.~Besson, I.~Kravchenko and K.~Nivedita, \emph{{Polarization angle dependence
  of vertically propagating radio-frequency signals in South Polar ice}},
  {\emph{Astroparticle Physics} {\bfseries 144} (2023) 102766},
  [\href{https://arxiv.org/abs/2110.13353}{{\ttfamily 2110.13353}}].

\bibitem{Jordan2020}
T.~M. {Jordan} et~al., \emph{{Modelling ice birefringence and oblique radio
  wave propagation for neutrino detection at the South Pole}},
  \href{http://dx.doi.org/10.1017/aog.2020.18}{\emph{Annals of Glaciology, 1-8}
  (2019) }, [\href{https://arxiv.org/abs/1910.01471}{{\ttfamily 1910.01471}}].

\bibitem{Connolly:2021cum}
A.~Connolly, \emph{{Impact of biaxial birefringence in polar ice at radio
  frequencies on signal polarizations in ultrahigh energy neutrino detection}},
  \href{http://dx.doi.org/10.1103/PhysRevD.105.123012}{\emph{Phys. Rev. D}
  {\bfseries 105} (2022) 123012},
  [\href{https://arxiv.org/abs/2110.09015}{{\ttfamily 2110.09015}}].

\bibitem{Heyer:2022ttn}
N.~Heyer and C.~Glaser, \emph{{First-principle calculation of birefringence
  effects for in-ice radio detection of neutrinos}},
  \href{https://arxiv.org/abs/2205.06169}{{\ttfamily 2205.06169}}.

\bibitem{RadarEchoTelescope:2020nhe}
{\scshape Radar Echo Telescope} collaboration, S.~Prohira et~al.,
  \emph{{Modeling in-ice radio propagation with parabolic equation methods}},
  \href{http://dx.doi.org/10.1103/PhysRevD.103.103007}{\emph{Phys. Rev. D}
  {\bfseries 103} (2021) 103007},
  [\href{https://arxiv.org/abs/2011.05997}{{\ttfamily 2011.05997}}].

\bibitem{NuRadioReco2019}
C.~Glaser, A.~Nelles, I.~Plaisier, C.~Welling, S.~W. Barwick,
  D.~García-Fernández et~al., \emph{{NuRadioReco:} a reconstruction framework
  for radio neutrino detectors},
  \href{http://dx.doi.org/10.1140/epjc/s10052-019-6971-5}{\emph{The European
  Physical Journal C} {\bfseries 79} (2019) 464},
  [\href{https://arxiv.org/abs/1903.07023}{{\ttfamily 1903.07023}}].

\bibitem{Glaser2020}
C.~Glaser and S.~W. Barwick, \emph{An improved trigger for askaryan radio
  detectors},
  \href{http://dx.doi.org/10.1088/1748-0221/16/05/t05001}{\emph{Journal of
  Instrumentation} {\bfseries 16} (2021) T05001},
  [\href{https://arxiv.org/abs/2011.12997}{{\ttfamily 2011.12997}}].

\bibitem{ARIANNATimeDomain2015}
S.~W. Barwick, E.~C. Berg, D.~Z. Besson, T.~Duffin, J.~C. Hanson, S.~R. Klein
  et~al., \emph{{Time-domain response of the {ARIANNA} detector}},
  \href{http://dx.doi.org/10.1016/j.astropartphys.2014.09.002}{\emph{Astropart.
  Phys.} {\bfseries 62} (2015) 139--151}.

\bibitem{ARA2019-PA}
P.~Allison, S.~Archambault, R.~Bard, J.~J. Beatty, M.~Beheler-Amass, D.~Z.
  Besson et~al., \emph{Design and performance of an interferometric trigger
  array for radio detection of high-energy neutrinos},
  \href{http://dx.doi.org/10.1016/j.nima.2019.01.067}{\emph{NIM-A} {\bfseries
  930} (2019) 112--125}, [\href{https://arxiv.org/abs/1809.04573}{{\ttfamily
  1809.04573}}].

\bibitem{GlaserICRC2019}
{C. Glaser for the ARIANNA Collaboration}, \emph{{Neutrino direction and energy
  resolution of Askaryan detectors}}, {\emph{Proc. 3th ICRC 2019, Madison,
  Wisconsin, USA, PoS(ICRC2019)899} },
  [\href{https://arxiv.org/abs/1911.02093}{{\ttfamily 1911.02093}}].

\bibitem{Stjarnholm:2021xpj}
S.~Stj\"arnholm, O.~Ericsson and C.~Glaser, \emph{{Neutrino direction and
  flavor reconstruction from radio detector data using deep convolutional
  neural networks}}, \href{http://dx.doi.org/10.22323/1.395.1055}{\emph{PoS}
  {\bfseries ICRC2021} (2021) 1055}.

\bibitem{DnR2019}
A.~Anker, S.~W. Barwick, H.~Bernhoff, D.~Z. Besson, N.~Bingefors,
  D.~García-Fernández et~al., \emph{Neutrino vertex reconstruction with
  in-ice radio detectors using surface reflections and implications for the
  neutrino energy resolution},
  \href{http://dx.doi.org/10.1088/1475-7516/2019/11/030}{\emph{Journal of
  Cosmology and Astroparticle Physics} {\bfseries 11} (2019) 030},
  [\href{https://arxiv.org/abs/1909.02677}{{\ttfamily 1909.02677}}].

\bibitem{ARIANNAPolarization2020}
{ARIANNA Collaboration}, A.~Anker, S.~W. Barwick, H.~Bernhoff, D.~Z. Besson,
  N.~Bingefors et~al., \emph{Probing the angular and polarization
  reconstruction of the arianna detector at the south pole},
  \href{https://arxiv.org/abs/2006.03027}{{\ttfamily 2006.03027}}.

\bibitem{Schoorlemmer:2015afa}
H.~Schoorlemmer et~al., \emph{{Energy and Flux Measurements of Ultra-High
  Energy Cosmic Rays Observed During the First ANITA Flight}},
  \href{http://dx.doi.org/10.1016/j.astropartphys.2016.01.001}{\emph{Astropart.
  Phys.} {\bfseries 77} (2016) 32--43},
  [\href{https://arxiv.org/abs/1506.05396}{{\ttfamily 1506.05396}}].

\bibitem{Welling:2019scz}
C.~Welling, C.~Glaser and A.~Nelles, \emph{{Reconstructing the cosmic-ray
  energy from the radio signal measured in one single station}},
  \href{http://dx.doi.org/10.1088/1475-7516/2019/10/075}{\emph{JCAP} {\bfseries
  10} (2019) 075}, [\href{https://arxiv.org/abs/1905.11185}{{\ttfamily
  1905.11185}}].

\bibitem{Aguilar:2021uzt}
J.~A. Aguilar et~al., \emph{{Reconstructing the neutrino energy for in-ice
  radio detectors: A study for the Radio Neutrino Observatory Greenland
  (RNO-G)}},
  \href{http://dx.doi.org/10.1140/epjc/s10052-022-10034-4}{\emph{Eur. Phys. J.
  C} {\bfseries 82} (2022) 147},
  [\href{https://arxiv.org/abs/2107.02604}{{\ttfamily 2107.02604}}].

\bibitem{ARIANNA:2019scz}
{\scshape ARIANNA} collaboration, A.~Anker et~al., \emph{{Targeting ultra-high
  energy neutrinos with the ARIANNA experiment}},
  \href{http://dx.doi.org/10.1016/j.asr.2019.06.016}{\emph{Adv. Space Res.}
  {\bfseries 64} (2019) 2595--2609},
  [\href{https://arxiv.org/abs/1903.01609}{{\ttfamily 1903.01609}}].

\bibitem{GGaswintPhD}
G.~Gaswint, \emph{Ph.d dissertation, {University of California-Irvine}},  2021.

\bibitem{WellingIFT2021}
C.~Welling, P.~Frank, T.~A. En\ss{}lin and A.~Nelles, \emph{{Reconstructing
  non-repeating radio pulses with Information Field Theory}},
  \href{http://dx.doi.org/10.1088/1475-7516/2021/04/071}{\emph{JCAP} {\bfseries
  04} (2021) 071}, [\href{https://arxiv.org/abs/2102.00258}{{\ttfamily
  2102.00258}}].

\bibitem{Gaisser2019-AtmNu}
T.~K. Gaisser, \emph{{Atmospheric Neutrinos}},
  \href{https://arxiv.org/abs/1910.08851}{{\ttfamily 1910.08851}}.

\bibitem{Arianna:2021vcx}
{\scshape Arianna} collaboration, A.~Anker et~al., \emph{{Improving sensitivity
  of the ARIANNA detector by rejecting thermal noise with deep learning}},
  \href{http://dx.doi.org/10.1088/1748-0221/17/03/P03007}{\emph{JINST}
  {\bfseries 17} (2022) P03007},
  [\href{https://arxiv.org/abs/2112.01031}{{\ttfamily 2112.01031}}].

\bibitem{ICRC2021Anker}
{A. Anker for the ARIANNA collaboration}, \emph{A novel trigger based on neural
  networks for radio neutrino detectors},
  \href{http://dx.doi.org/10.22323/1.395.1069}{\emph{PoS(ICRC2021)1069} }.

\bibitem{ARA2020-limit}
{\scshape ARA} collaboration, P.~Allison et~al., \emph{{Constraints on the
  diffuse flux of ultrahigh energy neutrinos from four years of Askaryan Radio
  Array data in two stations}},
  \href{http://dx.doi.org/10.1103/PhysRevD.102.043021}{\emph{Phys. Rev. D}
  {\bfseries 102} (2020) 043021},
  [\href{https://arxiv.org/abs/1912.00987}{{\ttfamily 1912.00987}}].

\bibitem{ARA:2022rwq}
{\scshape ARA} collaboration, P.~Allison et~al., \emph{{Low-threshold
  ultrahigh-energy neutrino search with the Askaryan Radio Array}},
  \href{http://dx.doi.org/10.1103/PhysRevD.105.122006}{\emph{Phys. Rev. D}
  {\bfseries 105} (2022) 122006},
  [\href{https://arxiv.org/abs/2202.07080}{{\ttfamily 2202.07080}}].

\bibitem{Aguilar:2021voo}
J.~A. Aguilar et~al., \emph{{Triboelectric Backgrounds to radio-based UHE
  Neutrino Experiments}},  \href{https://arxiv.org/abs/2103.06079}{{\ttfamily
  2103.06079}}.

\bibitem{Anker:ARIANNAlimit2019}
A.~Anker et~al., \emph{{A search for cosmogenic neutrinos with the ARIANNA test
  bed using 4.5 years of data}},
  \href{http://dx.doi.org/10.1088/1475-7516/2020/03/053}{\emph{JCAP} {\bfseries
  03} (2020) 053}, [\href{https://arxiv.org/abs/1909.00840}{{\ttfamily
  1909.00840}}].

\bibitem{ARAICRC2021}
{K. Hughes for the ARA collaboration}, \emph{Implementing a low-threshold
  analysis with the {Askaryan Radio Array (ARA)}},
  \href{http://dx.doi.org/10.22323/1.395.1153}{\emph{PoS(ICRC2021)1153} }.

\bibitem{leshan_zhao_2022_6785194}
{L. Zhao for the ARIANNA collaboration}, \emph{{Novel Background Rejection
  Techniques in a Search for Ultra-high Energy Neutrinos Using an ARIANNA
  Detector Station at the South Pole}},
  \href{http://dx.doi.org/10.5281/zenodo.6785194}{\emph{Neutrino 2022, Soul,
  South Korea} (2022) }.

\bibitem{Fujita-reflection}
S.~Fujita et~al., \emph{{A summary of the complex dielectric permittivity of
  ice in the megahertz range and its applications for radar sounding of polar
  ice sheets}}, {\emph{International Symposium on Physics of Ice Core Records.
  Shikotsukohan, Hokkaido, Japan, September 14-17, 1998.} (2000) 185--212}.

\bibitem{Besson2010-biref-reflect}
D.~Besson, I.~Kravchenko, A.~Ramos and J.~Remmers, \emph{{Radio Frequency
  Birefringence in South Polar Ice and Implications for Neutrino
  Reconstruction}},
  \href{http://dx.doi.org/10.1016/j.astropartphys.2011.01.008}{\emph{Astropart.
  Phys.} {\bfseries 34} (2011) 755--768},
  [\href{https://arxiv.org/abs/1005.4589}{{\ttfamily 1005.4589}}].

\bibitem{Bay:ReflectionLayers}
{\emph{\url{http://icecube.berkeley.edu/~bay/dustmap/}} }.

\bibitem{DeKockere:2022bto}
S.~De~Kockere, K.~D. de~Vries, N.~van Eijndhoven and U.~A. Latif,
  \emph{{Simulation of in-ice cosmic ray air shower induced particle
  cascades}},  \href{https://arxiv.org/abs/2202.09211}{{\ttfamily 2202.09211}}.

\bibitem{Price:RadioComparison}
P.~B. Price, \emph{{Comparison of optical, radio, and acoustical detectors for
  ultrahigh-energy neutrinos}},
  \href{http://dx.doi.org/10.1016/0927-6505(96)00004-7}{\emph{Astropart. Phys.}
  {\bfseries 5} (1996) 43--52},
  [\href{https://arxiv.org/abs/astro-ph/9510119}{{\ttfamily
  astro-ph/9510119}}].

\bibitem{RICE2003-limits}
I.~Kravchenko et~al., \emph{{Rice limits on the diffuse ultrahigh energy
  neutrino flux}},
  \href{http://dx.doi.org/10.1103/PhysRevD.73.082002}{\emph{Phys. Rev. D}
  {\bfseries 73} (2006) 082002},
  [\href{https://arxiv.org/abs/astro-ph/0601148}{{\ttfamily
  astro-ph/0601148}}].

\bibitem{ANITA2008-Design}
{\scshape ANITA} collaboration, P.~W. Gorham et~al., \emph{{The Antarctic
  Impulsive Transient Antenna Ultra-high Energy Neutrino Detector Design,
  Performance, and Sensitivity for 2006-2007 Balloon Flight}},
  \href{http://dx.doi.org/10.1016/j.astropartphys.2009.05.003}{\emph{Astropart.
  Phys.} {\bfseries 32} (2009) 10--41},
  [\href{https://arxiv.org/abs/0812.1920}{{\ttfamily 0812.1920}}].

\bibitem{ANITA-2ndFlight}
{\scshape ANITA} collaboration, P.~W. Gorham et~al., \emph{{Observational
  Constraints on the Ultra-high Energy Cosmic Neutrino Flux from the Second
  Flight of the ANITA Experiment}},
  \href{http://dx.doi.org/10.1103/PhysRevD.82.022004}{\emph{Phys. Rev. D}
  {\bfseries 82} (2010) 022004},
  [\href{https://arxiv.org/abs/1003.2961}{{\ttfamily 1003.2961}}].

\bibitem{Barwick:ARIANNA2014}
{\scshape ARIANNA} collaboration, S.~W. Barwick et~al., \emph{{A First Search
  for Cosmogenic Neutrinos with the ARIANNA Hexagonal Radio Array}},
  \href{http://dx.doi.org/10.1016/j.astropartphys.2015.04.002}{\emph{Astropart.
  Phys.} {\bfseries 70} (2015) 12--26},
  [\href{https://arxiv.org/abs/1410.7352}{{\ttfamily 1410.7352}}].

\bibitem{Design:ARIANNA2014}
{\scshape ARIANNA} collaboration, S.~W. Barwick et~al., \emph{{Design and
  Performance of the ARIANNA HRA-3 Neutrino Detector Systems}}, {\emph{IEEE
  Trans. Nucl. Sci.} {\bfseries 62} (2015) 2202},
  [\href{https://arxiv.org/abs/1410.7369}{{\ttfamily 1410.7369}}].

\bibitem{ARA2011}
P.~Allison et~al., \emph{{Design and Initial Performance of the Askaryan Radio
  Array Prototype EeV Neutrino Detector at the South Pole}},
  \href{http://dx.doi.org/10.1016/j.astropartphys.2011.11.010}{\emph{Astropart.
  Phys.} {\bfseries 35} (2012) 457--477},
  [\href{https://arxiv.org/abs/1105.2854}{{\ttfamily 1105.2854}}].

\bibitem{ANITA-4thFlight}
{\scshape ANITA} collaboration, P.~W. Gorham et~al., \emph{{Constraints on the
  ultrahigh-energy cosmic neutrino flux from the fourth flight of ANITA}},
  \href{http://dx.doi.org/10.1103/PhysRevD.99.122001}{\emph{Phys. Rev. D}
  {\bfseries 99} (2019) 122001},
  [\href{https://arxiv.org/abs/1902.04005}{{\ttfamily 1902.04005}}].

\bibitem{ANITA2010-CR}
{\scshape ANITA} collaboration, S.~Hoover et~al., \emph{{Observation of
  Ultra-high-energy Cosmic Rays with the ANITA Balloon-borne Radio
  Interferometer}},
  \href{http://dx.doi.org/10.1103/PhysRevLett.105.151101}{\emph{Phys. Rev.
  Lett.} {\bfseries 105} (2010) 151101},
  [\href{https://arxiv.org/abs/1005.0035}{{\ttfamily 1005.0035}}].

\bibitem{ANITA2016-CR}
H.~Schoorlemmer et~al., \emph{{Energy and Flux Measurements of Ultra-High
  Energy Cosmic Rays Observed During the First ANITA Flight}},
  \href{http://dx.doi.org/10.1016/j.astropartphys.2016.01.001}{\emph{Astropart.
  Phys.} {\bfseries 77} (2016) 32--43},
  [\href{https://arxiv.org/abs/1506.05396}{{\ttfamily 1506.05396}}].

\bibitem{ANITA2016-4up}
{\scshape ANITA} collaboration, P.~W. Gorham et~al., \emph{{Characteristics of
  Four Upward-pointing Cosmic-ray-like Events Observed with ANITA}},
  \href{http://dx.doi.org/10.1103/PhysRevLett.117.071101}{\emph{Phys. Rev.
  Lett.} {\bfseries 117} (2016) 071101},
  [\href{https://arxiv.org/abs/1603.05218}{{\ttfamily 1603.05218}}].

\bibitem{ANITA2019-upward}
{\scshape ANITA} collaboration, P.~W. Gorham et~al., \emph{{Observation of an
  Unusual Upward-going Cosmic-ray-like Event in the Third Flight of ANITA}},
  \href{http://dx.doi.org/10.1103/PhysRevLett.121.161102}{\emph{Phys. Rev.
  Lett.} {\bfseries 121} (2018) 161102},
  [\href{https://arxiv.org/abs/1803.05088}{{\ttfamily 1803.05088}}].

\bibitem{ANITA2019-noTau}
A.~Romero-Wolf et~al., \emph{{Comprehensive analysis of anomalous ANITA events
  disfavors a diffuse tau-neutrino flux origin}},
  \href{http://dx.doi.org/10.1103/PhysRevD.99.063011}{\emph{Phys. Rev. D}
  {\bfseries 99} (2019) 063011},
  [\href{https://arxiv.org/abs/1811.07261}{{\ttfamily 1811.07261}}].

\bibitem{IceCubeICRC2019-ANITA}
I.~Safa, A.~Pizzuto, C.~A. Arg\"uelles, F.~Halzen, R.~Hussain, A.~Kheirandish
  et~al., \emph{{Constraining anomalous EeV ANITA detections with PeV
  neutrinos}}, \href{http://dx.doi.org/10.22323/1.358.0995}{\emph{PoS}
  {\bfseries ICRC2019} (2019) 995}.

\bibitem{IceCube2020-ANITA}
{\scshape IceCube} collaboration, M.~G. Aartsen et~al., \emph{A search for
  {IceCube} events in the direction of {ANITA} neutrino candidates},
  \href{http://dx.doi.org/10.3847/1538-4357/ab791d}{\emph{The Astrophysical
  Journal} {\bfseries 892} (2020) 53},
  [\href{https://arxiv.org/abs/2001.01737}{{\ttfamily 2001.01737}}].

\bibitem{deVries2019}
K.~D. de~Vries and S.~Prohira, \emph{{Coherent transition radiation from the
  geomagnetically-induced current in cosmic-ray air showers: Implications for
  the anomalous events observed by ANITA}},
  \href{http://dx.doi.org/10.1103/PhysRevLett.123.091102}{\emph{Phys. Rev.
  Lett.} {\bfseries 123} (2019) 091102},
  [\href{https://arxiv.org/abs/1903.08750}{{\ttfamily 1903.08750}}].

\bibitem{Shoemaker2019}
I.~M. Shoemaker, A.~Kusenko, P.~K. Munneke, A.~Romero-Wolf, D.~M. Schroeder and
  M.~J. Siegert, \emph{{Reflections On the Anomalous ANITA Events: The
  Antarctic Subsurface as a Possible Explanation}},
  \href{http://dx.doi.org/10.1017/aog.2020.19}{\emph{Annals Glaciol.}
  {\bfseries 61} (2020) 92--98},
  [\href{https://arxiv.org/abs/1905.02846}{{\ttfamily 1905.02846}}].

\bibitem{Nelles:ICRC2019-WindTurbine}
{A. Nelles for the ARIANNA collaboration}, \emph{{A wind-turbine for autonomous
  stations for radio detection of neutrinos}},
  \href{http://dx.doi.org/10.22323/1.358.0968}{\emph{PoS} {\bfseries ICRC2019}
  (2019) 968}.

\bibitem{ARIANNAICRC2021CosmicRays}
{L. Zhao for the ARIANNA collaboration}, \emph{Polarization reconstruction of
  cosmic rays with the arianna neutrino radio detector},
  \href{http://dx.doi.org/10.22323/1.395.1156}{\emph{PoS(ICRC2021)1156} }.

\bibitem{ARIANNAICRC2021Direction}
{S. Barwick for the ARIANNA collaboration}, \emph{Capabilities of arianna:
  Neutrino pointing resolution and implications for future ultra-high energy
  neutrino astronomy},
  \href{http://dx.doi.org/10.22323/1.395.1151}{\emph{PoS(ICRC2021)1151} }.

\bibitem{SPICEhole}
K.~Casey et~al., \emph{{The 1500 m South Pole ice core: recovering a 40 ka
  environmental record}}, {\emph{Ann. of Glac.} {\bfseries 55} (2014)
  137--146}.

\bibitem{PUEOICRC2021}
{A.G. Vieregg for the PUEO collaboration}, \emph{Discovering the highest energy
  neutrinos with the payload for ultrahigh energy observations {(PUEO)}},
  \href{http://dx.doi.org/10.22323/1.395.1029}{\emph{PoS(ICRC2021)1029} }.

\bibitem{IceCubeGen2-2020}
{The IceCube-Gen2 Collaboration et al.}, \emph{{IceCube-Gen2: The Window to the
  Extreme Universe}},
  \href{http://dx.doi.org/10.1088/1361-6471/abbd48}{\emph{Journal of Physics G:
  Nuclear and Particle Physics} {\bfseries 48} (2021) 060501},
  [\href{https://arxiv.org/abs/2008.04323}{{\ttfamily 2008.04323}}].

\bibitem{icecube_spectrum}
{\scshape IceCube} collaboration, J.~Stettner, \emph{{Measurement of the
  Diffuse Astrophysical Muon-Neutrino Spectrum with Ten Years of IceCube
  Data}}, \href{http://dx.doi.org/10.22323/1.358.1017}{\emph{PoS} {\bfseries
  ICRC2019} 1017}, [\href{https://arxiv.org/abs/1908.09551}{{\ttfamily
  1908.09551}}].

\bibitem{AugerNuLimit2019}
{\scshape Auger} collaboration, A.~Aab et~al., \emph{Probing the origin of
  ultra-high-energy cosmic rays with neutrinos in the {EeV} energy range using
  the {Pierre Auger Observatory}},
  \href{http://dx.doi.org/10.1088/1475-7516/2019/10/022}{\emph{Journal of
  Cosmology and Astroparticle Physics} {\bfseries 10} (2019) 022},
  [\href{https://arxiv.org/abs/1906.07422}{{\ttfamily 1906.07422}}].

\bibitem{IceCubeFlux2018}
M.~G. Aartsen et~al., \emph{Differential limit on the extremely-high-energy
  cosmic neutrino flux in the presence of astrophysical background from nine
  years of {IceCube} data},
  \href{http://dx.doi.org/10.1103/PhysRevD.98.062003}{\emph{Physical Review D}
  {\bfseries 98} (2018) 062003},
  [\href{https://arxiv.org/abs/1807.01820}{{\ttfamily 1807.01820}}].

\bibitem{GrandWhitePaper2018}
{GRAND Collaboration}, J.~Alvarez-Muniz, R.~A. Batista, A.~B. V., J.~Bolmont,
  M.~Bustamante et~al., \emph{{The Giant Radio Array for Neutrino Detection
  (GRAND): Science and Design}},
  \href{https://arxiv.org/abs/1810.09994}{{\ttfamily 1810.09994}}.

\bibitem{Vliet2019}
A.~van Vliet, R.~A. Batista and J.~R. Hörandel, \emph{Determining the fraction
  of cosmic-ray protons at ultra-high energies with cosmogenic neutrinos},
  \href{http://dx.doi.org/10.1103/physrevd.100.021302}{\emph{Physical Review D}
  {\bfseries 100} 021302}, [\href{https://arxiv.org/abs/1901.01899}{{\ttfamily
  1901.01899}}].

\bibitem{BergmanICRC2021}
{D. Bergman for the Telescope Array collaboration}, \emph{{Telescope Array
  Combined Fit to Cosmic Ray Spectrum and Composition}},
  \href{http://dx.doi.org/10.22323/1.395.0338}{\emph{PoS(ICRC2021)0338} }.

\bibitem{Wang:2020bqr}
{\scshape TAROGE, ARIANNA} collaboration, S.-H. Wang, \emph{{Status,
  Calibration, and Cosmic Ray Detection of ARIANNA-HCR Station}},
  \href{http://dx.doi.org/10.22323/1.358.0462}{\emph{PoS} {\bfseries ICRC2019}
  (2020) 462}.

\bibitem{Liu:2020uyd}
T.~Liu, \emph{{The Status of the New Stations of Taiwan Astroparticle Radiowave
  Observatory for Geo-synchrotron Emissions(TAROGE)}},
  \href{http://dx.doi.org/10.22323/1.358.0341}{\emph{PoS} {\bfseries ICRC2019}
  (2020) 341}.

\bibitem{beacon2020}
S.~Wissel, A.~Romero-Wolf, H.~Schoorlemmer, W.~R.~C. Jr., J.~Alvarez-Muñiz,
  E.~Zas et~al., \emph{Prospects for high-elevation radio detection of $>$100
  {PeV} tau neutrinos},
  \href{http://dx.doi.org/10.1088/1475-7516/2020/11/065}{\emph{Journal of
  Cosmology and Astroparticle Physics} {\bfseries 11} (2020) 065},
  [\href{https://arxiv.org/abs/2004.12718}{{\ttfamily 2004.12718}}].

\bibitem{TAROGE:2022soh}
{\scshape TAROGE, ARIANNA} collaboration, A.~Anker et~al., \emph{{TAROGE-M:
  Radio Antenna Array on Antarctic High Mountain for Detecting Near-Horizontal
  Ultra-High Energy Air Showers}},
  \href{https://arxiv.org/abs/2207.10616}{{\ttfamily 2207.10616}}.

\bibitem{ARDOUIN2011717}
D.~Ardouin, C.~Cârloganu, D.~Charrier, Q.~Gou, H.~Hu, L.~Kai et~al.,
  \emph{First detection of extensive air showers by the trend self-triggering
  radio experiment},
  \href{http://dx.doi.org/https://doi.org/10.1016/j.astropartphys.2011.01.002}{\emph{Astroparticle
  Physics} {\bfseries 34} (2011) 717--731}.

\bibitem{Monroe2019}
R.~Monroe, A.~R. Wolf, G.~Hallinan, A.~Nelles, M.~Eastwood, M.~Anderson et~al.,
  \emph{Self-triggered radio detection and identification of cosmic air showers
  with the ovro-lwa},
  \href{http://dx.doi.org/10.1016/j.nima.2019.163086}{\emph{NIM-A} {\bfseries
  953} 163086}, [\href{https://arxiv.org/abs/1907.10193}{{\ttfamily
  1907.10193}}].

\bibitem{Prohira2020}
S.~Prohira et~al., \emph{Observation of radar echoes from high-energy particle
  cascades},
  \href{http://dx.doi.org/10.1103/PhysRevLett.124.091101}{\emph{Phys. Rev.
  Lett.} {\bfseries 124} (2020) 091101},
  [\href{https://arxiv.org/abs/1910.12830}{{\ttfamily 1910.12830}}].

\bibitem{Prohira2021}
S.~Prohira, K.~D. de~Vries, P.~Allison, J.~Beatty, D.~Besson, A.~Connolly
  et~al., \emph{The radar echo telescope for cosmic rays: Pathfinder experiment
  for a next-generation neutrino observatory},
  \href{http://dx.doi.org/10.1103/PhysRevD.104.102006}{\emph{Phys. Rev. D}
  {\bfseries 104} 102006}, [\href{https://arxiv.org/abs/2104.00459}{{\ttfamily
  2104.00459}}].

\bibitem{Miller:PRIDE}
T.~Miller et~al., \emph{{PRIDE (Passive Radio [frequency] Ice Depth
  Experiment): An instrument to passively measure ice depth from a Europan
  orbiter using neutrinos}}, {\emph{Icarus} {\bfseries 220} (2012) 877--888}.

\end{thebibliography}\endgroup
                      % to print subject index
\end{document}